\documentclass[runningheads]{llncs}

\usepackage{bm}
\usepackage{pgfplots}
\pgfplotsset{compat=1.18}
\usepgfplotslibrary{colorbrewer, statistics}
\usepackage{pgfplotstable}

\usepackage{tcolorbox}
\usepackage{xspace}
\usepackage{color}
\usepackage[table]{xcolor}
\usepackage{hyperref}
\usepackage{enumitem}
\hypersetup{linkcolor=black,citecolor=black,anchorcolor=black,filecolor=black,menucolor=black,runcolor=black,urlcolor=black,hidelinks}
\usepackage{url}
\usepackage{breakurl}
\usepackage{cite}
\usepackage[T1]{fontenc}
\usepackage{seqsplit}
\usepackage{graphicx}

\usepackage[font=small]{caption}
\usepackage{subcaption}
\usepackage{booktabs}
\usepackage{arydshln}
\usepackage{wrapfig}
\usepackage{pifont}
\usepackage{amssymb}
\usepackage[round-mode=places,round-precision=1,group-separator={,},group-minimum-digits={3},output-decimal-marker={.}, round-pad=false, exponent-mode = threshold, exponent-thresholds = -3:5, tight-spacing=true]{siunitx} %

\usepackage{tikz}
\tikzset{>=latex}
\usetikzlibrary{calc}
\usetikzlibrary{shapes.geometric}
\usetikzlibrary{decorations.pathreplacing}
\usetikzlibrary{positioning}
\usetikzlibrary{automata,positioning}

\newcommand*\circled[1]{\tikz[baseline=(char.base)]{
            \node[shape=circle,draw,fill=black,text=white,inner sep=0.4pt] (char) {#1};}}

\definecolor{editorGreen}{cmyk}{0.44, 0, 0.75, 0.25}
\definecolor{editorPink}{cmyk}{0, 1, 0, 0}
\definecolor{editorPurple}{cmyk}{0.5, 1, 0, 0}
\definecolor{orange}{cmyk}{1, .27, 0, 0}

\usepackage[]{algorithm}
\usepackage{algcompatible}
\newfloat{algorithm}{t}{lop}
\usepackage[noend]{algpseudocode}
\algrenewcommand\algorithmicindent{.5em}%
\newcommand{\AlgoComment}[1]{\Comment{{\small\color{blue}#1}}}

\algnewcommand{\IIf}[1]{\State\algorithmicif\ #1\ \algorithmicthen}
\algnewcommand{\EndIIf}{\unskip}
\algnewcommand{\ElsIIf}{\unskip\ \algorithmicelse\ \algorithmicif}
\algnewcommand{\ElseI}{\unskip\ \algorithmicelse\ \algorithmicif}

\newcommand{\NumProjectsFromAPISearch}{339}
\newcommand{\NumProjectsSearch}{1,124}

\newcommand{\NumTestsSum}{290,133}

\newcommand{\NumProjects}{1,463}

\newcommand{\NumSpecs}{73}
\newcommand{\NumPySpecs}{55}
\newcommand{\NumLibSpecs}{18}

\newcommand{\NumSpecsFromPyMOP}{69}
\newcommand{\NumSpecsFromOtherTools}{3}

\newcommand{\NumSpecsComparingTools}{22}

\newcommand{\NumSpecsMultObj}{12}

\newcommand{\NumTotalBugsFound}{93}

\newcommand{\NumTotalPRAccepted}{39}

\newcommand{\NumTotalIssuesAndPrAccepted}{44}

\newcommand{\NumTotalPRIssuesRejected}{12}

\newcommand{\NumTruePositive}{215}
\newcommand{\NumTruePositiveUniq}{121}
\newcommand{\PercTruePositiveUniq}{50.4\%}

\newcommand{\NumFalsePositive}{135}
\newcommand{\NumFalsePositiveUniq}{103}
\newcommand{\PercFalsePositiveUniq}{42.9\%}

\newcommand{\NumDifficultInspect}{16}
\newcommand{\NumDifficultInspectUniq}{16}
\newcommand{\PercDifficultInspectUniq}{6.7\%}

\newcommand{\NumUniqViolations}{240}
\newcommand{\NumAllViolations}{366}
\newcommand{\NumProjectInspection}{109}

\newcommand{\NumTrueBugsNotReported}{28}

\newcommand{\NumNoEffect}{6}
\newcommand{\NumCommittedOnPurpose}{10}
\newcommand{\NumDifficultToChange}{5}
\newcommand{\NumNotMaintained}{2}
\newcommand{\NumCodeChanged}{3}
\newcommand{\NumFixedInLatest}{2}

\newcommand{\Tool}{PyMOP\xspace}
\newcommand{\dynapyt}{DynaPyt\xspace}

\newcommand{\dylin}{DyLin\xspace}
\newcommand{\pythonrv}{PythonRV\xspace}
\newcommand{\vypr}{\textsc{VyPR2}\xspace}
\newcommand{\hatrv}{Hat-RV\xspace}
\newcommand{\javamop}{JavaMOP\xspace}
\newcommand{\rvmonitor}{RV-Monitor\xspace}
\newcommand{\RV}{RV\xspace}
\newcommand{\rv}{RV\xspace}
\newcommand{\specs}{specs\xspace}
\newcommand{\spec}{spec\xspace}
\newcommand{\Specs}{Specs\xspace}

\newcommand{\Fix}[1]{\textcolor{red}{[#1]}}
\newcommand{\raw}[1]{} %

\newcommand{\mh}[1]{\textcolor{editorGreen}{[Mohammed:~#1]}}

\newcommand{\Space}[1]{} %
\newcommand{\CodeIn}[1]{{\ifmmode{\mathtt{#1}}\else$\mathtt{#1}$\fi}}

\newcommand{\x}{x\xspace}
\newcommand{\Contrib}[1]{$\star$#1}
\newcommand{\MyPara}[1]{\vspace{1pt}\noindent\textbf{#1}.}
\newcommand{\MyParaOnly}[1]{\noindent\textbf{#1}}

\newcommand{\eg}{\textit{e.g.}\xspace}
\newcommand{\ie}{\textit{i.e.}\xspace}

\newcommand{\ArtifactLink}{\texttt{\url{https://github.com/allrob23/pymop-artifacts-rv}}}

\usepackage{listings}
\colorlet{punct}{red!60!black}
\definecolor{background}{HTML}{EEEEEE}
\definecolor{delim}{RGB}{20,105,176}
\colorlet{numb}{magenta!60!black}
\definecolor{blue}{rgb}{0.0, 0.0, 1.0}
\definecolor{red}{rgb}{1.0, 0.0, 0.0}
\definecolor{green}{rgb}{0.0, 0.5, 0.0}

\lstdefinelanguage{json}{
    basicstyle=\footnotesize\ttfamily,
    numbers=left,
    numberstyle=\scriptsize,
    stepnumber=1,
    numbersep=8pt,
    showstringspaces=false,
    breaklines=true,
    frame=lines,
    backgroundcolor=\color{background},
    keywords=[1]{true,false,null},
    keywordstyle=[1]\color{blue}\bfseries,
    string=[s]{"}{"},
    literate=
      *{0}{{{\color{numb}0}}}{1}
       {1}{{{\color{numb}1}}}{1}
       {2}{{{\color{numb}2}}}{1}
       {3}{{{\color{numb}3}}}{1}
       {4}{{{\color{numb}4}}}{1}
       {5}{{{\color{numb}5}}}{1}
       {6}{{{\color{numb}6}}}{1}
       {7}{{{\color{numb}7}}}{1}
       {8}{{{\color{numb}8}}}{1}
       {9}{{{\color{numb}9}}}{1}
       {:}{{{\color{punct}{:}}}}{1}
       {,}{{{\color{punct}{,}}}}{1}
       {\{}{{{\color{delim}{\{}}}}{1}
       {\}}{{{\color{delim}{\}}}}}{1}
       {[}{{{\color{delim}{[}}}}{1}
       {]}{{{\color{delim}{]}}}}{1},
}

\lstdefinelanguage{Python}{
    basicstyle=\footnotesize\ttfamily,
    numbers=left,
    numberstyle=\scriptsize,
    stepnumber=1,
    numbersep=8pt,
    showstringspaces=false,
    breaklines=true,
    frame=lines,
    backgroundcolor=\color{background},
    keywords=[1]{def,class,if,else,elif,for,while,try,except,with,return,import,from,as,in,not,and,or,True,False,None},
    keywordstyle=[1]\color{blue}\bfseries,
    string=[s]{"}{"},
    stringstyle=\color{red},
    comment=[l]{\#},
    commentstyle=\color{green}\itshape,
    literate=
      *{0}{{{\color{numb}0}}}{1}
       {1}{{{\color{numb}1}}}{1}
       {2}{{{\color{numb}2}}}{1}
       {3}{{{\color{numb}3}}}{1}
       {4}{{{\color{numb}4}}}{1}
       {5}{{{\color{numb}5}}}{1}
       {6}{{{\color{numb}6}}}{1}
       {7}{{{\color{numb}7}}}{1}
       {8}{{{\color{numb}8}}}{1}
       {9}{{{\color{numb}9}}}{1}
       {:}{{{\color{punct}{:}}}}{1}
       {,}{{{\color{punct}{,}}}}{1}
       {(}{{{\color{delim}{(}}}}{1}
       {)}{{{\color{delim}{)}}}}{1}
       {[}{{{\color{delim}{[}}}}{1}
       {]}{{{\color{delim}{]}}}}{1},
}

\newcommand{\oss}{open-source projects\xspace}
\newcommand{\MonitorCodeGenerator}{Monitor Synthesizer\xspace}
\newcommand{\InstrumentationEngine}{Instrumenter\xspace}
\newcommand{\MonitoringEngine}{Monitoring Engine\xspace}
\newcommand{\toctou}{TOCTOU\xspace}

\newcommand{\Dom}{\textsf{\footnotesize Dom}}

\newcommand{\A}{\mathbb{A}\langle X \rangle}
\newcommand{\T}{\mathbb{T}}
\newcommand{\B}{\mathbb{B}\langle X \rangle}
\newcommand{\CX}{\mathbb{C}\langle X \rangle}
\newcommand{\CXX}{\mathbb{C}^+\langle X \rangle}
\newcommand{\DX}{\mathbb{D}\langle X \rangle}

\newcommand{\iLMCS}{\iota}

\renewcommand{\i}{\iota}

\newcommand{\DefMacro}[2]{\expandafter\newcommand\csname #1\endcsname{#2}}
\newcommand{\UseMacro}[1]{\csname #1\endcsname}

\DefMacro{PydocsMustShutdownBeforeCloseSocket}{PydocsMustShutBeforeClose}
\DefMacro{PydocsShouldUseStreamWriterCorrectly}{PydocsUseWriterCorrectly}
\DefMacro{PydocsShouldNotInstantiateStreamWriter}{PydocsMakeNoWriter}
\DefMacro{PydocsMustShutdownThreadPoolExecutor}{PydocsShutPoolExecutor}

\newcommand{\rqAlgos}{RQ1\xspace}
\newcommand{\rqInstrumentation}{RQ2\xspace}
\newcommand{\rqEffectiveness}{RQ3\xspace}
\newcommand{\rqEfficiency}{RQ4\xspace}

\DefMacro{algo-boxplot-b-min}{0.370}
\DefMacro{algo-boxplot-b-Q1}{1.165}
\DefMacro{algo-boxplot-b-median}{3.630}
\DefMacro{algo-boxplot-b-Q3}{12.415}
\DefMacro{algo-boxplot-b-max}{82.320}
\DefMacro{algo-boxplot-b-mean}{10.250}
\DefMacro{algo-boxplot-b-sum}{13110.060}

\DefMacro{algo-boxplot-c-min}{0.370}
\DefMacro{algo-boxplot-c-Q1}{1.120}
\DefMacro{algo-boxplot-c-median}{2.980}
\DefMacro{algo-boxplot-c-Q3}{9.530}
\DefMacro{algo-boxplot-c-max}{57.120}
\DefMacro{algo-boxplot-c-mean}{7.917}
\DefMacro{algo-boxplot-c-sum}{10125.980}

\DefMacro{algo-boxplot-cplus-min}{0.370}
\DefMacro{algo-boxplot-cplus-Q1}{1.110}
\DefMacro{algo-boxplot-cplus-median}{2.980}
\DefMacro{algo-boxplot-cplus-Q3}{9.585}
\DefMacro{algo-boxplot-cplus-max}{55.130}
\DefMacro{algo-boxplot-cplus-mean}{7.953}
\DefMacro{algo-boxplot-cplus-sum}{10171.350}

\DefMacro{algo-boxplot-d-min}{0.360}
\DefMacro{algo-boxplot-d-Q1}{1.080}
\DefMacro{algo-boxplot-d-median}{2.810}
\DefMacro{algo-boxplot-d-Q3}{8.465}
\DefMacro{algo-boxplot-d-max}{50.080}
\DefMacro{algo-boxplot-d-mean}{6.978}
\DefMacro{algo-boxplot-d-sum}{8925.240}

\DefMacro{algo-boxplot-no-iqr-b-min}{0.370}
\DefMacro{algo-boxplot-no-iqr-b-Q1}{1.480}
\DefMacro{algo-boxplot-no-iqr-b-median}{6.080}
\DefMacro{algo-boxplot-no-iqr-b-Q3}{34.330}
\DefMacro{algo-boxplot-no-iqr-b-max}{17405.390}
\DefMacro{algo-boxplot-no-iqr-b-mean}{275.726}
\DefMacro{algo-boxplot-no-iqr-b-sum}{430408.960}

\DefMacro{algo-boxplot-no-iqr-c-min}{0.370}
\DefMacro{algo-boxplot-no-iqr-c-Q1}{1.340}
\DefMacro{algo-boxplot-no-iqr-c-median}{5.000}
\DefMacro{algo-boxplot-no-iqr-c-Q3}{25.320}
\DefMacro{algo-boxplot-no-iqr-c-max}{12353.700}
\DefMacro{algo-boxplot-no-iqr-c-mean}{106.284}
\DefMacro{algo-boxplot-no-iqr-c-sum}{165909.390}

\DefMacro{algo-boxplot-no-iqr-cplus-min}{0.370}
\DefMacro{algo-boxplot-no-iqr-cplus-Q1}{1.360}
\DefMacro{algo-boxplot-no-iqr-cplus-median}{5.050}
\DefMacro{algo-boxplot-no-iqr-cplus-Q3}{25.280}
\DefMacro{algo-boxplot-no-iqr-cplus-max}{12497.560}
\DefMacro{algo-boxplot-no-iqr-cplus-mean}{106.189}
\DefMacro{algo-boxplot-no-iqr-cplus-sum}{165761.270}

\DefMacro{algo-boxplot-no-iqr-d-min}{0.360}
\DefMacro{algo-boxplot-no-iqr-d-Q1}{1.330}
\DefMacro{algo-boxplot-no-iqr-d-median}{4.570}
\DefMacro{algo-boxplot-no-iqr-d-Q3}{20.840}
\DefMacro{algo-boxplot-no-iqr-d-max}{8050.600}
\DefMacro{algo-boxplot-no-iqr-d-mean}{71.733}
\DefMacro{algo-boxplot-no-iqr-d-sum}{111975.010}

\DefMacro{overhead-total-p1-b-rel}{1716.07}
\DefMacro{overhead-mean-p1-b-rel}{11.92}
\DefMacro{overhead-max-p1-b-rel}{36.5}
\DefMacro{overhead-min-p1-b-rel}{0.25}
\DefMacro{overhead-median-p1-b-rel}{9.89}
\DefMacro{overhead-total-p1-c-rel}{1577.79}
\DefMacro{overhead-mean-p1-c-rel}{10.96}
\DefMacro{overhead-max-p1-c-rel}{33.0}
\DefMacro{overhead-min-p1-c-rel}{0.2}
\DefMacro{overhead-median-p1-c-rel}{9.15}
\DefMacro{overhead-total-p1-cplus-rel}{1522.91}
\DefMacro{overhead-mean-p1-cplus-rel}{10.58}
\DefMacro{overhead-max-p1-cplus-rel}{32.0}
\DefMacro{overhead-min-p1-cplus-rel}{0.19}
\DefMacro{overhead-median-p1-cplus-rel}{8.71}
\DefMacro{overhead-total-p1-d-rel}{1520.31}
\DefMacro{overhead-mean-p1-d-rel}{10.56}
\DefMacro{overhead-max-p1-d-rel}{31.5}
\DefMacro{overhead-min-p1-d-rel}{0.18}
\DefMacro{overhead-median-p1-d-rel}{8.64}
\DefMacro{overhead-total-p2-b-rel}{1339.18}
\DefMacro{overhead-mean-p2-b-rel}{9.36}
\DefMacro{overhead-max-p2-b-rel}{34.0}
\DefMacro{overhead-min-p2-b-rel}{1.1}
\DefMacro{overhead-median-p2-b-rel}{7.75}
\DefMacro{overhead-total-p2-c-rel}{1220.62}
\DefMacro{overhead-mean-p2-c-rel}{8.54}
\DefMacro{overhead-max-p2-c-rel}{29.33}
\DefMacro{overhead-min-p2-c-rel}{1.06}
\DefMacro{overhead-median-p2-c-rel}{7.07}
\DefMacro{overhead-total-p2-cplus-rel}{1173.74}
\DefMacro{overhead-mean-p2-cplus-rel}{8.21}
\DefMacro{overhead-max-p2-cplus-rel}{28.67}
\DefMacro{overhead-min-p2-cplus-rel}{1.08}
\DefMacro{overhead-median-p2-cplus-rel}{6.77}
\DefMacro{overhead-total-p2-d-rel}{1200.81}
\DefMacro{overhead-mean-p2-d-rel}{8.4}
\DefMacro{overhead-max-p2-d-rel}{30.0}
\DefMacro{overhead-min-p2-d-rel}{1.08}
\DefMacro{overhead-median-p2-d-rel}{6.77}
\DefMacro{overhead-total-p3-b-rel}{1023.02}
\DefMacro{overhead-mean-p3-b-rel}{7.1}
\DefMacro{overhead-max-p3-b-rel}{33.0}
\DefMacro{overhead-min-p3-b-rel}{1.05}
\DefMacro{overhead-median-p3-b-rel}{5.76}
\DefMacro{overhead-total-p3-c-rel}{881.69}
\DefMacro{overhead-mean-p3-c-rel}{6.12}
\DefMacro{overhead-max-p3-c-rel}{29.0}
\DefMacro{overhead-min-p3-c-rel}{1.05}
\DefMacro{overhead-median-p3-c-rel}{4.75}
\DefMacro{overhead-total-p3-cplus-rel}{846.99}
\DefMacro{overhead-mean-p3-cplus-rel}{5.88}
\DefMacro{overhead-max-p3-cplus-rel}{30.2}
\DefMacro{overhead-min-p3-cplus-rel}{1.06}
\DefMacro{overhead-median-p3-cplus-rel}{4.62}
\DefMacro{overhead-total-p3-d-rel}{833.32}
\DefMacro{overhead-mean-p3-d-rel}{5.79}
\DefMacro{overhead-max-p3-d-rel}{29.0}
\DefMacro{overhead-min-p3-d-rel}{1.03}
\DefMacro{overhead-median-p3-d-rel}{4.44}
\DefMacro{overhead-total-p4-b-rel}{1273.68}
\DefMacro{overhead-mean-p4-b-rel}{8.85}
\DefMacro{overhead-max-p4-b-rel}{273.83}
\DefMacro{overhead-min-p4-b-rel}{1.05}
\DefMacro{overhead-median-p4-b-rel}{5.65}
\DefMacro{overhead-total-p4-c-rel}{760.2}
\DefMacro{overhead-mean-p4-c-rel}{5.28}
\DefMacro{overhead-max-p4-c-rel}{22.33}
\DefMacro{overhead-min-p4-c-rel}{1.04}
\DefMacro{overhead-median-p4-c-rel}{4.48}
\DefMacro{overhead-total-p4-cplus-rel}{756.83}
\DefMacro{overhead-mean-p4-cplus-rel}{5.26}
\DefMacro{overhead-max-p4-cplus-rel}{23.33}
\DefMacro{overhead-min-p4-cplus-rel}{1.04}
\DefMacro{overhead-median-p4-cplus-rel}{4.39}
\DefMacro{overhead-total-p4-d-rel}{734.26}
\DefMacro{overhead-mean-p4-d-rel}{5.1}
\DefMacro{overhead-max-p4-d-rel}{43.25}
\DefMacro{overhead-min-p4-d-rel}{1.04}
\DefMacro{overhead-median-p4-d-rel}{4.29}
\DefMacro{overhead-total-p5-b-rel}{1262.98}
\DefMacro{overhead-mean-p5-b-rel}{8.83}
\DefMacro{overhead-max-p5-b-rel}{129.04}
\DefMacro{overhead-min-p5-b-rel}{1.02}
\DefMacro{overhead-median-p5-b-rel}{6.07}
\DefMacro{overhead-total-p5-c-rel}{800.61}
\DefMacro{overhead-mean-p5-c-rel}{5.6}
\DefMacro{overhead-max-p5-c-rel}{36.5}
\DefMacro{overhead-min-p5-c-rel}{1.01}
\DefMacro{overhead-median-p5-c-rel}{4.52}
\DefMacro{overhead-total-p5-cplus-rel}{799.64}
\DefMacro{overhead-mean-p5-cplus-rel}{5.59}
\DefMacro{overhead-max-p5-cplus-rel}{35.88}
\DefMacro{overhead-min-p5-cplus-rel}{1.01}
\DefMacro{overhead-median-p5-cplus-rel}{4.45}
\DefMacro{overhead-total-p5-d-rel}{746.64}
\DefMacro{overhead-mean-p5-d-rel}{5.22}
\DefMacro{overhead-max-p5-d-rel}{34.25}
\DefMacro{overhead-min-p5-d-rel}{1.03}
\DefMacro{overhead-median-p5-d-rel}{4.15}
\DefMacro{overhead-total-p6-b-rel}{1889.25}
\DefMacro{overhead-mean-p6-b-rel}{13.12}
\DefMacro{overhead-max-p6-b-rel}{139.19}
\DefMacro{overhead-min-p6-b-rel}{1.01}
\DefMacro{overhead-median-p6-b-rel}{8.06}
\DefMacro{overhead-total-p6-c-rel}{1024.04}
\DefMacro{overhead-mean-p6-c-rel}{7.11}
\DefMacro{overhead-max-p6-c-rel}{33.9}
\DefMacro{overhead-min-p6-c-rel}{1.0}
\DefMacro{overhead-median-p6-c-rel}{5.42}
\DefMacro{overhead-total-p6-cplus-rel}{1026.38}
\DefMacro{overhead-mean-p6-cplus-rel}{7.13}
\DefMacro{overhead-max-p6-cplus-rel}{33.38}
\DefMacro{overhead-min-p6-cplus-rel}{1.0}
\DefMacro{overhead-median-p6-cplus-rel}{5.38}
\DefMacro{overhead-total-p6-d-rel}{914.6}
\DefMacro{overhead-mean-p6-d-rel}{6.35}
\DefMacro{overhead-max-p6-d-rel}{32.08}
\DefMacro{overhead-min-p6-d-rel}{1.0}
\DefMacro{overhead-median-p6-d-rel}{4.94}
\DefMacro{overhead-total-p7-b-rel}{2273.66}
\DefMacro{overhead-mean-p7-b-rel}{15.79}
\DefMacro{overhead-max-p7-b-rel}{165.76}
\DefMacro{overhead-min-p7-b-rel}{1.11}
\DefMacro{overhead-median-p7-b-rel}{10.57}
\DefMacro{overhead-total-p7-c-rel}{1167.57}
\DefMacro{overhead-mean-p7-c-rel}{8.11}
\DefMacro{overhead-max-p7-c-rel}{38.05}
\DefMacro{overhead-min-p7-c-rel}{1.27}
\DefMacro{overhead-median-p7-c-rel}{6.06}
\DefMacro{overhead-total-p7-cplus-rel}{1177.68}
\DefMacro{overhead-mean-p7-cplus-rel}{8.18}
\DefMacro{overhead-max-p7-cplus-rel}{38.32}
\DefMacro{overhead-min-p7-cplus-rel}{1.23}
\DefMacro{overhead-median-p7-cplus-rel}{6.56}
\DefMacro{overhead-total-p7-d-rel}{977.32}
\DefMacro{overhead-mean-p7-d-rel}{6.79}
\DefMacro{overhead-max-p7-d-rel}{37.05}
\DefMacro{overhead-min-p7-d-rel}{1.36}
\DefMacro{overhead-median-p7-d-rel}{5.46}
\DefMacro{overhead-total-p8-b-rel}{3714.35}
\DefMacro{overhead-mean-p8-b-rel}{25.79}
\DefMacro{overhead-max-p8-b-rel}{272.65}
\DefMacro{overhead-min-p8-b-rel}{1.02}
\DefMacro{overhead-median-p8-b-rel}{15.58}
\DefMacro{overhead-total-p8-c-rel}{1611.73}
\DefMacro{overhead-mean-p8-c-rel}{11.19}
\DefMacro{overhead-max-p8-c-rel}{51.52}
\DefMacro{overhead-min-p8-c-rel}{1.01}
\DefMacro{overhead-median-p8-c-rel}{8.38}
\DefMacro{overhead-total-p8-cplus-rel}{1630.54}
\DefMacro{overhead-mean-p8-cplus-rel}{11.32}
\DefMacro{overhead-max-p8-cplus-rel}{62.3}
\DefMacro{overhead-min-p8-cplus-rel}{1.02}
\DefMacro{overhead-median-p8-cplus-rel}{8.43}
\DefMacro{overhead-total-p8-d-rel}{1232.66}
\DefMacro{overhead-mean-p8-d-rel}{8.56}
\DefMacro{overhead-max-p8-d-rel}{40.47}
\DefMacro{overhead-min-p8-d-rel}{1.01}
\DefMacro{overhead-median-p8-d-rel}{6.03}
\DefMacro{overhead-total-p9-b-rel}{9564.22}
\DefMacro{overhead-mean-p9-b-rel}{66.88}
\DefMacro{overhead-max-p9-b-rel}{1442.69}
\DefMacro{overhead-min-p9-b-rel}{1.58}
\DefMacro{overhead-median-p9-b-rel}{31.63}
\DefMacro{overhead-total-p9-c-rel}{3354.03}
\DefMacro{overhead-mean-p9-c-rel}{23.45}
\DefMacro{overhead-max-p9-c-rel}{150.83}
\DefMacro{overhead-min-p9-c-rel}{1.26}
\DefMacro{overhead-median-p9-c-rel}{17.48}
\DefMacro{overhead-total-p9-cplus-rel}{3389.7}
\DefMacro{overhead-mean-p9-cplus-rel}{23.7}
\DefMacro{overhead-max-p9-cplus-rel}{145.33}
\DefMacro{overhead-min-p9-cplus-rel}{1.22}
\DefMacro{overhead-median-p9-cplus-rel}{18.38}
\DefMacro{overhead-total-p9-d-rel}{2156.69}
\DefMacro{overhead-mean-p9-d-rel}{15.08}
\DefMacro{overhead-max-p9-d-rel}{74.48}
\DefMacro{overhead-min-p9-d-rel}{1.17}
\DefMacro{overhead-median-p9-d-rel}{11.5}
\DefMacro{overhead-total-p10-b-rel}{24751.27}
\DefMacro{overhead-mean-p10-b-rel}{171.88}
\DefMacro{overhead-max-p10-b-rel}{5500.27}
\DefMacro{overhead-min-p10-b-rel}{1.09}
\DefMacro{overhead-median-p10-b-rel}{58.59}
\DefMacro{overhead-total-p10-c-rel}{9934.51}
\DefMacro{overhead-mean-p10-c-rel}{68.99}
\DefMacro{overhead-max-p10-c-rel}{870.71}
\DefMacro{overhead-min-p10-c-rel}{1.07}
\DefMacro{overhead-median-p10-c-rel}{24.67}
\DefMacro{overhead-total-p10-cplus-rel}{9577.36}
\DefMacro{overhead-mean-p10-cplus-rel}{66.51}
\DefMacro{overhead-max-p10-cplus-rel}{876.84}
\DefMacro{overhead-min-p10-cplus-rel}{1.09}
\DefMacro{overhead-median-p10-cplus-rel}{24.64}
\DefMacro{overhead-total-p10-d-rel}{6272.52}
\DefMacro{overhead-mean-p10-d-rel}{43.56}
\DefMacro{overhead-max-p10-d-rel}{848.76}
\DefMacro{overhead-min-p10-d-rel}{1.07}
\DefMacro{overhead-median-p10-d-rel}{12.1}
\DefMacro{fastest-count-p1-b}{3}
\DefMacro{fastest-count-p1-c}{7}
\DefMacro{fastest-count-p1-cplus}{75}
\DefMacro{fastest-count-p1-d}{59}
\DefMacro{fastest-count-p2-b}{0}
\DefMacro{fastest-count-p2-c}{15}
\DefMacro{fastest-count-p2-cplus}{73}
\DefMacro{fastest-count-p2-d}{55}
\DefMacro{fastest-count-p3-b}{0}
\DefMacro{fastest-count-p3-c}{13}
\DefMacro{fastest-count-p3-cplus}{54}
\DefMacro{fastest-count-p3-d}{77}
\DefMacro{fastest-count-p4-b}{0}
\DefMacro{fastest-count-p4-c}{24}
\DefMacro{fastest-count-p4-cplus}{25}
\DefMacro{fastest-count-p4-d}{95}
\DefMacro{fastest-count-p5-b}{0}
\DefMacro{fastest-count-p5-c}{18}
\DefMacro{fastest-count-p5-cplus}{24}
\DefMacro{fastest-count-p5-d}{101}
\DefMacro{fastest-count-p6-b}{0}
\DefMacro{fastest-count-p6-c}{17}
\DefMacro{fastest-count-p6-cplus}{22}
\DefMacro{fastest-count-p6-d}{105}
\DefMacro{fastest-count-p7-b}{1}
\DefMacro{fastest-count-p7-c}{17}
\DefMacro{fastest-count-p7-cplus}{9}
\DefMacro{fastest-count-p7-d}{117}
\DefMacro{fastest-count-p8-b}{0}
\DefMacro{fastest-count-p8-c}{8}
\DefMacro{fastest-count-p8-cplus}{7}
\DefMacro{fastest-count-p8-d}{129}
\DefMacro{fastest-count-p9-b}{0}
\DefMacro{fastest-count-p9-c}{11}
\DefMacro{fastest-count-p9-cplus}{3}
\DefMacro{fastest-count-p9-d}{129}
\DefMacro{fastest-count-p10-b}{1}
\DefMacro{fastest-count-p10-c}{11}
\DefMacro{fastest-count-p10-cplus}{3}
\DefMacro{fastest-count-p10-d}{129}
\DefMacro{fastest-count-1s-p1-b}{1}
\DefMacro{fastest-count-1s-p1-c}{1}
\DefMacro{fastest-count-1s-p1-cplus}{0}
\DefMacro{fastest-count-1s-p1-d}{5}
\DefMacro{fastest-count-1s-p1-all}{137}
\DefMacro{fastest-count-1s-p2-b}{0}
\DefMacro{fastest-count-1s-p2-c}{1}
\DefMacro{fastest-count-1s-p2-cplus}{0}
\DefMacro{fastest-count-1s-p2-d}{1}
\DefMacro{fastest-count-1s-p2-all}{141}
\DefMacro{fastest-count-1s-p3-b}{0}
\DefMacro{fastest-count-1s-p3-c}{0}
\DefMacro{fastest-count-1s-p3-cplus}{0}
\DefMacro{fastest-count-1s-p3-d}{2}
\DefMacro{fastest-count-1s-p3-all}{142}
\DefMacro{fastest-count-1s-p4-b}{0}
\DefMacro{fastest-count-1s-p4-c}{3}
\DefMacro{fastest-count-1s-p4-cplus}{0}
\DefMacro{fastest-count-1s-p4-d}{11}
\DefMacro{fastest-count-1s-p4-all}{130}
\DefMacro{fastest-count-1s-p5-b}{0}
\DefMacro{fastest-count-1s-p5-c}{4}
\DefMacro{fastest-count-1s-p5-cplus}{0}
\DefMacro{fastest-count-1s-p5-d}{35}
\DefMacro{fastest-count-1s-p5-all}{104}
\DefMacro{fastest-count-1s-p6-b}{0}
\DefMacro{fastest-count-1s-p6-c}{11}
\DefMacro{fastest-count-1s-p6-cplus}{0}
\DefMacro{fastest-count-1s-p6-d}{86}
\DefMacro{fastest-count-1s-p6-all}{47}
\DefMacro{fastest-count-1s-p7-b}{1}
\DefMacro{fastest-count-1s-p7-c}{13}
\DefMacro{fastest-count-1s-p7-cplus}{0}
\DefMacro{fastest-count-1s-p7-d}{110}
\DefMacro{fastest-count-1s-p7-all}{20}
\DefMacro{fastest-count-1s-p8-b}{0}
\DefMacro{fastest-count-1s-p8-c}{10}
\DefMacro{fastest-count-1s-p8-cplus}{0}
\DefMacro{fastest-count-1s-p8-d}{129}
\DefMacro{fastest-count-1s-p8-all}{5}
\DefMacro{fastest-count-1s-p9-b}{0}
\DefMacro{fastest-count-1s-p9-c}{12}
\DefMacro{fastest-count-1s-p9-cplus}{0}
\DefMacro{fastest-count-1s-p9-d}{130}
\DefMacro{fastest-count-1s-p9-all}{1}
\DefMacro{fastest-count-1s-p10-b}{1}
\DefMacro{fastest-count-1s-p10-c}{13}
\DefMacro{fastest-count-1s-p10-cplus}{0}
\DefMacro{fastest-count-1s-p10-d}{130}
\DefMacro{fastest-count-1s-p10-all}{0}
\DefMacro{fastest-count-2s-p1-b}{0}
\DefMacro{fastest-count-2s-p1-c}{1}
\DefMacro{fastest-count-2s-p1-cplus}{0}
\DefMacro{fastest-count-2s-p1-d}{3}
\DefMacro{fastest-count-2s-p1-all}{140}
\DefMacro{fastest-count-2s-p2-b}{0}
\DefMacro{fastest-count-2s-p2-c}{0}
\DefMacro{fastest-count-2s-p2-cplus}{0}
\DefMacro{fastest-count-2s-p2-d}{1}
\DefMacro{fastest-count-2s-p2-all}{142}
\DefMacro{fastest-count-2s-p3-b}{0}
\DefMacro{fastest-count-2s-p3-c}{0}
\DefMacro{fastest-count-2s-p3-cplus}{0}
\DefMacro{fastest-count-2s-p3-d}{1}
\DefMacro{fastest-count-2s-p3-all}{143}
\DefMacro{fastest-count-2s-p4-b}{0}
\DefMacro{fastest-count-2s-p4-c}{0}
\DefMacro{fastest-count-2s-p4-cplus}{0}
\DefMacro{fastest-count-2s-p4-d}{4}
\DefMacro{fastest-count-2s-p4-all}{140}
\DefMacro{fastest-count-2s-p5-b}{0}
\DefMacro{fastest-count-2s-p5-c}{1}
\DefMacro{fastest-count-2s-p5-cplus}{0}
\DefMacro{fastest-count-2s-p5-d}{6}
\DefMacro{fastest-count-2s-p5-all}{136}
\DefMacro{fastest-count-2s-p6-b}{0}
\DefMacro{fastest-count-2s-p6-c}{5}
\DefMacro{fastest-count-2s-p6-cplus}{0}
\DefMacro{fastest-count-2s-p6-d}{44}
\DefMacro{fastest-count-2s-p6-all}{95}
\DefMacro{fastest-count-2s-p7-b}{1}
\DefMacro{fastest-count-2s-p7-c}{6}
\DefMacro{fastest-count-2s-p7-cplus}{0}
\DefMacro{fastest-count-2s-p7-d}{98}
\DefMacro{fastest-count-2s-p7-all}{39}
\DefMacro{fastest-count-2s-p8-b}{0}
\DefMacro{fastest-count-2s-p8-c}{8}
\DefMacro{fastest-count-2s-p8-cplus}{0}
\DefMacro{fastest-count-2s-p8-d}{125}
\DefMacro{fastest-count-2s-p8-all}{11}
\DefMacro{fastest-count-2s-p9-b}{0}
\DefMacro{fastest-count-2s-p9-c}{10}
\DefMacro{fastest-count-2s-p9-cplus}{0}
\DefMacro{fastest-count-2s-p9-d}{130}
\DefMacro{fastest-count-2s-p9-all}{3}
\DefMacro{fastest-count-2s-p10-b}{1}
\DefMacro{fastest-count-2s-p10-c}{13}
\DefMacro{fastest-count-2s-p10-cplus}{0}
\DefMacro{fastest-count-2s-p10-d}{130}
\DefMacro{fastest-count-2s-p10-all}{0}
\DefMacro{fastest-count-5s-p1-b}{0}
\DefMacro{fastest-count-5s-p1-c}{1}
\DefMacro{fastest-count-5s-p1-cplus}{0}
\DefMacro{fastest-count-5s-p1-d}{3}
\DefMacro{fastest-count-5s-p1-all}{140}
\DefMacro{fastest-count-5s-p2-b}{0}
\DefMacro{fastest-count-5s-p2-c}{0}
\DefMacro{fastest-count-5s-p2-cplus}{0}
\DefMacro{fastest-count-5s-p2-d}{1}
\DefMacro{fastest-count-5s-p2-all}{142}
\DefMacro{fastest-count-5s-p3-b}{0}
\DefMacro{fastest-count-5s-p3-c}{0}
\DefMacro{fastest-count-5s-p3-cplus}{0}
\DefMacro{fastest-count-5s-p3-d}{0}
\DefMacro{fastest-count-5s-p3-all}{144}
\DefMacro{fastest-count-5s-p4-b}{0}
\DefMacro{fastest-count-5s-p4-c}{0}
\DefMacro{fastest-count-5s-p4-cplus}{0}
\DefMacro{fastest-count-5s-p4-d}{1}
\DefMacro{fastest-count-5s-p4-all}{143}
\DefMacro{fastest-count-5s-p5-b}{0}
\DefMacro{fastest-count-5s-p5-c}{0}
\DefMacro{fastest-count-5s-p5-cplus}{0}
\DefMacro{fastest-count-5s-p5-d}{2}
\DefMacro{fastest-count-5s-p5-all}{141}
\DefMacro{fastest-count-5s-p6-b}{0}
\DefMacro{fastest-count-5s-p6-c}{4}
\DefMacro{fastest-count-5s-p6-cplus}{0}
\DefMacro{fastest-count-5s-p6-d}{14}
\DefMacro{fastest-count-5s-p6-all}{126}
\DefMacro{fastest-count-5s-p7-b}{0}
\DefMacro{fastest-count-5s-p7-c}{0}
\DefMacro{fastest-count-5s-p7-cplus}{0}
\DefMacro{fastest-count-5s-p7-d}{53}
\DefMacro{fastest-count-5s-p7-all}{91}
\DefMacro{fastest-count-5s-p8-b}{0}
\DefMacro{fastest-count-5s-p8-c}{3}
\DefMacro{fastest-count-5s-p8-cplus}{0}
\DefMacro{fastest-count-5s-p8-d}{101}
\DefMacro{fastest-count-5s-p8-all}{40}
\DefMacro{fastest-count-5s-p9-b}{0}
\DefMacro{fastest-count-5s-p9-c}{5}
\DefMacro{fastest-count-5s-p9-cplus}{0}
\DefMacro{fastest-count-5s-p9-d}{126}
\DefMacro{fastest-count-5s-p9-all}{12}
\DefMacro{fastest-count-5s-p10-b}{1}
\DefMacro{fastest-count-5s-p10-c}{13}
\DefMacro{fastest-count-5s-p10-cplus}{0}
\DefMacro{fastest-count-5s-p10-d}{130}
\DefMacro{fastest-count-5s-p10-all}{0}
\DefMacro{fastest-all-count-b}{5}
\DefMacro{fastest-all-count-c}{141}
\DefMacro{fastest-all-count-cplus}{295}
\DefMacro{fastest-all-count-d}{996}
\DefMacro{fastest-all-percent-b}{0.3}
\DefMacro{fastest-all-percent-c}{9.8}
\DefMacro{fastest-all-percent-c+}{20.5}
\DefMacro{fastest-all-percent-d}{69.3}
\DefMacro{fastest-count-2s-all-b}{2}
\DefMacro{fastest-count-2s-all-c}{44}
\DefMacro{fastest-count-2s-all-d}{542}
\DefMacro{fastest-count-2s-all-all}{849}
\DefMacro{fastest-count-2s-all-percent-b}{0.1}
\DefMacro{fastest-count-2s-all-percent-c}{3.1}
\DefMacro{fastest-count-2s-all-percent-d}{37.7}
\DefMacro{fastest-count-2s-all-percent-all}{59.1}
\DefMacro{fastest-count-5s-all-b}{1}
\DefMacro{fastest-count-5s-all-c}{26}
\DefMacro{fastest-count-5s-all-d}{431}
\DefMacro{fastest-count-5s-all-all}{979}
\DefMacro{fastest-count-5s-all-percent-b}{0.1}
\DefMacro{fastest-count-5s-all-percent-c}{1.8}
\DefMacro{fastest-count-5s-all-percent-d}{30.0}
\DefMacro{fastest-count-5s-all-percent-all}{68.1}
\DefMacro{fastest-D-vs-second-fastest-diff-mean}{58.50}
\DefMacro{fastest-D-vs-second-fastest-diff-median}{0.98}

\DefMacro{Comparison-1-algo-D-p1-overhead}{1.51438}
\DefMacro{Comparison-1-algo-D-p2-overhead}{2.00722}
\DefMacro{Comparison-1-algo-D-p3-overhead}{2.69618}
\DefMacro{Comparison-1-algo-D-p4-overhead}{3.30684}
\DefMacro{Comparison-1-algo-D-p5-overhead}{4.10525}
\DefMacro{Comparison-1-algo-D-p6-overhead}{5.06251}
\DefMacro{Comparison-1-algo-D-p7-overhead}{6.12772}
\DefMacro{Comparison-1-algo-D-p8-overhead}{7.77635}
\DefMacro{Comparison-1-algo-D-p9-overhead}{9.98024}
\DefMacro{Comparison-1-algo-D-p10-overhead}{37.85853}
\DefMacro{Comparison-1-algo-dynapyt-p1-overhead}{1.76290}
\DefMacro{Comparison-1-algo-dynapyt-p2-overhead}{2.82453}
\DefMacro{Comparison-1-algo-dynapyt-p3-overhead}{3.54646}
\DefMacro{Comparison-1-algo-dynapyt-p4-overhead}{4.50737}
\DefMacro{Comparison-1-algo-dynapyt-p5-overhead}{5.91141}
\DefMacro{Comparison-1-algo-dynapyt-p6-overhead}{7.83800}
\DefMacro{Comparison-1-algo-dynapyt-p7-overhead}{11.95469}
\DefMacro{Comparison-1-algo-dynapyt-p8-overhead}{20.74028}
\DefMacro{Comparison-1-algo-dynapyt-p9-overhead}{36.88630}
\DefMacro{Comparison-1-algo-dynapyt-p10-overhead}{3357.94859}
\DefMacro{Comparison-1-algo-dynapyt_libs-p1-overhead}{233.36537}
\DefMacro{Comparison-1-algo-dynapyt_libs-p2-overhead}{606.19613}
\DefMacro{Comparison-1-algo-dynapyt_libs-p3-overhead}{1073.51579}
\DefMacro{Comparison-1-algo-dynapyt_libs-p4-overhead}{1572.36627}
\DefMacro{Comparison-1-algo-dynapyt_libs-p5-overhead}{2275.94528}
\DefMacro{Comparison-1-algo-dynapyt_libs-p6-overhead}{2927.57547}
\DefMacro{Comparison-1-algo-dynapyt_libs-p7-overhead}{3718.55353}
\DefMacro{Comparison-1-algo-dynapyt_libs-p8-overhead}{4407.78851}
\DefMacro{Comparison-1-algo-dynapyt_libs-p9-overhead}{5306.20355}
\DefMacro{Comparison-1-algo-dynapyt_libs-p10-overhead}{9723.70073}
\DefMacro{Comparison-1-algo-dylin-p1-overhead}{1.89308}
\DefMacro{Comparison-1-algo-dylin-p2-overhead}{3.26343}
\DefMacro{Comparison-1-algo-dylin-p3-overhead}{4.12123}
\DefMacro{Comparison-1-algo-dylin-p4-overhead}{5.08584}
\DefMacro{Comparison-1-algo-dylin-p5-overhead}{6.48344}
\DefMacro{Comparison-1-algo-dylin-p6-overhead}{8.64341}
\DefMacro{Comparison-1-algo-dylin-p7-overhead}{12.09603}
\DefMacro{Comparison-1-algo-dylin-p8-overhead}{19.97862}
\DefMacro{Comparison-1-algo-dylin-p9-overhead}{37.01067}
\DefMacro{Comparison-1-algo-dylin-p10-overhead}{3248.29561}
\DefMacro{Comparison-1-algo-dylin_libs-p1-overhead}{216.34956}
\DefMacro{Comparison-1-algo-dylin_libs-p2-overhead}{581.69851}
\DefMacro{Comparison-1-algo-dylin_libs-p3-overhead}{1032.36235}
\DefMacro{Comparison-1-algo-dylin_libs-p4-overhead}{1513.05081}
\DefMacro{Comparison-1-algo-dylin_libs-p5-overhead}{2172.03892}
\DefMacro{Comparison-1-algo-dylin_libs-p6-overhead}{2798.29906}
\DefMacro{Comparison-1-algo-dylin_libs-p7-overhead}{3574.80363}
\DefMacro{Comparison-1-algo-dylin_libs-p8-overhead}{4249.68938}
\DefMacro{Comparison-1-algo-dylin_libs-p9-overhead}{5144.02605}
\DefMacro{Comparison-1-algo-dylin_libs-p10-overhead}{9346.11177}
\DefMacro{Comparison-1-algo-D-p1-instrumentation-overhead}{0.07952}
\DefMacro{Comparison-1-algo-D-p2-instrumentation-overhead}{0.17543}
\DefMacro{Comparison-1-algo-D-p3-instrumentation-overhead}{0.34226}
\DefMacro{Comparison-1-algo-D-p4-instrumentation-overhead}{0.50875}
\DefMacro{Comparison-1-algo-D-p5-instrumentation-overhead}{0.73813}
\DefMacro{Comparison-1-algo-D-p6-instrumentation-overhead}{0.99236}
\DefMacro{Comparison-1-algo-D-p7-instrumentation-overhead}{1.26554}
\DefMacro{Comparison-1-algo-D-p8-instrumentation-overhead}{1.53455}
\DefMacro{Comparison-1-algo-D-p9-instrumentation-overhead}{2.11778}
\DefMacro{Comparison-1-algo-D-p10-instrumentation-overhead}{8.32600}
\DefMacro{Comparison-1-algo-dynapyt-p1-instrumentation-overhead}{0.79730}
\DefMacro{Comparison-1-algo-dynapyt-p2-instrumentation-overhead}{1.60757}
\DefMacro{Comparison-1-algo-dynapyt-p3-instrumentation-overhead}{2.29754}
\DefMacro{Comparison-1-algo-dynapyt-p4-instrumentation-overhead}{2.99526}
\DefMacro{Comparison-1-algo-dynapyt-p5-instrumentation-overhead}{4.31265}
\DefMacro{Comparison-1-algo-dynapyt-p6-instrumentation-overhead}{6.31386}
\DefMacro{Comparison-1-algo-dynapyt-p7-instrumentation-overhead}{9.80526}
\DefMacro{Comparison-1-algo-dynapyt-p8-instrumentation-overhead}{17.14012}
\DefMacro{Comparison-1-algo-dynapyt-p9-instrumentation-overhead}{32.91956}
\DefMacro{Comparison-1-algo-dynapyt-p10-instrumentation-overhead}{3356.71960}
\DefMacro{Comparison-1-algo-dynapyt_libs-p1-instrumentation-overhead}{222.98113}
\DefMacro{Comparison-1-algo-dynapyt_libs-p2-instrumentation-overhead}{600.88438}
\DefMacro{Comparison-1-algo-dynapyt_libs-p3-instrumentation-overhead}{1069.23804}
\DefMacro{Comparison-1-algo-dynapyt_libs-p4-instrumentation-overhead}{1562.31430}
\DefMacro{Comparison-1-algo-dynapyt_libs-p5-instrumentation-overhead}{2271.20508}
\DefMacro{Comparison-1-algo-dynapyt_libs-p6-instrumentation-overhead}{2924.79595}
\DefMacro{Comparison-1-algo-dynapyt_libs-p7-instrumentation-overhead}{3713.27790}
\DefMacro{Comparison-1-algo-dynapyt_libs-p8-instrumentation-overhead}{4403.79520}
\DefMacro{Comparison-1-algo-dynapyt_libs-p9-instrumentation-overhead}{5302.30099}
\DefMacro{Comparison-1-algo-dynapyt_libs-p10-instrumentation-overhead}{9718.50484}
\DefMacro{Comparison-1-algo-dylin-p1-instrumentation-overhead}{0.95342}
\DefMacro{Comparison-1-algo-dylin-p2-instrumentation-overhead}{2.11399}
\DefMacro{Comparison-1-algo-dylin-p3-instrumentation-overhead}{2.92519}
\DefMacro{Comparison-1-algo-dylin-p4-instrumentation-overhead}{3.61655}
\DefMacro{Comparison-1-algo-dylin-p5-instrumentation-overhead}{4.90360}
\DefMacro{Comparison-1-algo-dylin-p6-instrumentation-overhead}{6.86044}
\DefMacro{Comparison-1-algo-dylin-p7-instrumentation-overhead}{10.25648}
\DefMacro{Comparison-1-algo-dylin-p8-instrumentation-overhead}{17.48590}
\DefMacro{Comparison-1-algo-dylin-p9-instrumentation-overhead}{33.33326}
\DefMacro{Comparison-1-algo-dylin-p10-instrumentation-overhead}{3247.03251}
\DefMacro{Comparison-1-algo-dylin_libs-p1-instrumentation-overhead}{214.74757}
\DefMacro{Comparison-1-algo-dylin_libs-p2-instrumentation-overhead}{580.39906}
\DefMacro{Comparison-1-algo-dylin_libs-p3-instrumentation-overhead}{1030.61308}
\DefMacro{Comparison-1-algo-dylin_libs-p4-instrumentation-overhead}{1511.39067}
\DefMacro{Comparison-1-algo-dylin_libs-p5-instrumentation-overhead}{2168.20801}
\DefMacro{Comparison-1-algo-dylin_libs-p6-instrumentation-overhead}{2797.02216}
\DefMacro{Comparison-1-algo-dylin_libs-p7-instrumentation-overhead}{3573.03230}
\DefMacro{Comparison-1-algo-dylin_libs-p8-instrumentation-overhead}{4248.11765}
\DefMacro{Comparison-1-algo-dylin_libs-p9-instrumentation-overhead}{5142.27862}
\DefMacro{Comparison-1-algo-dylin_libs-p10-instrumentation-overhead}{9344.22557}
\DefMacro{Comparison-1-algo-D-p1-monitoring-overhead}{1.40210}
\DefMacro{Comparison-1-algo-D-p2-monitoring-overhead}{1.80047}
\DefMacro{Comparison-1-algo-D-p3-monitoring-overhead}{2.26467}
\DefMacro{Comparison-1-algo-D-p4-monitoring-overhead}{2.74474}
\DefMacro{Comparison-1-algo-D-p5-monitoring-overhead}{3.34250}
\DefMacro{Comparison-1-algo-D-p6-monitoring-overhead}{4.04535}
\DefMacro{Comparison-1-algo-D-p7-monitoring-overhead}{4.91860}
\DefMacro{Comparison-1-algo-D-p8-monitoring-overhead}{6.10443}
\DefMacro{Comparison-1-algo-D-p9-monitoring-overhead}{8.04444}
\DefMacro{Comparison-1-algo-D-p10-monitoring-overhead}{29.53253}
\DefMacro{Comparison-1-algo-dynapyt-p1-monitoring-overhead}{0.91359}
\DefMacro{Comparison-1-algo-dynapyt-p2-monitoring-overhead}{1.04504}
\DefMacro{Comparison-1-algo-dynapyt-p3-monitoring-overhead}{1.13283}
\DefMacro{Comparison-1-algo-dynapyt-p4-monitoring-overhead}{1.22753}
\DefMacro{Comparison-1-algo-dynapyt-p5-monitoring-overhead}{1.31776}
\DefMacro{Comparison-1-algo-dynapyt-p6-monitoring-overhead}{1.42001}
\DefMacro{Comparison-1-algo-dynapyt-p7-monitoring-overhead}{1.54163}
\DefMacro{Comparison-1-algo-dynapyt-p8-monitoring-overhead}{1.77482}
\DefMacro{Comparison-1-algo-dynapyt-p9-monitoring-overhead}{2.37554}
\DefMacro{Comparison-1-algo-dynapyt-p10-monitoring-overhead}{162.00942}
\DefMacro{Comparison-1-algo-dynapyt_libs-p1-monitoring-overhead}{2.68714}
\DefMacro{Comparison-1-algo-dynapyt_libs-p2-monitoring-overhead}{3.41371}
\DefMacro{Comparison-1-algo-dynapyt_libs-p3-monitoring-overhead}{3.84773}
\DefMacro{Comparison-1-algo-dynapyt_libs-p4-monitoring-overhead}{4.21139}
\DefMacro{Comparison-1-algo-dynapyt_libs-p5-monitoring-overhead}{4.79635}
\DefMacro{Comparison-1-algo-dynapyt_libs-p6-monitoring-overhead}{5.55713}
\DefMacro{Comparison-1-algo-dynapyt_libs-p7-monitoring-overhead}{6.55852}
\DefMacro{Comparison-1-algo-dynapyt_libs-p8-monitoring-overhead}{8.22790}
\DefMacro{Comparison-1-algo-dynapyt_libs-p9-monitoring-overhead}{11.79250}
\DefMacro{Comparison-1-algo-dynapyt_libs-p10-monitoring-overhead}{249.37173}
\DefMacro{Comparison-1-algo-dylin-p1-monitoring-overhead}{0.74826}
\DefMacro{Comparison-1-algo-dylin-p2-monitoring-overhead}{0.98152}
\DefMacro{Comparison-1-algo-dylin-p3-monitoring-overhead}{1.09151}
\DefMacro{Comparison-1-algo-dylin-p4-monitoring-overhead}{1.19388}
\DefMacro{Comparison-1-algo-dylin-p5-monitoring-overhead}{1.30169}
\DefMacro{Comparison-1-algo-dylin-p6-monitoring-overhead}{1.39636}
\DefMacro{Comparison-1-algo-dylin-p7-monitoring-overhead}{1.54244}
\DefMacro{Comparison-1-algo-dylin-p8-monitoring-overhead}{1.82770}
\DefMacro{Comparison-1-algo-dylin-p9-monitoring-overhead}{2.67339}
\DefMacro{Comparison-1-algo-dylin-p10-monitoring-overhead}{57.31545}
\DefMacro{Comparison-1-algo-dylin_libs-p1-monitoring-overhead}{1.03141}
\DefMacro{Comparison-1-algo-dylin_libs-p2-monitoring-overhead}{1.23303}
\DefMacro{Comparison-1-algo-dylin_libs-p3-monitoring-overhead}{1.38440}
\DefMacro{Comparison-1-algo-dylin_libs-p4-monitoring-overhead}{1.51604}
\DefMacro{Comparison-1-algo-dylin_libs-p5-monitoring-overhead}{1.63089}
\DefMacro{Comparison-1-algo-dylin_libs-p6-monitoring-overhead}{1.76586}
\DefMacro{Comparison-1-algo-dylin_libs-p7-monitoring-overhead}{1.90745}
\DefMacro{Comparison-1-algo-dylin_libs-p8-monitoring-overhead}{2.16112}
\DefMacro{Comparison-1-algo-dylin_libs-p9-monitoring-overhead}{2.60889}
\DefMacro{Comparison-1-algo-dylin_libs-p10-monitoring-overhead}{54.09450}
\DefMacro{Comparison-1-total-projects}{755}

\DefMacro{Comparison-2-algo-D-p1-overhead}{1.20876}
\DefMacro{Comparison-2-algo-D-p2-overhead}{1.68625}
\DefMacro{Comparison-2-algo-D-p3-overhead}{2.23895}
\DefMacro{Comparison-2-algo-D-p4-overhead}{2.80430}
\DefMacro{Comparison-2-algo-D-p5-overhead}{3.50266}
\DefMacro{Comparison-2-algo-D-p6-overhead}{4.39982}
\DefMacro{Comparison-2-algo-D-p7-overhead}{5.35174}
\DefMacro{Comparison-2-algo-D-p8-overhead}{6.78735}
\DefMacro{Comparison-2-algo-D-p9-overhead}{9.11257}
\DefMacro{Comparison-2-algo-D-p10-overhead}{37.85853}
\DefMacro{Comparison-2-algo-dynapyt-p1-overhead}{1.18670}
\DefMacro{Comparison-2-algo-dynapyt-p2-overhead}{2.00334}
\DefMacro{Comparison-2-algo-dynapyt-p3-overhead}{2.77636}
\DefMacro{Comparison-2-algo-dynapyt-p4-overhead}{3.54614}
\DefMacro{Comparison-2-algo-dynapyt-p5-overhead}{4.54044}
\DefMacro{Comparison-2-algo-dynapyt-p6-overhead}{6.26111}
\DefMacro{Comparison-2-algo-dynapyt-p7-overhead}{9.15736}
\DefMacro{Comparison-2-algo-dynapyt-p8-overhead}{14.39557}
\DefMacro{Comparison-2-algo-dynapyt-p9-overhead}{30.59867}
\DefMacro{Comparison-2-algo-dynapyt-p10-overhead}{3357.94859}
\DefMacro{Comparison-2-algo-dylin-p1-overhead}{1.25320}
\DefMacro{Comparison-2-algo-dylin-p2-overhead}{2.13027}
\DefMacro{Comparison-2-algo-dylin-p3-overhead}{3.13554}
\DefMacro{Comparison-2-algo-dylin-p4-overhead}{4.09705}
\DefMacro{Comparison-2-algo-dylin-p5-overhead}{5.08383}
\DefMacro{Comparison-2-algo-dylin-p6-overhead}{6.76633}
\DefMacro{Comparison-2-algo-dylin-p7-overhead}{9.10890}
\DefMacro{Comparison-2-algo-dylin-p8-overhead}{14.27124}
\DefMacro{Comparison-2-algo-dylin-p9-overhead}{30.06146}
\DefMacro{Comparison-2-algo-dylin-p10-overhead}{3248.29561}
\DefMacro{Comparison-2-algo-D-p1-instrumentation-overhead}{0.04771}
\DefMacro{Comparison-2-algo-D-p2-instrumentation-overhead}{0.12336}
\DefMacro{Comparison-2-algo-D-p3-instrumentation-overhead}{0.23297}
\DefMacro{Comparison-2-algo-D-p4-instrumentation-overhead}{0.38281}
\DefMacro{Comparison-2-algo-D-p5-instrumentation-overhead}{0.57202}
\DefMacro{Comparison-2-algo-D-p6-instrumentation-overhead}{0.80472}
\DefMacro{Comparison-2-algo-D-p7-instrumentation-overhead}{1.04444}
\DefMacro{Comparison-2-algo-D-p8-instrumentation-overhead}{1.37188}
\DefMacro{Comparison-2-algo-D-p9-instrumentation-overhead}{1.88254}
\DefMacro{Comparison-2-algo-D-p10-instrumentation-overhead}{8.32600}
\DefMacro{Comparison-2-algo-dynapyt-p1-instrumentation-overhead}{0.34757}
\DefMacro{Comparison-2-algo-dynapyt-p2-instrumentation-overhead}{0.87653}
\DefMacro{Comparison-2-algo-dynapyt-p3-instrumentation-overhead}{1.56558}
\DefMacro{Comparison-2-algo-dynapyt-p4-instrumentation-overhead}{2.26971}
\DefMacro{Comparison-2-algo-dynapyt-p5-instrumentation-overhead}{3.02166}
\DefMacro{Comparison-2-algo-dynapyt-p6-instrumentation-overhead}{4.43162}
\DefMacro{Comparison-2-algo-dynapyt-p7-instrumentation-overhead}{6.73595}
\DefMacro{Comparison-2-algo-dynapyt-p8-instrumentation-overhead}{11.54640}
\DefMacro{Comparison-2-algo-dynapyt-p9-instrumentation-overhead}{26.57241}
\DefMacro{Comparison-2-algo-dynapyt-p10-instrumentation-overhead}{3356.71960}
\DefMacro{Comparison-2-algo-dylin-p1-instrumentation-overhead}{0.44597}
\DefMacro{Comparison-2-algo-dylin-p2-instrumentation-overhead}{1.08487}
\DefMacro{Comparison-2-algo-dylin-p3-instrumentation-overhead}{1.97336}
\DefMacro{Comparison-2-algo-dylin-p4-instrumentation-overhead}{2.83952}
\DefMacro{Comparison-2-algo-dylin-p5-instrumentation-overhead}{3.63451}
\DefMacro{Comparison-2-algo-dylin-p6-instrumentation-overhead}{5.12193}
\DefMacro{Comparison-2-algo-dylin-p7-instrumentation-overhead}{7.36884}
\DefMacro{Comparison-2-algo-dylin-p8-instrumentation-overhead}{11.79616}
\DefMacro{Comparison-2-algo-dylin-p9-instrumentation-overhead}{26.46678}
\DefMacro{Comparison-2-algo-dylin-p10-instrumentation-overhead}{3247.03251}
\DefMacro{Comparison-2-algo-D-p1-monitoring-overhead}{1.10327}
\DefMacro{Comparison-2-algo-D-p2-monitoring-overhead}{1.55116}
\DefMacro{Comparison-2-algo-D-p3-monitoring-overhead}{1.96656}
\DefMacro{Comparison-2-algo-D-p4-monitoring-overhead}{2.40122}
\DefMacro{Comparison-2-algo-D-p5-monitoring-overhead}{2.87631}
\DefMacro{Comparison-2-algo-D-p6-monitoring-overhead}{3.50277}
\DefMacro{Comparison-2-algo-D-p7-monitoring-overhead}{4.23600}
\DefMacro{Comparison-2-algo-D-p8-monitoring-overhead}{5.49010}
\DefMacro{Comparison-2-algo-D-p9-monitoring-overhead}{7.32049}
\DefMacro{Comparison-2-algo-D-p10-monitoring-overhead}{35.45053}
\DefMacro{Comparison-2-algo-dynapyt-p1-monitoring-overhead}{0.72048}
\DefMacro{Comparison-2-algo-dynapyt-p2-monitoring-overhead}{0.93876}
\DefMacro{Comparison-2-algo-dynapyt-p3-monitoring-overhead}{1.04574}
\DefMacro{Comparison-2-algo-dynapyt-p4-monitoring-overhead}{1.12892}
\DefMacro{Comparison-2-algo-dynapyt-p5-monitoring-overhead}{1.21331}
\DefMacro{Comparison-2-algo-dynapyt-p6-monitoring-overhead}{1.31825}
\DefMacro{Comparison-2-algo-dynapyt-p7-monitoring-overhead}{1.47091}
\DefMacro{Comparison-2-algo-dynapyt-p8-monitoring-overhead}{1.64301}
\DefMacro{Comparison-2-algo-dynapyt-p9-monitoring-overhead}{2.28154}
\DefMacro{Comparison-2-algo-dynapyt-p10-monitoring-overhead}{209.92949}
\DefMacro{Comparison-2-algo-dylin-p1-monitoring-overhead}{0.66577}
\DefMacro{Comparison-2-algo-dylin-p2-monitoring-overhead}{0.85955}
\DefMacro{Comparison-2-algo-dylin-p3-monitoring-overhead}{1.00232}
\DefMacro{Comparison-2-algo-dylin-p4-monitoring-overhead}{1.09160}
\DefMacro{Comparison-2-algo-dylin-p5-monitoring-overhead}{1.19223}
\DefMacro{Comparison-2-algo-dylin-p6-monitoring-overhead}{1.30423}
\DefMacro{Comparison-2-algo-dylin-p7-monitoring-overhead}{1.43718}
\DefMacro{Comparison-2-algo-dylin-p8-monitoring-overhead}{1.65895}
\DefMacro{Comparison-2-algo-dylin-p9-monitoring-overhead}{2.29361}
\DefMacro{Comparison-2-algo-dylin-p10-monitoring-overhead}{57.31545}
\DefMacro{Comparison-2-total-projects}{1201}

\DefMacro{Comparison-3-algo-D-p1-overhead}{1.51490}
\DefMacro{Comparison-3-algo-D-p2-overhead}{1.99830}
\DefMacro{Comparison-3-algo-D-p3-overhead}{2.69724}
\DefMacro{Comparison-3-algo-D-p4-overhead}{3.32792}
\DefMacro{Comparison-3-algo-D-p5-overhead}{4.27116}
\DefMacro{Comparison-3-algo-D-p6-overhead}{5.19989}
\DefMacro{Comparison-3-algo-D-p7-overhead}{6.33330}
\DefMacro{Comparison-3-algo-D-p8-overhead}{7.91207}
\DefMacro{Comparison-3-algo-D-p9-overhead}{10.33945}
\DefMacro{Comparison-3-algo-D-p10-overhead}{37.85853}
\DefMacro{Comparison-3-algo-dynapyt_libs-p1-overhead}{233.24614}
\DefMacro{Comparison-3-algo-dynapyt_libs-p2-overhead}{599.78705}
\DefMacro{Comparison-3-algo-dynapyt_libs-p3-overhead}{1086.13349}
\DefMacro{Comparison-3-algo-dynapyt_libs-p4-overhead}{1592.53769}
\DefMacro{Comparison-3-algo-dynapyt_libs-p5-overhead}{2343.49064}
\DefMacro{Comparison-3-algo-dynapyt_libs-p6-overhead}{3021.15538}
\DefMacro{Comparison-3-algo-dynapyt_libs-p7-overhead}{3818.01683}
\DefMacro{Comparison-3-algo-dynapyt_libs-p8-overhead}{4607.82491}
\DefMacro{Comparison-3-algo-dynapyt_libs-p9-overhead}{5566.19192}
\DefMacro{Comparison-3-algo-dynapyt_libs-p10-overhead}{9723.70073}
\DefMacro{Comparison-3-algo-dylin_libs-p1-overhead}{215.85211}
\DefMacro{Comparison-3-algo-dylin_libs-p2-overhead}{568.59951}
\DefMacro{Comparison-3-algo-dylin_libs-p3-overhead}{1044.32885}
\DefMacro{Comparison-3-algo-dylin_libs-p4-overhead}{1529.98940}
\DefMacro{Comparison-3-algo-dylin_libs-p5-overhead}{2232.94982}
\DefMacro{Comparison-3-algo-dylin_libs-p6-overhead}{2897.46427}
\DefMacro{Comparison-3-algo-dylin_libs-p7-overhead}{3656.56090}
\DefMacro{Comparison-3-algo-dylin_libs-p8-overhead}{4418.22653}
\DefMacro{Comparison-3-algo-dylin_libs-p9-overhead}{5376.17033}
\DefMacro{Comparison-3-algo-dylin_libs-p10-overhead}{9346.11177}
\DefMacro{Comparison-3-algo-D-p1-instrumentation-overhead}{0.07936}
\DefMacro{Comparison-3-algo-D-p2-instrumentation-overhead}{0.17476}
\DefMacro{Comparison-3-algo-D-p3-instrumentation-overhead}{0.35041}
\DefMacro{Comparison-3-algo-D-p4-instrumentation-overhead}{0.51805}
\DefMacro{Comparison-3-algo-D-p5-instrumentation-overhead}{0.75841}
\DefMacro{Comparison-3-algo-D-p6-instrumentation-overhead}{1.02340}
\DefMacro{Comparison-3-algo-D-p7-instrumentation-overhead}{1.29354}
\DefMacro{Comparison-3-algo-D-p8-instrumentation-overhead}{1.59730}
\DefMacro{Comparison-3-algo-D-p9-instrumentation-overhead}{2.18091}
\DefMacro{Comparison-3-algo-D-p10-instrumentation-overhead}{8.32600}
\DefMacro{Comparison-3-algo-dynapyt_libs-p1-instrumentation-overhead}{221.81336}
\DefMacro{Comparison-3-algo-dynapyt_libs-p2-instrumentation-overhead}{588.93835}
\DefMacro{Comparison-3-algo-dynapyt_libs-p3-instrumentation-overhead}{1077.76023}
\DefMacro{Comparison-3-algo-dynapyt_libs-p4-instrumentation-overhead}{1577.94457}
\DefMacro{Comparison-3-algo-dynapyt_libs-p5-instrumentation-overhead}{2331.89682}
\DefMacro{Comparison-3-algo-dynapyt_libs-p6-instrumentation-overhead}{3014.50010}
\DefMacro{Comparison-3-algo-dynapyt_libs-p7-instrumentation-overhead}{3813.58478}
\DefMacro{Comparison-3-algo-dynapyt_libs-p8-instrumentation-overhead}{4604.70581}
\DefMacro{Comparison-3-algo-dynapyt_libs-p9-instrumentation-overhead}{5562.98494}
\DefMacro{Comparison-3-algo-dynapyt_libs-p10-instrumentation-overhead}{9718.50484}
\DefMacro{Comparison-3-algo-dylin_libs-p1-instrumentation-overhead}{213.76316}
\DefMacro{Comparison-3-algo-dylin_libs-p2-instrumentation-overhead}{566.43944}
\DefMacro{Comparison-3-algo-dylin_libs-p3-instrumentation-overhead}{1042.85428}
\DefMacro{Comparison-3-algo-dylin_libs-p4-instrumentation-overhead}{1520.04446}
\DefMacro{Comparison-3-algo-dylin_libs-p5-instrumentation-overhead}{2229.18855}
\DefMacro{Comparison-3-algo-dylin_libs-p6-instrumentation-overhead}{2896.04726}
\DefMacro{Comparison-3-algo-dylin_libs-p7-instrumentation-overhead}{3653.77459}
\DefMacro{Comparison-3-algo-dylin_libs-p8-instrumentation-overhead}{4416.32111}
\DefMacro{Comparison-3-algo-dylin_libs-p9-instrumentation-overhead}{5374.50632}
\DefMacro{Comparison-3-algo-dylin_libs-p10-instrumentation-overhead}{9344.22557}
\DefMacro{Comparison-3-algo-D-p1-monitoring-overhead}{1.40541}
\DefMacro{Comparison-3-algo-D-p2-monitoring-overhead}{1.79986}
\DefMacro{Comparison-3-algo-D-p3-monitoring-overhead}{2.26503}
\DefMacro{Comparison-3-algo-D-p4-monitoring-overhead}{2.75764}
\DefMacro{Comparison-3-algo-D-p5-monitoring-overhead}{3.37644}
\DefMacro{Comparison-3-algo-D-p6-monitoring-overhead}{4.08571}
\DefMacro{Comparison-3-algo-D-p7-monitoring-overhead}{5.05555}
\DefMacro{Comparison-3-algo-D-p8-monitoring-overhead}{6.33282}
\DefMacro{Comparison-3-algo-D-p9-monitoring-overhead}{8.20561}
\DefMacro{Comparison-3-algo-D-p10-monitoring-overhead}{29.53253}
\DefMacro{Comparison-3-algo-dynapyt_libs-p1-monitoring-overhead}{2.67697}
\DefMacro{Comparison-3-algo-dynapyt_libs-p2-monitoring-overhead}{3.41371}
\DefMacro{Comparison-3-algo-dynapyt_libs-p3-monitoring-overhead}{3.84256}
\DefMacro{Comparison-3-algo-dynapyt_libs-p4-monitoring-overhead}{4.18461}
\DefMacro{Comparison-3-algo-dynapyt_libs-p5-monitoring-overhead}{4.67730}
\DefMacro{Comparison-3-algo-dynapyt_libs-p6-monitoring-overhead}{5.50791}
\DefMacro{Comparison-3-algo-dynapyt_libs-p7-monitoring-overhead}{6.48192}
\DefMacro{Comparison-3-algo-dynapyt_libs-p8-monitoring-overhead}{8.20628}
\DefMacro{Comparison-3-algo-dynapyt_libs-p9-monitoring-overhead}{11.62536}
\DefMacro{Comparison-3-algo-dynapyt_libs-p10-monitoring-overhead}{249.37173}
\DefMacro{Comparison-3-algo-dylin_libs-p1-monitoring-overhead}{1.03208}
\DefMacro{Comparison-3-algo-dylin_libs-p2-monitoring-overhead}{1.24077}
\DefMacro{Comparison-3-algo-dylin_libs-p3-monitoring-overhead}{1.39443}
\DefMacro{Comparison-3-algo-dylin_libs-p4-monitoring-overhead}{1.51867}
\DefMacro{Comparison-3-algo-dylin_libs-p5-monitoring-overhead}{1.63272}
\DefMacro{Comparison-3-algo-dylin_libs-p6-monitoring-overhead}{1.76586}
\DefMacro{Comparison-3-algo-dylin_libs-p7-monitoring-overhead}{1.90866}
\DefMacro{Comparison-3-algo-dylin_libs-p8-monitoring-overhead}{2.18053}
\DefMacro{Comparison-3-algo-dylin_libs-p9-monitoring-overhead}{2.70591}
\DefMacro{Comparison-3-algo-dylin_libs-p10-monitoring-overhead}{54.09450}
\DefMacro{Comparison-3-total-projects}{820}

\DefMacro{Comparison-4-algo-D-p1-overhead}{1.46509}
\DefMacro{Comparison-4-algo-D-p2-overhead}{1.92825}
\DefMacro{Comparison-4-algo-D-p3-overhead}{2.55532}
\DefMacro{Comparison-4-algo-D-p4-overhead}{3.18901}
\DefMacro{Comparison-4-algo-D-p5-overhead}{4.07953}
\DefMacro{Comparison-4-algo-D-p6-overhead}{5.00346}
\DefMacro{Comparison-4-algo-D-p7-overhead}{6.10659}
\DefMacro{Comparison-4-algo-D-p8-overhead}{7.69329}
\DefMacro{Comparison-4-algo-D-p9-overhead}{10.21125}
\DefMacro{Comparison-4-algo-D-p10-overhead}{37.85853}
\DefMacro{Comparison-4-algo-dylin-p1-overhead}{1.62615}
\DefMacro{Comparison-4-algo-dylin-p2-overhead}{2.95598}
\DefMacro{Comparison-4-algo-dylin-p3-overhead}{3.90633}
\DefMacro{Comparison-4-algo-dylin-p4-overhead}{4.81114}
\DefMacro{Comparison-4-algo-dylin-p5-overhead}{6.11107}
\DefMacro{Comparison-4-algo-dylin-p6-overhead}{8.12370}
\DefMacro{Comparison-4-algo-dylin-p7-overhead}{11.04700}
\DefMacro{Comparison-4-algo-dylin-p8-overhead}{18.53331}
\DefMacro{Comparison-4-algo-dylin-p9-overhead}{35.54248}
\DefMacro{Comparison-4-algo-dylin-p10-overhead}{3248.29561}
\DefMacro{Comparison-4-algo-dylin_libs-p1-overhead}{191.88853}
\DefMacro{Comparison-4-algo-dylin_libs-p2-overhead}{522.72264}
\DefMacro{Comparison-4-algo-dylin_libs-p3-overhead}{970.35131}
\DefMacro{Comparison-4-algo-dylin_libs-p4-overhead}{1438.03868}
\DefMacro{Comparison-4-algo-dylin_libs-p5-overhead}{2070.23770}
\DefMacro{Comparison-4-algo-dylin_libs-p6-overhead}{2746.32949}
\DefMacro{Comparison-4-algo-dylin_libs-p7-overhead}{3574.80363}
\DefMacro{Comparison-4-algo-dylin_libs-p8-overhead}{4284.05678}
\DefMacro{Comparison-4-algo-dylin_libs-p9-overhead}{5231.84763}
\DefMacro{Comparison-4-algo-dylin_libs-p10-overhead}{9346.11177}
\DefMacro{Comparison-4-algo-D-p1-instrumentation-overhead}{0.07096}
\DefMacro{Comparison-4-algo-D-p2-instrumentation-overhead}{0.15945}
\DefMacro{Comparison-4-algo-D-p3-instrumentation-overhead}{0.31381}
\DefMacro{Comparison-4-algo-D-p4-instrumentation-overhead}{0.49053}
\DefMacro{Comparison-4-algo-D-p5-instrumentation-overhead}{0.71098}
\DefMacro{Comparison-4-algo-D-p6-instrumentation-overhead}{0.97219}
\DefMacro{Comparison-4-algo-D-p7-instrumentation-overhead}{1.24286}
\DefMacro{Comparison-4-algo-D-p8-instrumentation-overhead}{1.54432}
\DefMacro{Comparison-4-algo-D-p9-instrumentation-overhead}{2.14544}
\DefMacro{Comparison-4-algo-D-p10-instrumentation-overhead}{8.32600}
\DefMacro{Comparison-4-algo-dylin-p1-instrumentation-overhead}{0.71667}
\DefMacro{Comparison-4-algo-dylin-p2-instrumentation-overhead}{1.73109}
\DefMacro{Comparison-4-algo-dylin-p3-instrumentation-overhead}{2.72561}
\DefMacro{Comparison-4-algo-dylin-p4-instrumentation-overhead}{3.45373}
\DefMacro{Comparison-4-algo-dylin-p5-instrumentation-overhead}{4.56602}
\DefMacro{Comparison-4-algo-dylin-p6-instrumentation-overhead}{6.54516}
\DefMacro{Comparison-4-algo-dylin-p7-instrumentation-overhead}{9.42283}
\DefMacro{Comparison-4-algo-dylin-p8-instrumentation-overhead}{15.45484}
\DefMacro{Comparison-4-algo-dylin-p9-instrumentation-overhead}{32.07119}
\DefMacro{Comparison-4-algo-dylin-p10-instrumentation-overhead}{3247.03251}
\DefMacro{Comparison-4-algo-dylin_libs-p1-instrumentation-overhead}{189.56276}
\DefMacro{Comparison-4-algo-dylin_libs-p2-instrumentation-overhead}{521.48373}
\DefMacro{Comparison-4-algo-dylin_libs-p3-instrumentation-overhead}{966.41750}
\DefMacro{Comparison-4-algo-dylin_libs-p4-instrumentation-overhead}{1435.55613}
\DefMacro{Comparison-4-algo-dylin_libs-p5-instrumentation-overhead}{2065.82435}
\DefMacro{Comparison-4-algo-dylin_libs-p6-instrumentation-overhead}{2744.58589}
\DefMacro{Comparison-4-algo-dylin_libs-p7-instrumentation-overhead}{3573.03230}
\DefMacro{Comparison-4-algo-dylin_libs-p8-instrumentation-overhead}{4282.32308}
\DefMacro{Comparison-4-algo-dylin_libs-p9-instrumentation-overhead}{5230.35645}
\DefMacro{Comparison-4-algo-dylin_libs-p10-instrumentation-overhead}{9344.22557}
\DefMacro{Comparison-4-algo-D-p1-monitoring-overhead}{1.35091}
\DefMacro{Comparison-4-algo-D-p2-monitoring-overhead}{1.75233}
\DefMacro{Comparison-4-algo-D-p3-monitoring-overhead}{2.15039}
\DefMacro{Comparison-4-algo-D-p4-monitoring-overhead}{2.68706}
\DefMacro{Comparison-4-algo-D-p5-monitoring-overhead}{3.31505}
\DefMacro{Comparison-4-algo-D-p6-monitoring-overhead}{3.95642}
\DefMacro{Comparison-4-algo-D-p7-monitoring-overhead}{4.86009}
\DefMacro{Comparison-4-algo-D-p8-monitoring-overhead}{6.08135}
\DefMacro{Comparison-4-algo-D-p9-monitoring-overhead}{8.11661}
\DefMacro{Comparison-4-algo-D-p10-monitoring-overhead}{29.53253}
\DefMacro{Comparison-4-algo-dylin-p1-monitoring-overhead}{0.67670}
\DefMacro{Comparison-4-algo-dylin-p2-monitoring-overhead}{0.88084}
\DefMacro{Comparison-4-algo-dylin-p3-monitoring-overhead}{1.02744}
\DefMacro{Comparison-4-algo-dylin-p4-monitoring-overhead}{1.13129}
\DefMacro{Comparison-4-algo-dylin-p5-monitoring-overhead}{1.24254}
\DefMacro{Comparison-4-algo-dylin-p6-monitoring-overhead}{1.36011}
\DefMacro{Comparison-4-algo-dylin-p7-monitoring-overhead}{1.50152}
\DefMacro{Comparison-4-algo-dylin-p8-monitoring-overhead}{1.77058}
\DefMacro{Comparison-4-algo-dylin-p9-monitoring-overhead}{2.46255}
\DefMacro{Comparison-4-algo-dylin-p10-monitoring-overhead}{57.31545}
\DefMacro{Comparison-4-algo-dylin_libs-p1-monitoring-overhead}{1.01784}
\DefMacro{Comparison-4-algo-dylin_libs-p2-monitoring-overhead}{1.22095}
\DefMacro{Comparison-4-algo-dylin_libs-p3-monitoring-overhead}{1.37446}
\DefMacro{Comparison-4-algo-dylin_libs-p4-monitoring-overhead}{1.51637}
\DefMacro{Comparison-4-algo-dylin_libs-p5-monitoring-overhead}{1.63090}
\DefMacro{Comparison-4-algo-dylin_libs-p6-monitoring-overhead}{1.76586}
\DefMacro{Comparison-4-algo-dylin_libs-p7-monitoring-overhead}{1.91459}
\DefMacro{Comparison-4-algo-dylin_libs-p8-monitoring-overhead}{2.21334}
\DefMacro{Comparison-4-algo-dylin_libs-p9-monitoring-overhead}{2.72280}
\DefMacro{Comparison-4-algo-dylin_libs-p10-monitoring-overhead}{93.79731}
\DefMacro{Comparison-4-total-projects}{886}

\DefMacro{Comparison-5-algo-D-p1-overhead}{1.21324}
\DefMacro{Comparison-5-algo-D-p2-overhead}{1.72169}
\DefMacro{Comparison-5-algo-D-p3-overhead}{2.27144}
\DefMacro{Comparison-5-algo-D-p4-overhead}{2.88556}
\DefMacro{Comparison-5-algo-D-p5-overhead}{3.60163}
\DefMacro{Comparison-5-algo-D-p6-overhead}{4.52072}
\DefMacro{Comparison-5-algo-D-p7-overhead}{5.59863}
\DefMacro{Comparison-5-algo-D-p8-overhead}{7.01140}
\DefMacro{Comparison-5-algo-D-p9-overhead}{9.44030}
\DefMacro{Comparison-5-algo-D-p10-overhead}{37.85853}
\DefMacro{Comparison-5-algo-dylin-p1-overhead}{1.30166}
\DefMacro{Comparison-5-algo-dylin-p2-overhead}{2.16539}
\DefMacro{Comparison-5-algo-dylin-p3-overhead}{3.18921}
\DefMacro{Comparison-5-algo-dylin-p4-overhead}{4.13178}
\DefMacro{Comparison-5-algo-dylin-p5-overhead}{5.08383}
\DefMacro{Comparison-5-algo-dylin-p6-overhead}{6.72137}
\DefMacro{Comparison-5-algo-dylin-p7-overhead}{9.03143}
\DefMacro{Comparison-5-algo-dylin-p8-overhead}{14.09883}
\DefMacro{Comparison-5-algo-dylin-p9-overhead}{29.93986}
\DefMacro{Comparison-5-algo-dylin-p10-overhead}{3248.29561}
\DefMacro{Comparison-5-algo-D-p1-instrumentation-overhead}{0.05139}
\DefMacro{Comparison-5-algo-D-p2-instrumentation-overhead}{0.12611}
\DefMacro{Comparison-5-algo-D-p3-instrumentation-overhead}{0.24800}
\DefMacro{Comparison-5-algo-D-p4-instrumentation-overhead}{0.40525}
\DefMacro{Comparison-5-algo-D-p5-instrumentation-overhead}{0.61842}
\DefMacro{Comparison-5-algo-D-p6-instrumentation-overhead}{0.86475}
\DefMacro{Comparison-5-algo-D-p7-instrumentation-overhead}{1.10362}
\DefMacro{Comparison-5-algo-D-p8-instrumentation-overhead}{1.43743}
\DefMacro{Comparison-5-algo-D-p9-instrumentation-overhead}{2.01511}
\DefMacro{Comparison-5-algo-D-p10-instrumentation-overhead}{8.32600}
\DefMacro{Comparison-5-algo-dylin-p1-instrumentation-overhead}{0.45932}
\DefMacro{Comparison-5-algo-dylin-p2-instrumentation-overhead}{1.12371}
\DefMacro{Comparison-5-algo-dylin-p3-instrumentation-overhead}{2.04399}
\DefMacro{Comparison-5-algo-dylin-p4-instrumentation-overhead}{2.89607}
\DefMacro{Comparison-5-algo-dylin-p5-instrumentation-overhead}{3.67134}
\DefMacro{Comparison-5-algo-dylin-p6-instrumentation-overhead}{5.08678}
\DefMacro{Comparison-5-algo-dylin-p7-instrumentation-overhead}{7.31630}
\DefMacro{Comparison-5-algo-dylin-p8-instrumentation-overhead}{11.62099}
\DefMacro{Comparison-5-algo-dylin-p9-instrumentation-overhead}{26.44863}
\DefMacro{Comparison-5-algo-dylin-p10-instrumentation-overhead}{3247.03251}
\DefMacro{Comparison-5-algo-D-p1-monitoring-overhead}{1.11717}
\DefMacro{Comparison-5-algo-D-p2-monitoring-overhead}{1.57487}
\DefMacro{Comparison-5-algo-D-p3-monitoring-overhead}{1.99288}
\DefMacro{Comparison-5-algo-D-p4-monitoring-overhead}{2.45078}
\DefMacro{Comparison-5-algo-D-p5-monitoring-overhead}{2.96167}
\DefMacro{Comparison-5-algo-D-p6-monitoring-overhead}{3.65218}
\DefMacro{Comparison-5-algo-D-p7-monitoring-overhead}{4.38733}
\DefMacro{Comparison-5-algo-D-p8-monitoring-overhead}{5.59795}
\DefMacro{Comparison-5-algo-D-p9-monitoring-overhead}{7.57945}
\DefMacro{Comparison-5-algo-D-p10-monitoring-overhead}{35.45053}
\DefMacro{Comparison-5-algo-dylin-p1-monitoring-overhead}{0.64040}
\DefMacro{Comparison-5-algo-dylin-p2-monitoring-overhead}{0.79960}
\DefMacro{Comparison-5-algo-dylin-p3-monitoring-overhead}{0.96961}
\DefMacro{Comparison-5-algo-dylin-p4-monitoring-overhead}{1.05632}
\DefMacro{Comparison-5-algo-dylin-p5-monitoring-overhead}{1.16679}
\DefMacro{Comparison-5-algo-dylin-p6-monitoring-overhead}{1.28085}
\DefMacro{Comparison-5-algo-dylin-p7-monitoring-overhead}{1.41516}
\DefMacro{Comparison-5-algo-dylin-p8-monitoring-overhead}{1.64536}
\DefMacro{Comparison-5-algo-dylin-p9-monitoring-overhead}{2.24799}
\DefMacro{Comparison-5-algo-dylin-p10-monitoring-overhead}{57.31545}
\DefMacro{Comparison-5-total-projects}{1275}

\DefMacro{Comparison-6-algo-D-p1-overhead}{1.43711}
\DefMacro{Comparison-6-algo-D-p2-overhead}{1.85880}
\DefMacro{Comparison-6-algo-D-p3-overhead}{2.44989}
\DefMacro{Comparison-6-algo-D-p4-overhead}{3.10815}
\DefMacro{Comparison-6-algo-D-p5-overhead}{3.93554}
\DefMacro{Comparison-6-algo-D-p6-overhead}{4.86744}
\DefMacro{Comparison-6-algo-D-p7-overhead}{6.04852}
\DefMacro{Comparison-6-algo-D-p8-overhead}{7.53047}
\DefMacro{Comparison-6-algo-D-p9-overhead}{10.02416}
\DefMacro{Comparison-6-algo-D-p10-overhead}{37.85853}
\DefMacro{Comparison-6-algo-dylin_libs-p1-overhead}{177.71310}
\DefMacro{Comparison-6-algo-dylin_libs-p2-overhead}{434.86953}
\DefMacro{Comparison-6-algo-dylin_libs-p3-overhead}{922.99682}
\DefMacro{Comparison-6-algo-dylin_libs-p4-overhead}{1332.56041}
\DefMacro{Comparison-6-algo-dylin_libs-p5-overhead}{1974.69573}
\DefMacro{Comparison-6-algo-dylin_libs-p6-overhead}{2664.92587}
\DefMacro{Comparison-6-algo-dylin_libs-p7-overhead}{3489.10200}
\DefMacro{Comparison-6-algo-dylin_libs-p8-overhead}{4239.46788}
\DefMacro{Comparison-6-algo-dylin_libs-p9-overhead}{5201.29858}
\DefMacro{Comparison-6-algo-dylin_libs-p10-overhead}{9346.11177}
\DefMacro{Comparison-6-algo-D-p1-instrumentation-overhead}{0.06363}
\DefMacro{Comparison-6-algo-D-p2-instrumentation-overhead}{0.13836}
\DefMacro{Comparison-6-algo-D-p3-instrumentation-overhead}{0.28807}
\DefMacro{Comparison-6-algo-D-p4-instrumentation-overhead}{0.46429}
\DefMacro{Comparison-6-algo-D-p5-instrumentation-overhead}{0.68978}
\DefMacro{Comparison-6-algo-D-p6-instrumentation-overhead}{0.95456}
\DefMacro{Comparison-6-algo-D-p7-instrumentation-overhead}{1.20262}
\DefMacro{Comparison-6-algo-D-p8-instrumentation-overhead}{1.51772}
\DefMacro{Comparison-6-algo-D-p9-instrumentation-overhead}{2.12302}
\DefMacro{Comparison-6-algo-D-p10-instrumentation-overhead}{8.32600}
\DefMacro{Comparison-6-algo-dylin_libs-p1-instrumentation-overhead}{177.00418}
\DefMacro{Comparison-6-algo-dylin_libs-p2-instrumentation-overhead}{432.69266}
\DefMacro{Comparison-6-algo-dylin_libs-p3-instrumentation-overhead}{916.24412}
\DefMacro{Comparison-6-algo-dylin_libs-p4-instrumentation-overhead}{1330.99773}
\DefMacro{Comparison-6-algo-dylin_libs-p5-instrumentation-overhead}{1972.70907}
\DefMacro{Comparison-6-algo-dylin_libs-p6-instrumentation-overhead}{2662.88834}
\DefMacro{Comparison-6-algo-dylin_libs-p7-instrumentation-overhead}{3487.81480}
\DefMacro{Comparison-6-algo-dylin_libs-p8-instrumentation-overhead}{4236.32898}
\DefMacro{Comparison-6-algo-dylin_libs-p9-instrumentation-overhead}{5199.88576}
\DefMacro{Comparison-6-algo-dylin_libs-p10-instrumentation-overhead}{9344.22557}
\DefMacro{Comparison-6-algo-D-p1-monitoring-overhead}{1.32679}
\DefMacro{Comparison-6-algo-D-p2-monitoring-overhead}{1.71437}
\DefMacro{Comparison-6-algo-D-p3-monitoring-overhead}{2.11097}
\DefMacro{Comparison-6-algo-D-p4-monitoring-overhead}{2.61930}
\DefMacro{Comparison-6-algo-D-p5-monitoring-overhead}{3.22359}
\DefMacro{Comparison-6-algo-D-p6-monitoring-overhead}{3.89487}
\DefMacro{Comparison-6-algo-D-p7-monitoring-overhead}{4.79626}
\DefMacro{Comparison-6-algo-D-p8-monitoring-overhead}{5.94614}
\DefMacro{Comparison-6-algo-D-p9-monitoring-overhead}{8.05111}
\DefMacro{Comparison-6-algo-D-p10-monitoring-overhead}{29.53253}
\DefMacro{Comparison-6-algo-dylin_libs-p1-monitoring-overhead}{1.01578}
\DefMacro{Comparison-6-algo-dylin_libs-p2-monitoring-overhead}{1.21199}
\DefMacro{Comparison-6-algo-dylin_libs-p3-monitoring-overhead}{1.37210}
\DefMacro{Comparison-6-algo-dylin_libs-p4-monitoring-overhead}{1.51716}
\DefMacro{Comparison-6-algo-dylin_libs-p5-monitoring-overhead}{1.63896}
\DefMacro{Comparison-6-algo-dylin_libs-p6-monitoring-overhead}{1.77214}
\DefMacro{Comparison-6-algo-dylin_libs-p7-monitoring-overhead}{1.94849}
\DefMacro{Comparison-6-algo-dylin_libs-p8-monitoring-overhead}{2.26922}
\DefMacro{Comparison-6-algo-dylin_libs-p9-monitoring-overhead}{2.86070}
\DefMacro{Comparison-6-algo-dylin_libs-p10-monitoring-overhead}{93.79731}
\DefMacro{Comparison-6-total-projects}{928}

\DefMacro{Comparison-7-algo-D-p1-overhead-min}{-17.531935691833496}
\DefMacro{Comparison-7-algo-D-p1-overhead-max}{6.539158821105955}
\DefMacro{Comparison-7-algo-D-p1-overhead-mean}{2.748768604937054}
\DefMacro{Comparison-7-algo-D-p1-overhead-median}{3.0437655448913574}
\DefMacro{Comparison-7-algo-D-p1-overhead-sum}{230.89656281471252}
\DefMacro{Comparison-7-algo-D-p2-overhead-min}{0.9138510227203367}
\DefMacro{Comparison-7-algo-D-p2-overhead-max}{7.8316168785095215}
\DefMacro{Comparison-7-algo-D-p2-overhead-mean}{2.934550112202054}
\DefMacro{Comparison-7-algo-D-p2-overhead-median}{2.9226174354553223}
\DefMacro{Comparison-7-algo-D-p2-overhead-sum}{246.50220942497253}
\DefMacro{Comparison-7-algo-D-p3-overhead-min}{1.3851821422576904}
\DefMacro{Comparison-7-algo-D-p3-overhead-max}{21.77379465103149}
\DefMacro{Comparison-7-algo-D-p3-overhead-mean}{3.624111842541468}
\DefMacro{Comparison-7-algo-D-p3-overhead-median}{3.237674832344055}
\DefMacro{Comparison-7-algo-D-p3-overhead-sum}{304.4253947734833}
\DefMacro{Comparison-7-algo-D-p4-overhead-min}{-0.6688289642333984}
\DefMacro{Comparison-7-algo-D-p4-overhead-max}{28.310580492019657}
\DefMacro{Comparison-7-algo-D-p4-overhead-mean}{3.9994160589717684}
\DefMacro{Comparison-7-algo-D-p4-overhead-median}{3.3517943620681763}
\DefMacro{Comparison-7-algo-D-p4-overhead-sum}{335.95094895362854}
\DefMacro{Comparison-7-algo-D-p5-overhead-min}{1.1801040172576904}
\DefMacro{Comparison-7-algo-D-p5-overhead-max}{26.05061149597168}
\DefMacro{Comparison-7-algo-D-p5-overhead-mean}{4.823712981882549}
\DefMacro{Comparison-7-algo-D-p5-overhead-median}{3.679613471031189}
\DefMacro{Comparison-7-algo-D-p5-overhead-sum}{405.19189047813416}
\DefMacro{Comparison-7-algo-D-p6-overhead-min}{-21.840582132339478}
\DefMacro{Comparison-7-algo-D-p6-overhead-max}{18.39743661880493}
\DefMacro{Comparison-7-algo-D-p6-overhead-mean}{3.7276109911146618}
\DefMacro{Comparison-7-algo-D-p6-overhead-median}{3.2490559816360474}
\DefMacro{Comparison-7-algo-D-p6-overhead-sum}{313.1193232536316}
\DefMacro{Comparison-7-algo-D-p7-overhead-min}{-4.346597671508789}
\DefMacro{Comparison-7-algo-D-p7-overhead-max}{44.480105638504014}
\DefMacro{Comparison-7-algo-D-p7-overhead-mean}{5.86995065779913}
\DefMacro{Comparison-7-algo-D-p7-overhead-median}{3.741107225418091}
\DefMacro{Comparison-7-algo-D-p7-overhead-sum}{493.07585525512695}
\DefMacro{Comparison-7-algo-D-p8-overhead-min}{1.5324716567993164}
\DefMacro{Comparison-7-algo-D-p8-overhead-max}{50.66503167152405}
\DefMacro{Comparison-7-algo-D-p8-overhead-mean}{5.829648494720459}
\DefMacro{Comparison-7-algo-D-p8-overhead-median}{3.601914882659912}
\DefMacro{Comparison-7-algo-D-p8-overhead-sum}{489.69047355651855}
\DefMacro{Comparison-7-algo-D-p9-overhead-min}{-0.3995933532714844}
\DefMacro{Comparison-7-algo-D-p9-overhead-max}{31.639625072479248}
\DefMacro{Comparison-7-algo-D-p9-overhead-mean}{5.945580945128486}
\DefMacro{Comparison-7-algo-D-p9-overhead-median}{4.042598366737366}
\DefMacro{Comparison-7-algo-D-p9-overhead-sum}{499.42879939079285}
\DefMacro{Comparison-7-algo-D-p10-overhead-min}{-6.343046188354492}
\DefMacro{Comparison-7-algo-D-p10-overhead-max}{873.9850316047668}
\DefMacro{Comparison-7-algo-D-p10-overhead-mean}{18.661349296569824}
\DefMacro{Comparison-7-algo-D-p10-overhead-median}{4.156611204147338}
\DefMacro{Comparison-7-algo-D-p10-overhead-sum}{1567.5533409118652}
\DefMacro{Comparison-7-algo-dynapyt-p1-overhead-min}{-17.354395627975464}
\DefMacro{Comparison-7-algo-dynapyt-p1-overhead-max}{46.38714909553528}
\DefMacro{Comparison-7-algo-dynapyt-p1-overhead-mean}{7.982053035781497}
\DefMacro{Comparison-7-algo-dynapyt-p1-overhead-median}{5.111110329627991}
\DefMacro{Comparison-7-algo-dynapyt-p1-overhead-sum}{670.4924550056458}
\DefMacro{Comparison-7-algo-dynapyt-p2-overhead-min}{0.6275177001953125}
\DefMacro{Comparison-7-algo-dynapyt-p2-overhead-max}{193.69263195991513}
\DefMacro{Comparison-7-algo-dynapyt-p2-overhead-mean}{12.373524526755014}
\DefMacro{Comparison-7-algo-dynapyt-p2-overhead-median}{3.896029472351074}
\DefMacro{Comparison-7-algo-dynapyt-p2-overhead-sum}{1039.3760602474213}
\DefMacro{Comparison-7-algo-dynapyt-p3-overhead-min}{0.7138285636901855}
\DefMacro{Comparison-7-algo-dynapyt-p3-overhead-max}{115.97745347023012}
\DefMacro{Comparison-7-algo-dynapyt-p3-overhead-mean}{11.648647819246564}
\DefMacro{Comparison-7-algo-dynapyt-p3-overhead-median}{3.357727289199829}
\DefMacro{Comparison-7-algo-dynapyt-p3-overhead-sum}{978.4864168167114}
\DefMacro{Comparison-7-algo-dynapyt-p4-overhead-min}{-1.3809013366699219}
\DefMacro{Comparison-7-algo-dynapyt-p4-overhead-max}{686.9892148971558}
\DefMacro{Comparison-7-algo-dynapyt-p4-overhead-mean}{24.77929675579071}
\DefMacro{Comparison-7-algo-dynapyt-p4-overhead-median}{5.377018213272095}
\DefMacro{Comparison-7-algo-dynapyt-p4-overhead-sum}{2081.4609274864197}
\DefMacro{Comparison-7-algo-dynapyt-p5-overhead-min}{0.8122794628143308}
\DefMacro{Comparison-7-algo-dynapyt-p5-overhead-max}{187.0938799381256}
\DefMacro{Comparison-7-algo-dynapyt-p5-overhead-mean}{23.791187822818756}
\DefMacro{Comparison-7-algo-dynapyt-p5-overhead-median}{6.243066787719727}
\DefMacro{Comparison-7-algo-dynapyt-p5-overhead-sum}{1998.4597771167755}
\DefMacro{Comparison-7-algo-dynapyt-p6-overhead-min}{-18.877495288848877}
\DefMacro{Comparison-7-algo-dynapyt-p6-overhead-max}{348.60596799850464}
\DefMacro{Comparison-7-algo-dynapyt-p6-overhead-mean}{19.29909351893834}
\DefMacro{Comparison-7-algo-dynapyt-p6-overhead-median}{3.084717869758606}
\DefMacro{Comparison-7-algo-dynapyt-p6-overhead-sum}{1621.1238555908203}
\DefMacro{Comparison-7-algo-dynapyt-p7-overhead-min}{-1.0980651378631592}
\DefMacro{Comparison-7-algo-dynapyt-p7-overhead-max}{851.6508932113647}
\DefMacro{Comparison-7-algo-dynapyt-p7-overhead-mean}{48.972306958266664}
\DefMacro{Comparison-7-algo-dynapyt-p7-overhead-median}{6.707386493682861}
\DefMacro{Comparison-7-algo-dynapyt-p7-overhead-sum}{4113.6737844944}
\DefMacro{Comparison-7-algo-dynapyt-p8-overhead-min}{0.6019384860992432}
\DefMacro{Comparison-7-algo-dynapyt-p8-overhead-max}{390.67218804359436}
\DefMacro{Comparison-7-algo-dynapyt-p8-overhead-mean}{40.10726450454621}
\DefMacro{Comparison-7-algo-dynapyt-p8-overhead-median}{4.346783757209778}
\DefMacro{Comparison-7-algo-dynapyt-p8-overhead-sum}{3369.0102183818817}
\DefMacro{Comparison-7-algo-dynapyt-p9-overhead-min}{-1.2658727169036865}
\DefMacro{Comparison-7-algo-dynapyt-p9-overhead-max}{1133.6775350570679}
\DefMacro{Comparison-7-algo-dynapyt-p9-overhead-mean}{77.37625916798909}
\DefMacro{Comparison-7-algo-dynapyt-p9-overhead-median}{5.644607901573181}
\DefMacro{Comparison-7-algo-dynapyt-p9-overhead-sum}{6499.605770111084}
\DefMacro{Comparison-7-algo-dynapyt-p10-overhead-min}{-88.25002694129944}
\DefMacro{Comparison-7-algo-dynapyt-p10-overhead-max}{1855.106014728546}
\DefMacro{Comparison-7-algo-dynapyt-p10-overhead-mean}{141.4320734143257}
\DefMacro{Comparison-7-algo-dynapyt-p10-overhead-median}{8.803974032402039}
\DefMacro{Comparison-7-algo-dynapyt-p10-overhead-sum}{11880.29416680336}
\DefMacro{Comparison-7-algo-dynapyt_libs-p1-overhead-min}{1778.2218751907349}
\DefMacro{Comparison-7-algo-dynapyt_libs-p1-overhead-max}{1814.4825513362885}
\DefMacro{Comparison-7-algo-dynapyt_libs-p1-overhead-mean}{1803.8509397591863}
\DefMacro{Comparison-7-algo-dynapyt_libs-p1-overhead-median}{1804.7899178266525}
\DefMacro{Comparison-7-algo-dynapyt_libs-p1-overhead-sum}{151523.47893977165}
\DefMacro{Comparison-7-algo-dynapyt_libs-p2-overhead-min}{1814.934898853302}
\DefMacro{Comparison-7-algo-dynapyt_libs-p2-overhead-max}{1829.078911781311}
\DefMacro{Comparison-7-algo-dynapyt_libs-p2-overhead-mean}{1822.4575737317402}
\DefMacro{Comparison-7-algo-dynapyt_libs-p2-overhead-median}{1822.028775215149}
\DefMacro{Comparison-7-algo-dynapyt_libs-p2-overhead-sum}{153086.4361934662}
\DefMacro{Comparison-7-algo-dynapyt_libs-p3-overhead-min}{1829.088473081589}
\DefMacro{Comparison-7-algo-dynapyt_libs-p3-overhead-max}{1847.0980684757235}
\DefMacro{Comparison-7-algo-dynapyt_libs-p3-overhead-mean}{1837.7581105203856}
\DefMacro{Comparison-7-algo-dynapyt_libs-p3-overhead-median}{1837.123016834259}
\DefMacro{Comparison-7-algo-dynapyt_libs-p3-overhead-sum}{154371.6812837124}
\DefMacro{Comparison-7-algo-dynapyt_libs-p4-overhead-min}{1847.2854421138766}
\DefMacro{Comparison-7-algo-dynapyt_libs-p4-overhead-max}{1880.8443763256073}
\DefMacro{Comparison-7-algo-dynapyt_libs-p4-overhead-mean}{1861.4011870651018}
\DefMacro{Comparison-7-algo-dynapyt_libs-p4-overhead-median}{1860.2003601789474}
\DefMacro{Comparison-7-algo-dynapyt_libs-p4-overhead-sum}{156357.69971346855}
\DefMacro{Comparison-7-algo-dynapyt_libs-p5-overhead-min}{1882.461578130722}
\DefMacro{Comparison-7-algo-dynapyt_libs-p5-overhead-max}{1930.8233478069303}
\DefMacro{Comparison-7-algo-dynapyt_libs-p5-overhead-mean}{1903.8122963649887}
\DefMacro{Comparison-7-algo-dynapyt_libs-p5-overhead-median}{1904.538880109787}
\DefMacro{Comparison-7-algo-dynapyt_libs-p5-overhead-sum}{159920.23289465904}
\DefMacro{Comparison-7-algo-dynapyt_libs-p6-overhead-min}{1931.5317380428312}
\DefMacro{Comparison-7-algo-dynapyt_libs-p6-overhead-max}{1999.1631028652193}
\DefMacro{Comparison-7-algo-dynapyt_libs-p6-overhead-mean}{1966.0250244651522}
\DefMacro{Comparison-7-algo-dynapyt_libs-p6-overhead-median}{1966.0271843671799}
\DefMacro{Comparison-7-algo-dynapyt_libs-p6-overhead-sum}{165146.10205507278}
\DefMacro{Comparison-7-algo-dynapyt_libs-p7-overhead-min}{1999.7099647521975}
\DefMacro{Comparison-7-algo-dynapyt_libs-p7-overhead-max}{2063.7888598442078}
\DefMacro{Comparison-7-algo-dynapyt_libs-p7-overhead-mean}{2032.6238339827173}
\DefMacro{Comparison-7-algo-dynapyt_libs-p7-overhead-median}{2031.1380642652512}
\DefMacro{Comparison-7-algo-dynapyt_libs-p7-overhead-sum}{170740.40205454826}
\DefMacro{Comparison-7-algo-dynapyt_libs-p8-overhead-min}{2064.281797170639}
\DefMacro{Comparison-7-algo-dynapyt_libs-p8-overhead-max}{2138.649395942688}
\DefMacro{Comparison-7-algo-dynapyt_libs-p8-overhead-mean}{2097.0919548415004}
\DefMacro{Comparison-7-algo-dynapyt_libs-p8-overhead-median}{2093.1091723442078}
\DefMacro{Comparison-7-algo-dynapyt_libs-p8-overhead-sum}{176155.72420668602}
\DefMacro{Comparison-7-algo-dynapyt_libs-p9-overhead-min}{2140.8399410247803}
\DefMacro{Comparison-7-algo-dynapyt_libs-p9-overhead-max}{2336.405560731888}
\DefMacro{Comparison-7-algo-dynapyt_libs-p9-overhead-mean}{2212.4681802023024}
\DefMacro{Comparison-7-algo-dynapyt_libs-p9-overhead-median}{2199.5894438028336}
\DefMacro{Comparison-7-algo-dynapyt_libs-p9-overhead-sum}{185847.3271369934}
\DefMacro{Comparison-7-algo-dynapyt_libs-p10-overhead-min}{2348.322142124176}
\DefMacro{Comparison-7-algo-dynapyt_libs-p10-overhead-max}{3487.683670282364}
\DefMacro{Comparison-7-algo-dynapyt_libs-p10-overhead-mean}{2683.108499685923}
\DefMacro{Comparison-7-algo-dynapyt_libs-p10-overhead-median}{2598.840780735016}
\DefMacro{Comparison-7-algo-dynapyt_libs-p10-overhead-sum}{225381.11397361755}
\DefMacro{Comparison-7-algo-dylin-p1-overhead-min}{-52.62245154380798}
\DefMacro{Comparison-7-algo-dylin-p1-overhead-max}{48.18282890319824}
\DefMacro{Comparison-7-algo-dylin-p1-overhead-mean}{8.041549708162036}
\DefMacro{Comparison-7-algo-dylin-p1-overhead-median}{5.1485068798065186}
\DefMacro{Comparison-7-algo-dylin-p1-overhead-sum}{675.490175485611}
\DefMacro{Comparison-7-algo-dylin-p2-overhead-min}{0.7730720043182373}
\DefMacro{Comparison-7-algo-dylin-p2-overhead-max}{189.53585863113403}
\DefMacro{Comparison-7-algo-dylin-p2-overhead-mean}{12.605377395947775}
\DefMacro{Comparison-7-algo-dylin-p2-overhead-median}{4.182110905647278}
\DefMacro{Comparison-7-algo-dylin-p2-overhead-sum}{1058.851701259613}
\DefMacro{Comparison-7-algo-dylin-p3-overhead-min}{1.1285121440887451}
\DefMacro{Comparison-7-algo-dylin-p3-overhead-max}{116.00199890136719}
\DefMacro{Comparison-7-algo-dylin-p3-overhead-mean}{11.520155381588708}
\DefMacro{Comparison-7-algo-dylin-p3-overhead-median}{3.5988510847091675}
\DefMacro{Comparison-7-algo-dylin-p3-overhead-sum}{967.6930520534515}
\DefMacro{Comparison-7-algo-dylin-p4-overhead-min}{1.2123441696166992}
\DefMacro{Comparison-7-algo-dylin-p4-overhead-max}{672.4576458930969}
\DefMacro{Comparison-7-algo-dylin-p4-overhead-mean}{24.327204979601362}
\DefMacro{Comparison-7-algo-dylin-p4-overhead-median}{5.782668709754944}
\DefMacro{Comparison-7-algo-dylin-p4-overhead-sum}{2043.4852182865143}
\DefMacro{Comparison-7-algo-dylin-p5-overhead-min}{1.1052250862121582}
\DefMacro{Comparison-7-algo-dylin-p5-overhead-max}{188.61834216117862}
\DefMacro{Comparison-7-algo-dylin-p5-overhead-mean}{22.22085242044358}
\DefMacro{Comparison-7-algo-dylin-p5-overhead-median}{6.360453248023987}
\DefMacro{Comparison-7-algo-dylin-p5-overhead-sum}{1866.5516033172607}
\DefMacro{Comparison-7-algo-dylin-p6-overhead-min}{-29.890096426010132}
\DefMacro{Comparison-7-algo-dylin-p6-overhead-max}{349.7316405773163}
\DefMacro{Comparison-7-algo-dylin-p6-overhead-mean}{17.88346186705998}
\DefMacro{Comparison-7-algo-dylin-p6-overhead-median}{3.29276967048645}
\DefMacro{Comparison-7-algo-dylin-p6-overhead-sum}{1502.2107968330383}
\DefMacro{Comparison-7-algo-dylin-p7-overhead-min}{-0.5282485485076922}
\DefMacro{Comparison-7-algo-dylin-p7-overhead-max}{870.0474996566772}
\DefMacro{Comparison-7-algo-dylin-p7-overhead-mean}{48.395632658685955}
\DefMacro{Comparison-7-algo-dylin-p7-overhead-median}{6.837208867073059}
\DefMacro{Comparison-7-algo-dylin-p7-overhead-sum}{4065.2331433296204}
\DefMacro{Comparison-7-algo-dylin-p8-overhead-min}{0.7406978607177734}
\DefMacro{Comparison-7-algo-dylin-p8-overhead-max}{388.0977547168731}
\DefMacro{Comparison-7-algo-dylin-p8-overhead-mean}{37.57715527500425}
\DefMacro{Comparison-7-algo-dylin-p8-overhead-median}{4.600078701972961}
\DefMacro{Comparison-7-algo-dylin-p8-overhead-sum}{3156.481043100357}
\DefMacro{Comparison-7-algo-dylin-p9-overhead-min}{-2.514890432357788}
\DefMacro{Comparison-7-algo-dylin-p9-overhead-max}{1116.1469230651855}
\DefMacro{Comparison-7-algo-dylin-p9-overhead-mean}{73.13088862101237}
\DefMacro{Comparison-7-algo-dylin-p9-overhead-median}{5.139993786811829}
\DefMacro{Comparison-7-algo-dylin-p9-overhead-sum}{6142.994644165039}
\DefMacro{Comparison-7-algo-dylin-p10-overhead-min}{-87.8480007648468}
\DefMacro{Comparison-7-algo-dylin-p10-overhead-max}{1821.0141563415527}
\DefMacro{Comparison-7-algo-dylin-p10-overhead-mean}{126.24815846624828}
\DefMacro{Comparison-7-algo-dylin-p10-overhead-median}{8.605026721954346}
\DefMacro{Comparison-7-algo-dylin-p10-overhead-sum}{10604.845311164856}
\DefMacro{Comparison-7-algo-D-p1-instrumentation-overhead-min}{-321.25085163116455}
\DefMacro{Comparison-7-algo-D-p1-instrumentation-overhead-max}{1.0217971801757812}
\DefMacro{Comparison-7-algo-D-p1-instrumentation-overhead-mean}{-4.140381154559908}
\DefMacro{Comparison-7-algo-D-p1-instrumentation-overhead-median}{0.003671526908874456}
\DefMacro{Comparison-7-algo-D-p1-instrumentation-overhead-sum}{-347.7920169830322}
\DefMacro{Comparison-7-algo-D-p2-instrumentation-overhead-min}{-166.1955668926239}
\DefMacro{Comparison-7-algo-D-p2-instrumentation-overhead-max}{1.1736414432525635}
\DefMacro{Comparison-7-algo-D-p2-instrumentation-overhead-mean}{-2.461139897505442}
\DefMacro{Comparison-7-algo-D-p2-instrumentation-overhead-median}{0.014834046363830483}
\DefMacro{Comparison-7-algo-D-p2-instrumentation-overhead-sum}{-206.73575139045715}
\DefMacro{Comparison-7-algo-D-p3-instrumentation-overhead-min}{-28.865823030471805}
\DefMacro{Comparison-7-algo-D-p3-instrumentation-overhead-max}{1.266205072402954}
\DefMacro{Comparison-7-algo-D-p3-instrumentation-overhead-mean}{-0.7888724406560262}
\DefMacro{Comparison-7-algo-D-p3-instrumentation-overhead-median}{-0.08094906806945812}
\DefMacro{Comparison-7-algo-D-p3-instrumentation-overhead-sum}{-66.2652850151062}
\DefMacro{Comparison-7-algo-D-p4-instrumentation-overhead-min}{-87.95799040794373}
\DefMacro{Comparison-7-algo-D-p4-instrumentation-overhead-max}{0.8044617176055908}
\DefMacro{Comparison-7-algo-D-p4-instrumentation-overhead-mean}{-2.3601964314778647}
\DefMacro{Comparison-7-algo-D-p4-instrumentation-overhead-median}{-0.04524827003479007}
\DefMacro{Comparison-7-algo-D-p4-instrumentation-overhead-sum}{-198.25650024414062}
\DefMacro{Comparison-7-algo-D-p5-instrumentation-overhead-min}{-44.64909362792969}
\DefMacro{Comparison-7-algo-D-p5-instrumentation-overhead-max}{5.849094390869141}
\DefMacro{Comparison-7-algo-D-p5-instrumentation-overhead-mean}{-1.7188466446740287}
\DefMacro{Comparison-7-algo-D-p5-instrumentation-overhead-median}{-0.4787886142730714}
\DefMacro{Comparison-7-algo-D-p5-instrumentation-overhead-sum}{-144.3831181526184}
\DefMacro{Comparison-7-algo-D-p6-instrumentation-overhead-min}{-359.54006338119507}
\DefMacro{Comparison-7-algo-D-p6-instrumentation-overhead-max}{1.9810686111450195}
\DefMacro{Comparison-7-algo-D-p6-instrumentation-overhead-mean}{-7.306610868090675}
\DefMacro{Comparison-7-algo-D-p6-instrumentation-overhead-median}{-0.3255630731582642}
\DefMacro{Comparison-7-algo-D-p6-instrumentation-overhead-sum}{-613.7553129196167}
\DefMacro{Comparison-7-algo-D-p7-instrumentation-overhead-min}{-81.11103677749634}
\DefMacro{Comparison-7-algo-D-p7-instrumentation-overhead-max}{0.8648204803466797}
\DefMacro{Comparison-7-algo-D-p7-instrumentation-overhead-mean}{-5.8393836333638145}
\DefMacro{Comparison-7-algo-D-p7-instrumentation-overhead-median}{-0.5857867002487183}
\DefMacro{Comparison-7-algo-D-p7-instrumentation-overhead-sum}{-490.5082252025604}
\DefMacro{Comparison-7-algo-D-p8-instrumentation-overhead-min}{-58.443073987960815}
\DefMacro{Comparison-7-algo-D-p8-instrumentation-overhead-max}{3.2871854305267334}
\DefMacro{Comparison-7-algo-D-p8-instrumentation-overhead-mean}{-3.2925430167289007}
\DefMacro{Comparison-7-algo-D-p8-instrumentation-overhead-median}{-0.6724045276641846}
\DefMacro{Comparison-7-algo-D-p8-instrumentation-overhead-sum}{-276.57361340522766}
\DefMacro{Comparison-7-algo-D-p9-instrumentation-overhead-min}{-56.04716968536377}
\DefMacro{Comparison-7-algo-D-p9-instrumentation-overhead-max}{3.6351406574249268}
\DefMacro{Comparison-7-algo-D-p9-instrumentation-overhead-mean}{-5.160634994506836}
\DefMacro{Comparison-7-algo-D-p9-instrumentation-overhead-median}{-0.896268367767334}
\DefMacro{Comparison-7-algo-D-p9-instrumentation-overhead-sum}{-433.4933395385742}
\DefMacro{Comparison-7-algo-D-p10-instrumentation-overhead-min}{-139.19865083694458}
\DefMacro{Comparison-7-algo-D-p10-instrumentation-overhead-max}{1.103107213973999}
\DefMacro{Comparison-7-algo-D-p10-instrumentation-overhead-mean}{-9.119358241558075}
\DefMacro{Comparison-7-algo-D-p10-instrumentation-overhead-median}{-0.9175297021865845}
\DefMacro{Comparison-7-algo-D-p10-instrumentation-overhead-sum}{-766.0260922908783}
\DefMacro{Comparison-7-algo-dynapyt-p1-instrumentation-overhead-min}{-308.469765663147}
\DefMacro{Comparison-7-algo-dynapyt-p1-instrumentation-overhead-max}{44.20897722244263}
\DefMacro{Comparison-7-algo-dynapyt-p1-instrumentation-overhead-mean}{2.634938662960416}
\DefMacro{Comparison-7-algo-dynapyt-p1-instrumentation-overhead-median}{2.9732086658477783}
\DefMacro{Comparison-7-algo-dynapyt-p1-instrumentation-overhead-sum}{221.33484768867493}
\DefMacro{Comparison-7-algo-dynapyt-p2-instrumentation-overhead-min}{-155.8150761127472}
\DefMacro{Comparison-7-algo-dynapyt-p2-instrumentation-overhead-max}{186.28737807273868}
\DefMacro{Comparison-7-algo-dynapyt-p2-instrumentation-overhead-mean}{8.200591320083255}
\DefMacro{Comparison-7-algo-dynapyt-p2-instrumentation-overhead-median}{2.421523094177246}
\DefMacro{Comparison-7-algo-dynapyt-p2-instrumentation-overhead-sum}{688.8496708869934}
\DefMacro{Comparison-7-algo-dynapyt-p3-instrumentation-overhead-min}{-11.58431601524353}
\DefMacro{Comparison-7-algo-dynapyt-p3-instrumentation-overhead-max}{114.22463703155518}
\DefMacro{Comparison-7-algo-dynapyt-p3-instrumentation-overhead-mean}{7.843346558866047}
\DefMacro{Comparison-7-algo-dynapyt-p3-instrumentation-overhead-median}{2.1374610662460327}
\DefMacro{Comparison-7-algo-dynapyt-p3-instrumentation-overhead-sum}{658.8411109447479}
\DefMacro{Comparison-7-algo-dynapyt-p4-instrumentation-overhead-min}{-87.73895525932312}
\DefMacro{Comparison-7-algo-dynapyt-p4-instrumentation-overhead-max}{686.4190127849579}
\DefMacro{Comparison-7-algo-dynapyt-p4-instrumentation-overhead-mean}{19.4920578229995}
\DefMacro{Comparison-7-algo-dynapyt-p4-instrumentation-overhead-median}{2.452626943588257}
\DefMacro{Comparison-7-algo-dynapyt-p4-instrumentation-overhead-sum}{1637.332857131958}
\DefMacro{Comparison-7-algo-dynapyt-p5-instrumentation-overhead-min}{-44.709306955337524}
\DefMacro{Comparison-7-algo-dynapyt-p5-instrumentation-overhead-max}{173.2667052745819}
\DefMacro{Comparison-7-algo-dynapyt-p5-instrumentation-overhead-mean}{17.42510690007891}
\DefMacro{Comparison-7-algo-dynapyt-p5-instrumentation-overhead-median}{2.759092092514038}
\DefMacro{Comparison-7-algo-dynapyt-p5-instrumentation-overhead-sum}{1463.7089796066284}
\DefMacro{Comparison-7-algo-dynapyt-p6-instrumentation-overhead-min}{-353.9785542488098}
\DefMacro{Comparison-7-algo-dynapyt-p6-instrumentation-overhead-max}{347.5933165550232}
\DefMacro{Comparison-7-algo-dynapyt-p6-instrumentation-overhead-mean}{8.431617189021338}
\DefMacro{Comparison-7-algo-dynapyt-p6-instrumentation-overhead-median}{1.0644084215164185}
\DefMacro{Comparison-7-algo-dynapyt-p6-instrumentation-overhead-sum}{708.2558438777924}
\DefMacro{Comparison-7-algo-dynapyt-p7-instrumentation-overhead-min}{-80.24102663993835}
\DefMacro{Comparison-7-algo-dynapyt-p7-instrumentation-overhead-max}{848.7959411144257}
\DefMacro{Comparison-7-algo-dynapyt-p7-instrumentation-overhead-mean}{38.05153015681675}
\DefMacro{Comparison-7-algo-dynapyt-p7-instrumentation-overhead-median}{2.1279326677322388}
\DefMacro{Comparison-7-algo-dynapyt-p7-instrumentation-overhead-sum}{3196.3285331726074}
\DefMacro{Comparison-7-algo-dynapyt-p8-instrumentation-overhead-min}{-16.515295028686523}
\DefMacro{Comparison-7-algo-dynapyt-p8-instrumentation-overhead-max}{384.5600051879883}
\DefMacro{Comparison-7-algo-dynapyt-p8-instrumentation-overhead-mean}{28.78119315419878}
\DefMacro{Comparison-7-algo-dynapyt-p8-instrumentation-overhead-median}{1.6272201538085938}
\DefMacro{Comparison-7-algo-dynapyt-p8-instrumentation-overhead-sum}{2417.6202249526978}
\DefMacro{Comparison-7-algo-dynapyt-p9-instrumentation-overhead-min}{-22.814985752105713}
\DefMacro{Comparison-7-algo-dynapyt-p9-instrumentation-overhead-max}{1126.3075969219208}
\DefMacro{Comparison-7-algo-dynapyt-p9-instrumentation-overhead-mean}{63.323083077158245}
\DefMacro{Comparison-7-algo-dynapyt-p9-instrumentation-overhead-median}{1.7338212728500366}
\DefMacro{Comparison-7-algo-dynapyt-p9-instrumentation-overhead-sum}{5319.138978481293}
\DefMacro{Comparison-7-algo-dynapyt-p10-instrumentation-overhead-min}{-88.86849045753479}
\DefMacro{Comparison-7-algo-dynapyt-p10-instrumentation-overhead-max}{1854.1805913448331}
\DefMacro{Comparison-7-algo-dynapyt-p10-instrumentation-overhead-mean}{106.39562572184063}
\DefMacro{Comparison-7-algo-dynapyt-p10-instrumentation-overhead-median}{2.2840335369110107}
\DefMacro{Comparison-7-algo-dynapyt-p10-instrumentation-overhead-sum}{8937.232560634613}
\DefMacro{Comparison-7-algo-dynapyt_libs-p1-instrumentation-overhead-min}{1530.3257734775543}
\DefMacro{Comparison-7-algo-dynapyt_libs-p1-instrumentation-overhead-max}{1812.0203564167023}
\DefMacro{Comparison-7-algo-dynapyt_libs-p1-instrumentation-overhead-mean}{1795.920425352596}
\DefMacro{Comparison-7-algo-dynapyt_libs-p1-instrumentation-overhead-median}{1800.4789643287659}
\DefMacro{Comparison-7-algo-dynapyt_libs-p1-instrumentation-overhead-sum}{150857.31572961807}
\DefMacro{Comparison-7-algo-dynapyt_libs-p2-instrumentation-overhead-min}{1650.174827337265}
\DefMacro{Comparison-7-algo-dynapyt_libs-p2-instrumentation-overhead-max}{1827.019795179367}
\DefMacro{Comparison-7-algo-dynapyt_libs-p2-instrumentation-overhead-mean}{1815.896051960332}
\DefMacro{Comparison-7-algo-dynapyt_libs-p2-instrumentation-overhead-median}{1818.5565999746323}
\DefMacro{Comparison-7-algo-dynapyt_libs-p2-instrumentation-overhead-sum}{152535.2683646679}
\DefMacro{Comparison-7-algo-dynapyt_libs-p3-instrumentation-overhead-min}{1746.7550020217893}
\DefMacro{Comparison-7-algo-dynapyt_libs-p3-instrumentation-overhead-max}{1844.6919624805448}
\DefMacro{Comparison-7-algo-dynapyt_libs-p3-instrumentation-overhead-mean}{1829.5629129296258}
\DefMacro{Comparison-7-algo-dynapyt_libs-p3-instrumentation-overhead-median}{1830.6823464632034}
\DefMacro{Comparison-7-algo-dynapyt_libs-p3-instrumentation-overhead-sum}{153683.28468608856}
\DefMacro{Comparison-7-algo-dynapyt_libs-p4-instrumentation-overhead-min}{1754.4613277912142}
\DefMacro{Comparison-7-algo-dynapyt_libs-p4-instrumentation-overhead-max}{1877.7798187732697}
\DefMacro{Comparison-7-algo-dynapyt_libs-p4-instrumentation-overhead-mean}{1850.1196627758798}
\DefMacro{Comparison-7-algo-dynapyt_libs-p4-instrumentation-overhead-median}{1853.715552330017}
\DefMacro{Comparison-7-algo-dynapyt_libs-p4-instrumentation-overhead-sum}{155410.0516731739}
\DefMacro{Comparison-7-algo-dynapyt_libs-p5-instrumentation-overhead-min}{1803.269667387009}
\DefMacro{Comparison-7-algo-dynapyt_libs-p5-instrumentation-overhead-max}{1924.2359721660614}
\DefMacro{Comparison-7-algo-dynapyt_libs-p5-instrumentation-overhead-mean}{1885.0449880844071}
\DefMacro{Comparison-7-algo-dynapyt_libs-p5-instrumentation-overhead-median}{1889.8985019922256}
\DefMacro{Comparison-7-algo-dynapyt_libs-p5-instrumentation-overhead-sum}{158343.7789990902}
\DefMacro{Comparison-7-algo-dynapyt_libs-p6-instrumentation-overhead-min}{1532.250590801239}
\DefMacro{Comparison-7-algo-dynapyt_libs-p6-instrumentation-overhead-max}{1997.5151724815369}
\DefMacro{Comparison-7-algo-dynapyt_libs-p6-instrumentation-overhead-mean}{1942.6008326382864}
\DefMacro{Comparison-7-algo-dynapyt_libs-p6-instrumentation-overhead-median}{1950.7845579385757}
\DefMacro{Comparison-7-algo-dynapyt_libs-p6-instrumentation-overhead-sum}{163178.46994161606}
\DefMacro{Comparison-7-algo-dynapyt_libs-p7-instrumentation-overhead-min}{1790.6737489700317}
\DefMacro{Comparison-7-algo-dynapyt_libs-p7-instrumentation-overhead-max}{2061.0032844543457}
\DefMacro{Comparison-7-algo-dynapyt_libs-p7-instrumentation-overhead-mean}{2001.521221711522}
\DefMacro{Comparison-7-algo-dynapyt_libs-p7-instrumentation-overhead-median}{2014.3923954963684}
\DefMacro{Comparison-7-algo-dynapyt_libs-p7-instrumentation-overhead-sum}{168127.78262376785}
\DefMacro{Comparison-7-algo-dynapyt_libs-p8-instrumentation-overhead-min}{1872.6211462020874}
\DefMacro{Comparison-7-algo-dynapyt_libs-p8-instrumentation-overhead-max}{2133.524235010147}
\DefMacro{Comparison-7-algo-dynapyt_libs-p8-instrumentation-overhead-mean}{2063.3280513655573}
\DefMacro{Comparison-7-algo-dynapyt_libs-p8-instrumentation-overhead-median}{2072.511999130249}
\DefMacro{Comparison-7-algo-dynapyt_libs-p8-instrumentation-overhead-sum}{173319.5563147068}
\DefMacro{Comparison-7-algo-dynapyt_libs-p9-instrumentation-overhead-min}{1839.8296110630033}
\DefMacro{Comparison-7-algo-dynapyt_libs-p9-instrumentation-overhead-max}{2317.4503769874573}
\DefMacro{Comparison-7-algo-dynapyt_libs-p9-instrumentation-overhead-mean}{2143.7460478714534}
\DefMacro{Comparison-7-algo-dynapyt_libs-p9-instrumentation-overhead-median}{2152.017005085945}
\DefMacro{Comparison-7-algo-dynapyt_libs-p9-instrumentation-overhead-sum}{180074.6680212021}
\DefMacro{Comparison-7-algo-dynapyt_libs-p10-instrumentation-overhead-min}{1726.624150753021}
\DefMacro{Comparison-7-algo-dynapyt_libs-p10-instrumentation-overhead-max}{3482.6261785030365}
\DefMacro{Comparison-7-algo-dynapyt_libs-p10-instrumentation-overhead-mean}{2487.5308292252676}
\DefMacro{Comparison-7-algo-dynapyt_libs-p10-instrumentation-overhead-median}{2475.9798144102097}
\DefMacro{Comparison-7-algo-dynapyt_libs-p10-instrumentation-overhead-sum}{208952.58965492249}
\DefMacro{Comparison-7-algo-dylin-p1-instrumentation-overhead-min}{-307.84759640693665}
\DefMacro{Comparison-7-algo-dylin-p1-instrumentation-overhead-max}{44.580864906311035}
\DefMacro{Comparison-7-algo-dylin-p1-instrumentation-overhead-mean}{2.9660558814094182}
\DefMacro{Comparison-7-algo-dylin-p1-instrumentation-overhead-median}{3.2777488231658936}
\DefMacro{Comparison-7-algo-dylin-p1-instrumentation-overhead-sum}{249.1486940383911}
\DefMacro{Comparison-7-algo-dylin-p2-instrumentation-overhead-min}{-155.38018488883972}
\DefMacro{Comparison-7-algo-dylin-p2-instrumentation-overhead-max}{182.21632432937622}
\DefMacro{Comparison-7-algo-dylin-p2-instrumentation-overhead-mean}{8.48475695507867}
\DefMacro{Comparison-7-algo-dylin-p2-instrumentation-overhead-median}{2.870360255241394}
\DefMacro{Comparison-7-algo-dylin-p2-instrumentation-overhead-sum}{712.7195842266083}
\DefMacro{Comparison-7-algo-dylin-p3-instrumentation-overhead-min}{-11.174869537353516}
\DefMacro{Comparison-7-algo-dylin-p3-instrumentation-overhead-max}{114.3472979068756}
\DefMacro{Comparison-7-algo-dylin-p3-instrumentation-overhead-mean}{8.125854608558473}
\DefMacro{Comparison-7-algo-dylin-p3-instrumentation-overhead-median}{2.445104241371155}
\DefMacro{Comparison-7-algo-dylin-p3-instrumentation-overhead-sum}{682.5717871189117}
\DefMacro{Comparison-7-algo-dylin-p4-instrumentation-overhead-min}{-87.33548426628113}
\DefMacro{Comparison-7-algo-dylin-p4-instrumentation-overhead-max}{671.8593668937683}
\DefMacro{Comparison-7-algo-dylin-p4-instrumentation-overhead-mean}{19.4077601035436}
\DefMacro{Comparison-7-algo-dylin-p4-instrumentation-overhead-median}{2.850053548812866}
\DefMacro{Comparison-7-algo-dylin-p4-instrumentation-overhead-sum}{1630.2518486976624}
\DefMacro{Comparison-7-algo-dylin-p5-instrumentation-overhead-min}{-44.326610803604126}
\DefMacro{Comparison-7-algo-dylin-p5-instrumentation-overhead-max}{171.45688128471375}
\DefMacro{Comparison-7-algo-dylin-p5-instrumentation-overhead-mean}{17.51139297939482}
\DefMacro{Comparison-7-algo-dylin-p5-instrumentation-overhead-median}{3.2135894298553467}
\DefMacro{Comparison-7-algo-dylin-p5-instrumentation-overhead-sum}{1470.957010269165}
\DefMacro{Comparison-7-algo-dylin-p6-instrumentation-overhead-min}{-353.94976234436035}
\DefMacro{Comparison-7-algo-dylin-p6-instrumentation-overhead-max}{348.68772196769714}
\DefMacro{Comparison-7-algo-dylin-p6-instrumentation-overhead-mean}{8.67625322512218}
\DefMacro{Comparison-7-algo-dylin-p6-instrumentation-overhead-median}{1.4119921922683716}
\DefMacro{Comparison-7-algo-dylin-p6-instrumentation-overhead-sum}{728.8052709102631}
\DefMacro{Comparison-7-algo-dylin-p7-instrumentation-overhead-min}{-79.85382652282715}
\DefMacro{Comparison-7-algo-dylin-p7-instrumentation-overhead-max}{866.6253387928009}
\DefMacro{Comparison-7-algo-dylin-p7-instrumentation-overhead-mean}{38.29351228475571}
\DefMacro{Comparison-7-algo-dylin-p7-instrumentation-overhead-median}{2.4618674516677856}
\DefMacro{Comparison-7-algo-dylin-p7-instrumentation-overhead-sum}{3216.6550319194794}
\DefMacro{Comparison-7-algo-dylin-p8-instrumentation-overhead-min}{-16.115252256393433}
\DefMacro{Comparison-7-algo-dylin-p8-instrumentation-overhead-max}{381.94793152809143}
\DefMacro{Comparison-7-algo-dylin-p8-instrumentation-overhead-mean}{28.93229142824809}
\DefMacro{Comparison-7-algo-dylin-p8-instrumentation-overhead-median}{2.0223416090011597}
\DefMacro{Comparison-7-algo-dylin-p8-instrumentation-overhead-sum}{2430.3124799728394}
\DefMacro{Comparison-7-algo-dylin-p9-instrumentation-overhead-min}{-22.45859956741333}
\DefMacro{Comparison-7-algo-dylin-p9-instrumentation-overhead-max}{1108.8055164813995}
\DefMacro{Comparison-7-algo-dylin-p9-instrumentation-overhead-mean}{62.86916653882889}
\DefMacro{Comparison-7-algo-dylin-p9-instrumentation-overhead-median}{2.1508160829544067}
\DefMacro{Comparison-7-algo-dylin-p9-instrumentation-overhead-sum}{5281.009989261627}
\DefMacro{Comparison-7-algo-dylin-p10-instrumentation-overhead-min}{-88.47398257255554}
\DefMacro{Comparison-7-algo-dylin-p10-instrumentation-overhead-max}{1820.0720798969269}
\DefMacro{Comparison-7-algo-dylin-p10-instrumentation-overhead-mean}{105.15950918765296}
\DefMacro{Comparison-7-algo-dylin-p10-instrumentation-overhead-median}{2.6504260301589966}
\DefMacro{Comparison-7-algo-dylin-p10-instrumentation-overhead-sum}{8833.398771762848}
\DefMacro{Comparison-7-algo-D-p1-monitoring-overhead-min}{-18.122225522994995}
\DefMacro{Comparison-7-algo-D-p1-monitoring-overhead-max}{5.967567682266235}
\DefMacro{Comparison-7-algo-D-p1-monitoring-overhead-mean}{2.145373489175524}
\DefMacro{Comparison-7-algo-D-p1-monitoring-overhead-median}{2.4238299131393433}
\DefMacro{Comparison-7-algo-D-p1-monitoring-overhead-sum}{180.21137309074402}
\DefMacro{Comparison-7-algo-D-p2-monitoring-overhead-min}{0.5654840469360352}
\DefMacro{Comparison-7-algo-D-p2-monitoring-overhead-max}{6.178576707839966}
\DefMacro{Comparison-7-algo-D-p2-monitoring-overhead-mean}{2.3055027780078707}
\DefMacro{Comparison-7-algo-D-p2-monitoring-overhead-median}{2.2573262453079224}
\DefMacro{Comparison-7-algo-D-p2-monitoring-overhead-sum}{193.66223335266113}
\DefMacro{Comparison-7-algo-D-p3-monitoring-overhead-min}{1.0317635536193848}
\DefMacro{Comparison-7-algo-D-p3-monitoring-overhead-max}{21.097189664840695}
\DefMacro{Comparison-7-algo-D-p3-monitoring-overhead-mean}{2.9622250426383245}
\DefMacro{Comparison-7-algo-D-p3-monitoring-overhead-median}{2.6499102115631104}
\DefMacro{Comparison-7-algo-D-p3-monitoring-overhead-sum}{248.82690358161926}
\DefMacro{Comparison-7-algo-D-p4-monitoring-overhead-min}{-1.3770387172698975}
\DefMacro{Comparison-7-algo-D-p4-monitoring-overhead-max}{27.516993761062622}
\DefMacro{Comparison-7-algo-D-p4-monitoring-overhead-mean}{3.3233223926453364}
\DefMacro{Comparison-7-algo-D-p4-monitoring-overhead-median}{2.6685476303100586}
\DefMacro{Comparison-7-algo-D-p4-monitoring-overhead-sum}{279.15908098220825}
\DefMacro{Comparison-7-algo-D-p5-monitoring-overhead-min}{0.8142914772033691}
\DefMacro{Comparison-7-algo-D-p5-monitoring-overhead-max}{25.463481187820435}
\DefMacro{Comparison-7-algo-D-p5-monitoring-overhead-mean}{4.045404956454322}
\DefMacro{Comparison-7-algo-D-p5-monitoring-overhead-median}{2.8634328842163086}
\DefMacro{Comparison-7-algo-D-p5-monitoring-overhead-sum}{339.8140163421631}
\DefMacro{Comparison-7-algo-D-p6-monitoring-overhead-min}{-22.659925937652588}
\DefMacro{Comparison-7-algo-D-p6-monitoring-overhead-max}{17.558505058288574}
\DefMacro{Comparison-7-algo-D-p6-monitoring-overhead-mean}{2.9748317201932273}
\DefMacro{Comparison-7-algo-D-p6-monitoring-overhead-median}{2.5767662525177}
\DefMacro{Comparison-7-algo-D-p6-monitoring-overhead-sum}{249.88586449623108}
\DefMacro{Comparison-7-algo-D-p7-monitoring-overhead-min}{-4.9940345287323}
\DefMacro{Comparison-7-algo-D-p7-monitoring-overhead-max}{43.48754382133484}
\DefMacro{Comparison-7-algo-D-p7-monitoring-overhead-mean}{5.166057711555844}
\DefMacro{Comparison-7-algo-D-p7-monitoring-overhead-median}{2.9078503847122192}
\DefMacro{Comparison-7-algo-D-p7-monitoring-overhead-sum}{433.9488477706909}
\DefMacro{Comparison-7-algo-D-p8-monitoring-overhead-min}{1.1675071716308594}
\DefMacro{Comparison-7-algo-D-p8-monitoring-overhead-max}{49.972089529037476}
\DefMacro{Comparison-7-algo-D-p8-monitoring-overhead-mean}{5.0951843829382035}
\DefMacro{Comparison-7-algo-D-p8-monitoring-overhead-median}{2.9085569381713867}
\DefMacro{Comparison-7-algo-D-p8-monitoring-overhead-sum}{427.9954881668091}
\DefMacro{Comparison-7-algo-D-p9-monitoring-overhead-min}{-0.8606216907501256}
\DefMacro{Comparison-7-algo-D-p9-monitoring-overhead-max}{30.932053804397583}
\DefMacro{Comparison-7-algo-D-p9-monitoring-overhead-mean}{5.163510947000413}
\DefMacro{Comparison-7-algo-D-p9-monitoring-overhead-median}{3.2510071992874146}
\DefMacro{Comparison-7-algo-D-p9-monitoring-overhead-sum}{433.73491954803467}
\DefMacro{Comparison-7-algo-D-p10-monitoring-overhead-min}{-6.680638074874878}
\DefMacro{Comparison-7-algo-D-p10-monitoring-overhead-max}{873.4039697647095}
\DefMacro{Comparison-7-algo-D-p10-monitoring-overhead-mean}{17.96356452362878}
\DefMacro{Comparison-7-algo-D-p10-monitoring-overhead-median}{3.528520941734314}
\DefMacro{Comparison-7-algo-D-p10-monitoring-overhead-sum}{1508.9394199848175}
\DefMacro{Comparison-7-algo-dynapyt-p1-monitoring-overhead-min}{-30.72577142715454}
\DefMacro{Comparison-7-algo-dynapyt-p1-monitoring-overhead-max}{16.795938968658447}
\DefMacro{Comparison-7-algo-dynapyt-p1-monitoring-overhead-mean}{0.6033381024996439}
\DefMacro{Comparison-7-algo-dynapyt-p1-monitoring-overhead-median}{0.39203691482543945}
\DefMacro{Comparison-7-algo-dynapyt-p1-monitoring-overhead-sum}{50.68040060997009}
\DefMacro{Comparison-7-algo-dynapyt-p2-monitoring-overhead-min}{-0.2944149971008301}
\DefMacro{Comparison-7-algo-dynapyt-p2-monitoring-overhead-max}{27.354622364044186}
\DefMacro{Comparison-7-algo-dynapyt-p2-monitoring-overhead-mean}{1.0827459749721347}
\DefMacro{Comparison-7-algo-dynapyt-p2-monitoring-overhead-median}{0.41241562366485596}
\DefMacro{Comparison-7-algo-dynapyt-p2-monitoring-overhead-sum}{90.9506618976593}
\DefMacro{Comparison-7-algo-dynapyt-p3-monitoring-overhead-min}{0.0034863948822021484}
\DefMacro{Comparison-7-algo-dynapyt-p3-monitoring-overhead-max}{27.619719982147217}
\DefMacro{Comparison-7-algo-dynapyt-p3-monitoring-overhead-mean}{2.3545420198213485}
\DefMacro{Comparison-7-algo-dynapyt-p3-monitoring-overhead-median}{0.4005414247512817}
\DefMacro{Comparison-7-algo-dynapyt-p3-monitoring-overhead-sum}{197.7815296649933}
\DefMacro{Comparison-7-algo-dynapyt-p4-monitoring-overhead-min}{-2.3081462383270264}
\DefMacro{Comparison-7-algo-dynapyt-p4-monitoring-overhead-max}{50.30504250526428}
\DefMacro{Comparison-7-algo-dynapyt-p4-monitoring-overhead-mean}{2.2509488349869136}
\DefMacro{Comparison-7-algo-dynapyt-p4-monitoring-overhead-median}{0.4537639617919922}
\DefMacro{Comparison-7-algo-dynapyt-p4-monitoring-overhead-sum}{189.07970213890076}
\DefMacro{Comparison-7-algo-dynapyt-p5-monitoring-overhead-min}{-0.12554931640625}
\DefMacro{Comparison-7-algo-dynapyt-p5-monitoring-overhead-max}{47.264092445373535}
\DefMacro{Comparison-7-algo-dynapyt-p5-monitoring-overhead-mean}{3.8689262526375905}
\DefMacro{Comparison-7-algo-dynapyt-p5-monitoring-overhead-median}{0.5416988134384155}
\DefMacro{Comparison-7-algo-dynapyt-p5-monitoring-overhead-sum}{324.9898052215576}
\DefMacro{Comparison-7-algo-dynapyt-p6-monitoring-overhead-min}{-25.25834822654724}
\DefMacro{Comparison-7-algo-dynapyt-p6-monitoring-overhead-max}{137.1548614501953}
\DefMacro{Comparison-7-algo-dynapyt-p6-monitoring-overhead-mean}{2.80808619090489}
\DefMacro{Comparison-7-algo-dynapyt-p6-monitoring-overhead-median}{0.4522373676300049}
\DefMacro{Comparison-7-algo-dynapyt-p6-monitoring-overhead-sum}{235.87924003601074}
\DefMacro{Comparison-7-algo-dynapyt-p7-monitoring-overhead-min}{-1.822585582733156}
\DefMacro{Comparison-7-algo-dynapyt-p7-monitoring-overhead-max}{41.95569396018982}
\DefMacro{Comparison-7-algo-dynapyt-p7-monitoring-overhead-mean}{4.377500221842811}
\DefMacro{Comparison-7-algo-dynapyt-p7-monitoring-overhead-median}{0.6545093059539795}
\DefMacro{Comparison-7-algo-dynapyt-p7-monitoring-overhead-sum}{367.71001863479614}
\DefMacro{Comparison-7-algo-dynapyt-p8-monitoring-overhead-min}{-0.46936464309692383}
\DefMacro{Comparison-7-algo-dynapyt-p8-monitoring-overhead-max}{208.64656019210813}
\DefMacro{Comparison-7-algo-dynapyt-p8-monitoring-overhead-mean}{7.299064221836272}
\DefMacro{Comparison-7-algo-dynapyt-p8-monitoring-overhead-median}{0.5302292108535767}
\DefMacro{Comparison-7-algo-dynapyt-p8-monitoring-overhead-sum}{613.1213946342468}
\DefMacro{Comparison-7-algo-dynapyt-p9-monitoring-overhead-min}{-4.086374759674072}
\DefMacro{Comparison-7-algo-dynapyt-p9-monitoring-overhead-max}{146.4078826904297}
\DefMacro{Comparison-7-algo-dynapyt-p9-monitoring-overhead-mean}{8.110471098195939}
\DefMacro{Comparison-7-algo-dynapyt-p9-monitoring-overhead-median}{0.5901800394058228}
\DefMacro{Comparison-7-algo-dynapyt-p9-monitoring-overhead-sum}{681.2795722484589}
\DefMacro{Comparison-7-algo-dynapyt-p10-monitoring-overhead-min}{-89.07079648971558}
\DefMacro{Comparison-7-algo-dynapyt-p10-monitoring-overhead-max}{607.7421979904175}
\DefMacro{Comparison-7-algo-dynapyt-p10-monitoring-overhead-mean}{25.219304677985964}
\DefMacro{Comparison-7-algo-dynapyt-p10-monitoring-overhead-median}{0.49434661865234386}
\DefMacro{Comparison-7-algo-dynapyt-p10-monitoring-overhead-sum}{2118.421592950821}
\DefMacro{Comparison-7-algo-dynapyt_libs-p1-monitoring-overhead-min}{-46.46706438064575}
\DefMacro{Comparison-7-algo-dynapyt_libs-p1-monitoring-overhead-max}{35.74440622329712}
\DefMacro{Comparison-7-algo-dynapyt_libs-p1-monitoring-overhead-mean}{3.186738136268797}
\DefMacro{Comparison-7-algo-dynapyt_libs-p1-monitoring-overhead-median}{2.3802287578582764}
\DefMacro{Comparison-7-algo-dynapyt_libs-p1-monitoring-overhead-sum}{267.686003446579}
\DefMacro{Comparison-7-algo-dynapyt_libs-p2-monitoring-overhead-min}{0.6904194355010989}
\DefMacro{Comparison-7-algo-dynapyt_libs-p2-monitoring-overhead-max}{30.828357934951782}
\DefMacro{Comparison-7-algo-dynapyt_libs-p2-monitoring-overhead-mean}{3.471334539708637}
\DefMacro{Comparison-7-algo-dynapyt_libs-p2-monitoring-overhead-median}{2.2494852542877197}
\DefMacro{Comparison-7-algo-dynapyt_libs-p2-monitoring-overhead-sum}{291.5921013355255}
\DefMacro{Comparison-7-algo-dynapyt_libs-p3-monitoring-overhead-min}{1.1051921844482422}
\DefMacro{Comparison-7-algo-dynapyt_libs-p3-monitoring-overhead-max}{70.38386559486388}
\DefMacro{Comparison-7-algo-dynapyt_libs-p3-monitoring-overhead-mean}{6.744438350200653}
\DefMacro{Comparison-7-algo-dynapyt_libs-p3-monitoring-overhead-median}{3.3034695386886597}
\DefMacro{Comparison-7-algo-dynapyt_libs-p3-monitoring-overhead-sum}{566.5328214168549}
\DefMacro{Comparison-7-algo-dynapyt_libs-p4-monitoring-overhead-min}{0.7655391693115237}
\DefMacro{Comparison-7-algo-dynapyt_libs-p4-monitoring-overhead-max}{85.08454060554504}
\DefMacro{Comparison-7-algo-dynapyt_libs-p4-monitoring-overhead-mean}{8.245234191417694}
\DefMacro{Comparison-7-algo-dynapyt_libs-p4-monitoring-overhead-median}{3.1517568826675415}
\DefMacro{Comparison-7-algo-dynapyt_libs-p4-monitoring-overhead-sum}{692.5996720790863}
\DefMacro{Comparison-7-algo-dynapyt_libs-p5-monitoring-overhead-min}{1.190925121307373}
\DefMacro{Comparison-7-algo-dynapyt_libs-p5-monitoring-overhead-max}{81.89839506149292}
\DefMacro{Comparison-7-algo-dynapyt_libs-p5-monitoring-overhead-mean}{16.270153610479262}
\DefMacro{Comparison-7-algo-dynapyt_libs-p5-monitoring-overhead-median}{5.95211136341095}
\DefMacro{Comparison-7-algo-dynapyt_libs-p5-monitoring-overhead-sum}{1366.6929032802582}
\DefMacro{Comparison-7-algo-dynapyt_libs-p6-monitoring-overhead-min}{0.9284274578094482}
\DefMacro{Comparison-7-algo-dynapyt_libs-p6-monitoring-overhead-max}{142.43726301193237}
\DefMacro{Comparison-7-algo-dynapyt_libs-p6-monitoring-overhead-mean}{15.364801687853676}
\DefMacro{Comparison-7-algo-dynapyt_libs-p6-monitoring-overhead-median}{5.122107744216919}
\DefMacro{Comparison-7-algo-dynapyt_libs-p6-monitoring-overhead-sum}{1290.6433417797089}
\DefMacro{Comparison-7-algo-dynapyt_libs-p7-monitoring-overhead-min}{1.2031047344207764}
\DefMacro{Comparison-7-algo-dynapyt_libs-p7-monitoring-overhead-max}{201.38968443870542}
\DefMacro{Comparison-7-algo-dynapyt_libs-p7-monitoring-overhead-mean}{24.559335691588267}
\DefMacro{Comparison-7-algo-dynapyt_libs-p7-monitoring-overhead-median}{9.667845249176027}
\DefMacro{Comparison-7-algo-dynapyt_libs-p7-monitoring-overhead-sum}{2062.9841980934143}
\DefMacro{Comparison-7-algo-dynapyt_libs-p8-monitoring-overhead-min}{1.1405971050262453}
\DefMacro{Comparison-7-algo-dynapyt_libs-p8-monitoring-overhead-max}{216.17627930641174}
\DefMacro{Comparison-7-algo-dynapyt_libs-p8-monitoring-overhead-mean}{29.73689634743191}
\DefMacro{Comparison-7-algo-dynapyt_libs-p8-monitoring-overhead-median}{11.383071541786194}
\DefMacro{Comparison-7-algo-dynapyt_libs-p8-monitoring-overhead-sum}{2497.8992931842804}
\DefMacro{Comparison-7-algo-dynapyt_libs-p9-monitoring-overhead-min}{1.1929943561553955}
\DefMacro{Comparison-7-algo-dynapyt_libs-p9-monitoring-overhead-max}{374.01782512664795}
\DefMacro{Comparison-7-algo-dynapyt_libs-p9-monitoring-overhead-mean}{62.77942733821415}
\DefMacro{Comparison-7-algo-dynapyt_libs-p9-monitoring-overhead-median}{24.89861035346985}
\DefMacro{Comparison-7-algo-dynapyt_libs-p9-monitoring-overhead-sum}{5273.471896409988}
\DefMacro{Comparison-7-algo-dynapyt_libs-p10-monitoring-overhead-min}{1.26908278465271}
\DefMacro{Comparison-7-algo-dynapyt_libs-p10-monitoring-overhead-max}{1399.1100249290466}
\DefMacro{Comparison-7-algo-dynapyt_libs-p10-monitoring-overhead-mean}{185.76052744615646}
\DefMacro{Comparison-7-algo-dynapyt_libs-p10-monitoring-overhead-median}{14.369719505310059}
\DefMacro{Comparison-7-algo-dynapyt_libs-p10-monitoring-overhead-sum}{15603.884305477142}
\DefMacro{Comparison-7-algo-dylin-p1-monitoring-overhead-min}{-66.61599659919742}
\DefMacro{Comparison-7-algo-dylin-p1-monitoring-overhead-max}{19.04261016845703}
\DefMacro{Comparison-7-algo-dylin-p1-monitoring-overhead-mean}{0.3317175564311796}
\DefMacro{Comparison-7-algo-dylin-p1-monitoring-overhead-median}{0.40316808223724365}
\DefMacro{Comparison-7-algo-dylin-p1-monitoring-overhead-sum}{27.864274740219088}
\DefMacro{Comparison-7-algo-dylin-p2-monitoring-overhead-min}{-0.27536487579345703}
\DefMacro{Comparison-7-algo-dylin-p2-monitoring-overhead-max}{17.983885288238522}
\DefMacro{Comparison-7-algo-dylin-p2-monitoring-overhead-mean}{1.0304332091694786}
\DefMacro{Comparison-7-algo-dylin-p2-monitoring-overhead-median}{0.3855221271514893}
\DefMacro{Comparison-7-algo-dylin-p2-monitoring-overhead-sum}{86.5563895702362}
\DefMacro{Comparison-7-algo-dylin-p3-monitoring-overhead-min}{-0.041804313659668024}
\DefMacro{Comparison-7-algo-dylin-p3-monitoring-overhead-max}{18.47891092300415}
\DefMacro{Comparison-7-algo-dylin-p3-monitoring-overhead-mean}{1.9435415324710665}
\DefMacro{Comparison-7-algo-dylin-p3-monitoring-overhead-median}{0.3576345443725586}
\DefMacro{Comparison-7-algo-dylin-p3-monitoring-overhead-sum}{163.25748872756958}
\DefMacro{Comparison-7-algo-dylin-p4-monitoring-overhead-min}{-0.225799560546875}
\DefMacro{Comparison-7-algo-dylin-p4-monitoring-overhead-max}{35.859867572784424}
\DefMacro{Comparison-7-algo-dylin-p4-monitoring-overhead-mean}{1.8831547782534646}
\DefMacro{Comparison-7-algo-dylin-p4-monitoring-overhead-median}{0.4464733600616455}
\DefMacro{Comparison-7-algo-dylin-p4-monitoring-overhead-sum}{158.18500137329102}
\DefMacro{Comparison-7-algo-dylin-p5-monitoring-overhead-min}{-0.3104982376098633}
\DefMacro{Comparison-7-algo-dylin-p5-monitoring-overhead-max}{18.751136779785156}
\DefMacro{Comparison-7-algo-dylin-p5-monitoring-overhead-mean}{2.2123047709465027}
\DefMacro{Comparison-7-algo-dylin-p5-monitoring-overhead-median}{0.4896966218948362}
\DefMacro{Comparison-7-algo-dylin-p5-monitoring-overhead-sum}{185.83360075950623}
\DefMacro{Comparison-7-algo-dylin-p6-monitoring-overhead-min}{-36.29974126815796}
\DefMacro{Comparison-7-algo-dylin-p6-monitoring-overhead-max}{31.84864091873169}
\DefMacro{Comparison-7-algo-dylin-p6-monitoring-overhead-mean}{1.147818502925691}
\DefMacro{Comparison-7-algo-dylin-p6-monitoring-overhead-median}{0.3715916872024537}
\DefMacro{Comparison-7-algo-dylin-p6-monitoring-overhead-sum}{96.41675424575806}
\DefMacro{Comparison-7-algo-dylin-p7-monitoring-overhead-min}{-1.6249046325683594}
\DefMacro{Comparison-7-algo-dylin-p7-monitoring-overhead-max}{39.93595910072325}
\DefMacro{Comparison-7-algo-dylin-p7-monitoring-overhead-mean}{3.558843794323149}
\DefMacro{Comparison-7-algo-dylin-p7-monitoring-overhead-median}{0.6397218704223633}
\DefMacro{Comparison-7-algo-dylin-p7-monitoring-overhead-sum}{298.94287872314453}
\DefMacro{Comparison-7-algo-dylin-p8-monitoring-overhead-min}{-0.9048519134521484}
\DefMacro{Comparison-7-algo-dylin-p8-monitoring-overhead-max}{72.12075114250183}
\DefMacro{Comparison-7-algo-dylin-p8-monitoring-overhead-mean}{4.617856718245006}
\DefMacro{Comparison-7-algo-dylin-p8-monitoring-overhead-median}{0.48372721672058105}
\DefMacro{Comparison-7-algo-dylin-p8-monitoring-overhead-sum}{387.89996433258057}
\DefMacro{Comparison-7-algo-dylin-p9-monitoring-overhead-min}{-5.326150178909302}
\DefMacro{Comparison-7-algo-dylin-p9-monitoring-overhead-max}{62.948201417922974}
\DefMacro{Comparison-7-algo-dylin-p9-monitoring-overhead-mean}{4.319017089548565}
\DefMacro{Comparison-7-algo-dylin-p9-monitoring-overhead-median}{0.5088400840759277}
\DefMacro{Comparison-7-algo-dylin-p9-monitoring-overhead-sum}{362.79743552207947}
\DefMacro{Comparison-7-algo-dylin-p10-monitoring-overhead-min}{-89.06327819824219}
\DefMacro{Comparison-7-algo-dylin-p10-monitoring-overhead-max}{290.19198870658875}
\DefMacro{Comparison-7-algo-dylin-p10-monitoring-overhead-mean}{11.271506264096214}
\DefMacro{Comparison-7-algo-dylin-p10-monitoring-overhead-median}{0.3902170658111572}
\DefMacro{Comparison-7-algo-dylin-p10-monitoring-overhead-sum}{946.806526184082}
\DefMacro{Comparison-7-total-projects}{840}

\DefMacro{Comparison-7-algo-original-instrumentation_time-min}{0.0}
\DefMacro{Comparison-7-algo-original-instrumentation_time-mean}{0.0}
\DefMacro{Comparison-7-algo-original-instrumentation_time-median}{0.0}
\DefMacro{Comparison-7-algo-original-instrumentation_time-max}{0.0}
\DefMacro{Comparison-7-algo-original-instrumentation_time-sum}{0.0}
\DefMacro{Comparison-7-algo-D-instrumentation_time-min}{0.0}
\DefMacro{Comparison-7-algo-D-instrumentation_time-mean}{0.7019722041629609}
\DefMacro{Comparison-7-algo-D-instrumentation_time-median}{0.6569908857345581}
\DefMacro{Comparison-7-algo-D-instrumentation_time-max}{6.667206048965454}
\DefMacro{Comparison-7-algo-D-instrumentation_time-sum}{589.6566514968872}
\DefMacro{Comparison-7-algo-dynapyt-instrumentation_time-min}{0.5128700733184814}
\DefMacro{Comparison-7-algo-dynapyt-instrumentation_time-mean}{34.97867799287751}
\DefMacro{Comparison-7-algo-dynapyt-instrumentation_time-median}{3.488035798072815}
\DefMacro{Comparison-7-algo-dynapyt-instrumentation_time-max}{1854.8130316734312}
\DefMacro{Comparison-7-algo-dynapyt-instrumentation_time-sum}{29382.089514017105}
\DefMacro{Comparison-7-algo-dynapyt_libs-instrumentation_time-min}{1773.5088763237}
\DefMacro{Comparison-7-algo-dynapyt_libs-instrumentation_time-mean}{1986.447871327968}
\DefMacro{Comparison-7-algo-dynapyt_libs-instrumentation_time-median}{1908.6060490608215}
\DefMacro{Comparison-7-algo-dynapyt_libs-instrumentation_time-max}{3483.2586188316345}
\DefMacro{Comparison-7-algo-dynapyt_libs-instrumentation_time-sum}{1668616.211915493}
\DefMacro{Comparison-7-algo-dylin-instrumentation_time-min}{0.7418668270111084}
\DefMacro{Comparison-7-algo-dylin-instrumentation_time-mean}{34.963424255734395}
\DefMacro{Comparison-7-algo-dylin-instrumentation_time-median}{3.8214142322540283}
\DefMacro{Comparison-7-algo-dylin-instrumentation_time-max}{1820.704520225525}
\DefMacro{Comparison-7-algo-dylin-instrumentation_time-sum}{29369.276374816895}
\DefMacro{Comparison-7-algo-original-test_duration-min}{0.2880303859710693}
\DefMacro{Comparison-7-algo-original-test_duration-mean}{4.920768936475118}
\DefMacro{Comparison-7-algo-original-test_duration-median}{0.8439263105392456}
\DefMacro{Comparison-7-algo-original-test_duration-max}{360.3594071865082}
\DefMacro{Comparison-7-algo-original-test_duration-sum}{4133.445906639099}
\DefMacro{Comparison-7-algo-D-test_duration-min}{1.385106086730957}
\DefMacro{Comparison-7-algo-D-test_duration-mean}{10.035266730898902}
\DefMacro{Comparison-7-algo-D-test_duration-median}{3.7151962518692017}
\DefMacro{Comparison-7-algo-D-test_duration-max}{963.0932297706604}
\DefMacro{Comparison-7-algo-D-test_duration-sum}{8429.624053955078}
\DefMacro{Comparison-7-algo-dynapyt-test_duration-min}{0.5020890235900879}
\DefMacro{Comparison-7-algo-dynapyt-test_duration-mean}{10.718261696043468}
\DefMacro{Comparison-7-algo-dynapyt-test_duration-median}{1.5367985963821411}
\DefMacro{Comparison-7-algo-dynapyt-test_duration-max}{614.1870589256287}
\DefMacro{Comparison-7-algo-dynapyt-test_duration-sum}{9003.339824676514}
\DefMacro{Comparison-7-algo-dynapyt_libs-test_duration-min}{1.124687433242798}
\DefMacro{Comparison-7-algo-dynapyt_libs-test_duration-mean}{40.53265767040707}
\DefMacro{Comparison-7-algo-dynapyt_libs-test_duration-median}{5.99101996421814}
\DefMacro{Comparison-7-algo-dynapyt_libs-test_duration-max}{1404.7670166492462}
\DefMacro{Comparison-7-algo-dynapyt_libs-test_duration-sum}{34047.43244314194}
\DefMacro{Comparison-7-algo-dylin-test_duration-min}{0.2764022350311279}
\DefMacro{Comparison-7-algo-dylin-test_duration-mean}{8.15238835811615}
\DefMacro{Comparison-7-algo-dylin-test_duration-median}{1.5069177150726318}
\DefMacro{Comparison-7-algo-dylin-test_duration-max}{348.95076608657837}
\DefMacro{Comparison-7-algo-dylin-test_duration-sum}{6848.006220817566}
\DefMacro{Comparison-7-algo-original-end_to_end_time-min}{0.2880303859710693}
\DefMacro{Comparison-7-algo-original-end_to_end_time-mean}{4.920768936475118}
\DefMacro{Comparison-7-algo-original-end_to_end_time-median}{0.8439263105392456}
\DefMacro{Comparison-7-algo-original-end_to_end_time-max}{360.3594071865082}
\DefMacro{Comparison-7-algo-original-end_to_end_time-sum}{4133.445906639099}
\DefMacro{Comparison-7-algo-D-end_to_end_time-min}{1.6970252990722656}
\DefMacro{Comparison-7-algo-D-end_to_end_time-mean}{10.737238935061864}
\DefMacro{Comparison-7-algo-D-end_to_end_time-median}{4.525495529174805}
\DefMacro{Comparison-7-algo-D-end_to_end_time-max}{963.6742916107178}
\DefMacro{Comparison-7-algo-D-end_to_end_time-sum}{9020}  %
\DefMacro{Comparison-7-algo-dynapyt-end_to_end_time-min}{1.0149590969085691}
\DefMacro{Comparison-7-algo-dynapyt-end_to_end_time-mean}{45.69693968892098}
\DefMacro{Comparison-7-algo-dynapyt-end_to_end_time-median}{6.614723086357117}
\DefMacro{Comparison-7-algo-dynapyt-end_to_end_time-max}{1855.738455057144}
\DefMacro{Comparison-7-algo-dynapyt-end_to_end_time-sum}{38385.42933869362}
\DefMacro{Comparison-7-algo-dynapyt_libs-end_to_end_time-min}{1780.1695504188538}
\DefMacro{Comparison-7-algo-dynapyt_libs-end_to_end_time-mean}{2026.980528998375}
\DefMacro{Comparison-7-algo-dynapyt_libs-end_to_end_time-median}{1935.0204584598541}
\DefMacro{Comparison-7-algo-dynapyt_libs-end_to_end_time-max}{3488.316110610962}
\DefMacro{Comparison-7-algo-dynapyt_libs-end_to_end_time-sum}{1702663.644358635}
\DefMacro{Comparison-7-algo-dylin-end_to_end_time-min}{1.1570320129394531}
\DefMacro{Comparison-7-algo-dylin-end_to_end_time-mean}{43.11581261385055}
\DefMacro{Comparison-7-algo-dylin-end_to_end_time-median}{6.837506055831909}
\DefMacro{Comparison-7-algo-dylin-end_to_end_time-max}{1821.6465966701508}
\DefMacro{Comparison-7-algo-dylin-end_to_end_time-sum}{36217.28259563446}
\DefMacro{Comparison-7-D-DyLin-test-duration-mean-diff}{1.9}

\DefMacro{algo-fastest-p1-c}{73}
\DefMacro{algo-fastest-p1-cplus}{41}
\DefMacro{algo-fastest-p1-d}{45}
\DefMacro{algo-fastest-p2-c}{58}
\DefMacro{algo-fastest-p2-cplus}{44}
\DefMacro{algo-fastest-p2-d}{58}
\DefMacro{algo-fastest-p3-c}{34}
\DefMacro{algo-fastest-p3-cplus}{29}
\DefMacro{algo-fastest-p3-d}{88}
\DefMacro{algo-fastest-p4-c}{30}
\DefMacro{algo-fastest-p4-cplus}{32}
\DefMacro{algo-fastest-p4-d}{94}
\DefMacro{algo-fastest-p5-c}{27}
\DefMacro{algo-fastest-p5-cplus}{16}
\DefMacro{algo-fastest-p5-d}{114}
\DefMacro{algo-fastest-p6-c}{28}
\DefMacro{algo-fastest-p6-cplus}{19}
\DefMacro{algo-fastest-p6-d}{107}
\DefMacro{algo-fastest-p7-c}{17}
\DefMacro{algo-fastest-p7-cplus}{21}
\DefMacro{algo-fastest-p7-d}{119}
\DefMacro{algo-fastest-p8-c}{18}
\DefMacro{algo-fastest-p8-cplus}{20}
\DefMacro{algo-fastest-p8-d}{118}
\DefMacro{algo-fastest-p9-c}{16}
\DefMacro{algo-fastest-p9-cplus}{19}
\DefMacro{algo-fastest-p9-d}{120}
\DefMacro{algo-fastest-p10-c}{7}
\DefMacro{algo-fastest-p10-cplus}{16}
\DefMacro{algo-fastest-p10-d}{134}

\DefMacro{algo-fastest-1s-p1-cplus}{0}
\DefMacro{algo-fastest-1s-p1-d}{0}
\DefMacro{algo-fastest-1s-p1-same}{159}
\DefMacro{algo-fastest-1s-p2-cplus}{0}
\DefMacro{algo-fastest-1s-p2-d}{2}
\DefMacro{algo-fastest-1s-p2-same}{158}
\DefMacro{algo-fastest-1s-p3-cplus}{0}
\DefMacro{algo-fastest-1s-p3-d}{0}
\DefMacro{algo-fastest-1s-p3-same}{151}
\DefMacro{algo-fastest-1s-p4-cplus}{1}
\DefMacro{algo-fastest-1s-p4-d}{10}
\DefMacro{algo-fastest-1s-p4-same}{145}
\DefMacro{algo-fastest-1s-p5-cplus}{0}
\DefMacro{algo-fastest-1s-p5-d}{25}
\DefMacro{algo-fastest-1s-p5-same}{132}
\DefMacro{algo-fastest-1s-p6-cplus}{4}
\DefMacro{algo-fastest-1s-p6-d}{43}
\DefMacro{algo-fastest-1s-p6-same}{107}
\DefMacro{algo-fastest-1s-p7-cplus}{8}
\DefMacro{algo-fastest-1s-p7-d}{72}
\DefMacro{algo-fastest-1s-p7-same}{77}
\DefMacro{algo-fastest-1s-p8-cplus}{6}
\DefMacro{algo-fastest-1s-p8-d}{87}
\DefMacro{algo-fastest-1s-p8-same}{63}
\DefMacro{algo-fastest-1s-p9-cplus}{12}
\DefMacro{algo-fastest-1s-p9-d}{104}
\DefMacro{algo-fastest-1s-p9-same}{39}
\DefMacro{algo-fastest-1s-p10-cplus}{11}
\DefMacro{algo-fastest-1s-p10-d}{127}
\DefMacro{algo-fastest-1s-p10-same}{19}

\DefMacro{algo-fastest-2s-p1-cplus}{0}
\DefMacro{algo-fastest-2s-p1-d}{0}
\DefMacro{algo-fastest-2s-p1-same}{159}
\DefMacro{algo-fastest-2s-p2-cplus}{0}
\DefMacro{algo-fastest-2s-p2-d}{1}
\DefMacro{algo-fastest-2s-p2-same}{159}
\DefMacro{algo-fastest-2s-p3-cplus}{0}
\DefMacro{algo-fastest-2s-p3-d}{0}
\DefMacro{algo-fastest-2s-p3-same}{151}
\DefMacro{algo-fastest-2s-p4-cplus}{1}
\DefMacro{algo-fastest-2s-p4-d}{7}
\DefMacro{algo-fastest-2s-p4-same}{148}
\DefMacro{algo-fastest-2s-p5-cplus}{0}
\DefMacro{algo-fastest-2s-p5-d}{17}
\DefMacro{algo-fastest-2s-p5-same}{140}
\DefMacro{algo-fastest-2s-p6-cplus}{1}
\DefMacro{algo-fastest-2s-p6-d}{19}
\DefMacro{algo-fastest-2s-p6-same}{134}
\DefMacro{algo-fastest-2s-p7-cplus}{5}
\DefMacro{algo-fastest-2s-p7-d}{49}
\DefMacro{algo-fastest-2s-p7-same}{103}
\DefMacro{algo-fastest-2s-p8-cplus}{4}
\DefMacro{algo-fastest-2s-p8-d}{66}
\DefMacro{algo-fastest-2s-p8-same}{86}
\DefMacro{algo-fastest-2s-p9-cplus}{9}
\DefMacro{algo-fastest-2s-p9-d}{95}
\DefMacro{algo-fastest-2s-p9-same}{51}
\DefMacro{algo-fastest-2s-p10-cplus}{8}
\DefMacro{algo-fastest-2s-p10-d}{122}
\DefMacro{algo-fastest-2s-p10-same}{27}

\DefMacro{algo-fastest-5s-p1-cplus}{0}
\DefMacro{algo-fastest-5s-p1-d}{0}
\DefMacro{algo-fastest-5s-p1-same}{159}
\DefMacro{algo-fastest-5s-p2-cplus}{0}
\DefMacro{algo-fastest-5s-p2-d}{0}
\DefMacro{algo-fastest-5s-p2-same}{160}
\DefMacro{algo-fastest-5s-p3-cplus}{0}
\DefMacro{algo-fastest-5s-p3-d}{0}
\DefMacro{algo-fastest-5s-p3-same}{151}
\DefMacro{algo-fastest-5s-p4-cplus}{1}
\DefMacro{algo-fastest-5s-p4-d}{2}
\DefMacro{algo-fastest-5s-p4-same}{153}
\DefMacro{algo-fastest-5s-p5-cplus}{0}
\DefMacro{algo-fastest-5s-p5-d}{12}
\DefMacro{algo-fastest-5s-p5-same}{145}
\DefMacro{algo-fastest-5s-p6-cplus}{1}
\DefMacro{algo-fastest-5s-p6-d}{4}
\DefMacro{algo-fastest-5s-p6-same}{149}
\DefMacro{algo-fastest-5s-p7-cplus}{3}
\DefMacro{algo-fastest-5s-p7-d}{20}
\DefMacro{algo-fastest-5s-p7-same}{134}
\DefMacro{algo-fastest-5s-p8-cplus}{2}
\DefMacro{algo-fastest-5s-p8-d}{42}
\DefMacro{algo-fastest-5s-p8-same}{112}
\DefMacro{algo-fastest-5s-p9-cplus}{4}
\DefMacro{algo-fastest-5s-p9-d}{76}
\DefMacro{algo-fastest-5s-p9-same}{75}
\DefMacro{algo-fastest-5s-p10-cplus}{5}
\DefMacro{algo-fastest-5s-p10-d}{110}
\DefMacro{algo-fastest-5s-p10-same}{42}

\DefMacro{comparison-All_4-original-sum-end-to-end}{4133.45}
\DefMacro{comparison-All_4-original-sum-instrumentation}{0.00}
\DefMacro{comparison-All_4-original-sum-monitoring}{4133.45}
\DefMacro{comparison-All_4-D-sum-end-to-end}{9019.28}
\DefMacro{comparison-All_4-D-sum-instrumentation}{589.66}
\DefMacro{comparison-All_4-D-sum-monitoring}{8429.62}
\DefMacro{comparison-All_4-dylin-sum-end-to-end}{36217.28}
\DefMacro{comparison-All_4-dylin-sum-instrumentation}{29369.28}
\DefMacro{comparison-All_4-dylin-sum-monitoring}{6848.01}
\DefMacro{comparison-All_4-dynapyt-sum-end-to-end}{38385.43}
\DefMacro{comparison-All_4-dynapyt-sum-instrumentation}{29382.09}
\DefMacro{comparison-All_4-dynapyt-sum-monitoring}{9003.34}
\DefMacro{comparison-All_4-dynapyt_libs-sum-end-to-end}{1702663.64}
\DefMacro{comparison-All_4-dynapyt_libs-sum-instrumentation}{1668616.21}
\DefMacro{comparison-All_4-dynapyt_libs-sum-monitoring}{34047.43}
\DefMacro{comparison-No_Lib-original-sum-end-to-end}{12877.52}
\DefMacro{comparison-No_Lib-original-sum-instrumentation}{0.00}
\DefMacro{comparison-No_Lib-original-sum-monitoring}{12877.52}
\DefMacro{comparison-No_Lib-D-sum-end-to-end}{26636.82}
\DefMacro{comparison-No_Lib-D-sum-instrumentation}{801.96}
\DefMacro{comparison-No_Lib-D-sum-monitoring}{25834.87}
\DefMacro{comparison-No_Lib-dylin-sum-end-to-end}{62318.37}
\DefMacro{comparison-No_Lib-dylin-sum-instrumentation}{33903.15}
\DefMacro{comparison-No_Lib-dylin-sum-monitoring}{28415.22}
\DefMacro{comparison-No_Lib-dynapyt-sum-end-to-end}{73659.36}
\DefMacro{comparison-No_Lib-dynapyt-sum-instrumentation}{33852.50}
\DefMacro{comparison-No_Lib-dynapyt-sum-monitoring}{39806.86}
\DefMacro{comparison-With_Lib-original-sum-end-to-end}{4372.83}
\DefMacro{comparison-With_Lib-original-sum-instrumentation}{0.00}
\DefMacro{comparison-With_Lib-original-sum-monitoring}{4372.83}
\DefMacro{comparison-With_Lib-D-sum-end-to-end}{9779.96}
\DefMacro{comparison-With_Lib-D-sum-instrumentation}{653.91}
\DefMacro{comparison-With_Lib-D-sum-monitoring}{9126.05}
\DefMacro{comparison-With_Lib-dynapyt_libs-sum-end-to-end}{1894519.33}
\DefMacro{comparison-With_Lib-dynapyt_libs-sum-instrumentation}{1856949.94}
\DefMacro{comparison-With_Lib-dynapyt_libs-sum-monitoring}{37569.39}

\DefMacro{comparison-All_4-dylin-relative-speedup}{6.08}
\DefMacro{comparison-All_4-dylin-median-relative-speedup}{1.31}
\DefMacro{comparison-All_4-dylin-min-relative-speedup}{0.00}
\DefMacro{comparison-All_4-dylin-max-relative-speedup}{412.25}
\DefMacro{comparison-All_4-dylin-absolute-speedup}{32.38}
\DefMacro{comparison-All_4-dylin-median-absolute-speedup}{1.42}
\DefMacro{comparison-All_4-dylin-min-absolute-speedup}{-961.83}
\DefMacro{comparison-All_4-dylin-max-absolute-speedup}{1817.23}
\DefMacro{comparison-All_4-dynapyt-relative-speedup}{6.33}
\DefMacro{comparison-All_4-dynapyt-median-relative-speedup}{1.24}
\DefMacro{comparison-All_4-dynapyt-min-relative-speedup}{0.00}
\DefMacro{comparison-All_4-dynapyt-max-relative-speedup}{419.97}
\DefMacro{comparison-All_4-dynapyt-absolute-speedup}{34.96}
\DefMacro{comparison-All_4-dynapyt-median-absolute-speedup}{1.17}
\DefMacro{comparison-All_4-dynapyt-min-absolute-speedup}{-962.24}
\DefMacro{comparison-All_4-dynapyt-max-absolute-speedup}{1851.32}
\DefMacro{comparison-All_4-dynapyt_libs-relative-speedup}{456.86}
\DefMacro{comparison-All_4-dynapyt_libs-median-relative-speedup}{434.76}
\DefMacro{comparison-All_4-dynapyt_libs-min-relative-speedup}{2.76}
\DefMacro{comparison-All_4-dynapyt_libs-max-relative-speedup}{1168.32}
\DefMacro{comparison-All_4-dynapyt_libs-absolute-speedup}{2016.24}
\DefMacro{comparison-All_4-dynapyt_libs-median-absolute-speedup}{1924.93}
\DefMacro{comparison-All_4-dynapyt_libs-min-absolute-speedup}{1692.24}
\DefMacro{comparison-All_4-dynapyt_libs-max-absolute-speedup}{3483.90}
\DefMacro{comparison-No_Lib-dylin-relative-speedup}{5.65}
\DefMacro{comparison-No_Lib-dylin-median-relative-speedup}{1.23}
\DefMacro{comparison-No_Lib-dylin-min-relative-speedup}{0.00}
\DefMacro{comparison-No_Lib-dylin-max-relative-speedup}{412.25}
\DefMacro{comparison-No_Lib-dylin-absolute-speedup}{31.97}
\DefMacro{comparison-No_Lib-dylin-median-absolute-speedup}{1.22}
\DefMacro{comparison-No_Lib-dylin-min-absolute-speedup}{-1697.13}
\DefMacro{comparison-No_Lib-dylin-max-absolute-speedup}{1817.23}
\DefMacro{comparison-No_Lib-dynapyt-relative-speedup}{6.26}
\DefMacro{comparison-No_Lib-dynapyt-median-relative-speedup}{1.19}
\DefMacro{comparison-No_Lib-dynapyt-min-relative-speedup}{0.00}
\DefMacro{comparison-No_Lib-dynapyt-max-relative-speedup}{419.97}
\DefMacro{comparison-No_Lib-dynapyt-absolute-speedup}{42.13}
\DefMacro{comparison-No_Lib-dynapyt-median-absolute-speedup}{1.04}
\DefMacro{comparison-No_Lib-dynapyt-min-absolute-speedup}{-1697.50}
\DefMacro{comparison-No_Lib-dynapyt-max-absolute-speedup}{2499.58}
\DefMacro{comparison-With_Lib-dynapyt_libs-relative-speedup}{460.26}
\DefMacro{comparison-With_Lib-dynapyt_libs-median-relative-speedup}{435.86}
\DefMacro{comparison-With_Lib-dynapyt_libs-min-relative-speedup}{2.76}
\DefMacro{comparison-With_Lib-dynapyt_libs-max-relative-speedup}{1168.32}
\DefMacro{comparison-With_Lib-dynapyt_libs-absolute-speedup}{2015.76}
\DefMacro{comparison-With_Lib-dynapyt_libs-median-absolute-speedup}{1915.10}
\DefMacro{comparison-With_Lib-dynapyt_libs-min-absolute-speedup}{1692.24}
\DefMacro{comparison-With_Lib-dynapyt_libs-max-absolute-speedup}{3483.90}

\DefMacro{comparison-statistics-Original-total-projects}{1463}
\DefMacro{comparison-statistics-D-total-projects}{1463}
\DefMacro{comparison-statistics-Dylin-total-projects}{1189}
\DefMacro{comparison-statistics-Dynapyt-total-projects}{1160}
\DefMacro{comparison-statistics-Dynapyt_libs-total-projects}{935}

\DefMacro{comparison-statistics-D-Dynapyt-total-projects}{840}
\DefMacro{comparison-statistics-D-Dynapyt-same-projects}{831}
\DefMacro{comparison-statistics-D-Dynapyt-D-projects}{9}
\DefMacro{comparison-statistics-D-Dynapyt-DynaPyt-projects}{0}

\DefMacro{comparison-statistics-D-Dynapyt-total-projects-filtered}{840}
\DefMacro{comparison-statistics-D-Dynapyt-same-projects-filtered}{834}
\DefMacro{comparison-statistics-D-Dynapyt-D-projects-filtered}{6}
\DefMacro{comparison-statistics-D-Dynapyt-DynaPyt-projects-filtered}{0}

\DefMacro{comparison-statistics-D-Dynapyt-same-violations}{7}
\DefMacro{comparison-statistics-D-Dynapyt-D-violations}{11}
\DefMacro{comparison-statistics-D-Dynapyt-DynaPyt-violations}{0}

\DefMacro{comparison-statistics-D-Dynapyt-same-violations-filtered}{7}
\DefMacro{comparison-statistics-D-Dynapyt-D-violations-filtered}{7}
\DefMacro{comparison-statistics-D-Dynapyt-DynaPyt-violations-filtered}{0}

\DefMacro{comparison-statistics-D-Dynapyt_libs-total-projects}{840}
\DefMacro{comparison-statistics-D-Dynapyt_libs-same-projects}{824}
\DefMacro{comparison-statistics-D-Dynapyt_libs-D-projects}{10}
\DefMacro{comparison-statistics-D-Dynapyt_libs-DynaPyt_libs-projects}{6}

\DefMacro{comparison-statistics-D-Dynapyt_libs-total-projects-filtered}{840}
\DefMacro{comparison-statistics-D-Dynapyt_libs-same-projects-filtered}{827}
\DefMacro{comparison-statistics-D-Dynapyt_libs-D-projects-filtered}{7}
\DefMacro{comparison-statistics-D-Dynapyt_libs-DynaPyt_libs-projects-filtered}{6}

\DefMacro{comparison-statistics-D-Dynapyt_libs-same-violations}{6}
\DefMacro{comparison-statistics-D-Dynapyt_libs-D-violations}{12}
\DefMacro{comparison-statistics-D-Dynapyt_libs-DynaPyt_libs-violations}{9}

\DefMacro{comparison-statistics-D-Dynapyt_libs-same-violations-filtered}{6}
\DefMacro{comparison-statistics-D-Dynapyt_libs-D-violations-filtered}{8}
\DefMacro{comparison-statistics-D-Dynapyt_libs-DynaPyt_libs-violations-filtered}{9}

\DefMacro{comparison-statistics-D-Dylin-total-projects}{840}
\DefMacro{comparison-statistics-D-Dylin-same-projects}{831}
\DefMacro{comparison-statistics-D-Dylin-D-projects}{9}
\DefMacro{comparison-statistics-D-Dylin-Dylin-projects}{0}

\DefMacro{comparison-statistics-D-Dylin-total-projects-filtered}{840}
\DefMacro{comparison-statistics-D-Dylin-same-projects-filtered}{834}
\DefMacro{comparison-statistics-D-Dylin-D-projects-filtered}{6}
\DefMacro{comparison-statistics-D-Dylin-Dylin-projects-filtered}{0}

\DefMacro{comparison-statistics-D-Dylin-same-violations}{7}
\DefMacro{comparison-statistics-D-Dylin-D-violations}{11}
\DefMacro{comparison-statistics-D-Dylin-Dylin-violations}{0}

\DefMacro{comparison-statistics-D-Dylin-same-violations-filtered}{7}
\DefMacro{comparison-statistics-D-Dylin-D-violations-filtered}{7}
\DefMacro{comparison-statistics-D-Dylin-Dylin-violations-filtered}{0}

\DefMacro{InstrTimeMeanAst}{7.52}
\DefMacro{InstrTimeMedianAst}{4.77}
\DefMacro{InstrTimeMaxAst}{103.55}
\DefMacro{InstrTimeMinAst}{3.51}
\DefMacro{InstrTimeSumAst}{9204.04}

\DefMacro{InstrTimeMeanCurse}{0.47}
\DefMacro{InstrTimeMedianCurse}{0.42}
\DefMacro{InstrTimeMaxCurse}{3.94}
\DefMacro{InstrTimeMinCurse}{0.00}
\DefMacro{InstrTimeSumCurse}{576.66}

\DefMacro{InstrTimeMeanMonkeypatching}{0.52}
\DefMacro{InstrTimeMedianMonkeypatching}{0.47}
\DefMacro{InstrTimeMaxMonkeypatching}{4.06}
\DefMacro{InstrTimeMinMonkeypatching}{0.00}
\DefMacro{InstrTimeSumMonkeypatching}{641.21}

\DefMacro{InstrTimeMeanOriginal}{0.00}
\DefMacro{InstrTimeMedianOriginal}{0.00}
\DefMacro{InstrTimeMaxOriginal}{0.00}
\DefMacro{InstrTimeMinOriginal}{0.00}
\DefMacro{InstrTimeSumOriginal}{0.00}

\DefMacro{E2ETimeMeanAst}{35.65}
\DefMacro{E2ETimeMedianAst}{8.09}
\DefMacro{E2ETimeMaxAst}{2662.99}
\DefMacro{E2ETimeMinAst}{4.61}
\DefMacro{E2ETimeSumAst}{43635.25}

\DefMacro{E2ETimeMeanCurse}{12.89}
\DefMacro{E2ETimeMedianCurse}{2.40}
\DefMacro{E2ETimeMaxCurse}{2210.89}
\DefMacro{E2ETimeMinCurse}{1.46}
\DefMacro{E2ETimeSumCurse}{15760.25}

\DefMacro{E2ETimeMeanMonkeypatching}{23.73}
\DefMacro{E2ETimeMedianMonkeypatching}{2.80}
\DefMacro{E2ETimeMaxMonkeypatching}{2639.66}
\DefMacro{E2ETimeMinMonkeypatching}{1.55}
\DefMacro{E2ETimeSumMonkeypatching}{29024.62}

\DefMacro{E2ETimeMeanOriginal}{6.82}
\DefMacro{E2ETimeMedianOriginal}{0.58}
\DefMacro{E2ETimeMaxOriginal}{2162.10}
\DefMacro{E2ETimeMinOriginal}{0.16}
\DefMacro{E2ETimeSumOriginal}{8348.56}

\DefMacro{TotalInstrTimeAst}{9204.035723422638}
\DefMacro{TotalTestTimeAst}{34431.20929815901}

\DefMacro{TotalInstrTimeCurse}{576.6607174873351}
\DefMacro{TotalTestTimeCurse}{15183.592797756195}

\DefMacro{TotalInstrTimeMonkeypatching}{641.2139904499053}
\DefMacro{TotalTestTimeMonkeypatching}{28383.402066469193}

\DefMacro{TotalInstrTimeOriginal}{0.0}
\DefMacro{TotalTestTimeOriginal}{8348.563475131989}

\DefMacro{TotalUniqueViolationsAst}{5043}
\DefMacro{text_TotalUniqueViolationsAst}{5,043}

\DefMacro{TotalUniqueViolationsCurse}{11332}
\DefMacro{text_TotalUniqueViolationsCurse}{11,332}

\DefMacro{TotalUniqueViolationsMonkeypatching}{833}
\DefMacro{text_TotalUniqueViolationsMonkeypatching}{833}

\DefMacro{TotalPassedTestsAst}{192224}
\DefMacro{TotalFailedTestsAst}{11267}
\DefMacro{text_TotalFailedTestsAst}{11,267}

\DefMacro{TotalPassedTestsCurse}{196398}
\DefMacro{TotalFailedTestsCurse}{11252}
\DefMacro{text_TotalFailedTestsCurse}{11,252}

\DefMacro{TotalPassedTestsMonkeypatching}{196560}
\DefMacro{TotalFailedTestsMonkeypatching}{11101}
\DefMacro{text_TotalFailedTestsMonkeypatching}{11,101}

\DefMacro{TotalPassedTestsOriginal}{196438}
\DefMacro{TotalFailedTestsOriginal}{10821}
\DefMacro{text_TotalFailedTestsOriginal}{10,821}
\DefMacro{ProjectCountFilteredRQ2}{1,224}
\DefMacro{TotalViolationsNonFiltered}{9,052}

\DefMacro{TotalViolationsPythonSourceCurseNonFiltered}{70}
\DefMacro{TotalViolationsSitePackagesCurseNonFiltered}{5278}
\DefMacro{TotalViolationsProjectUnderTestCurseNonFiltered}{2233}
\DefMacro{text_TotalViolationsPythonSourceCurseNonFiltered}{70}
\DefMacro{text_TotalViolationsSitePackagesCurseNonFiltered}{5,278}
\DefMacro{text_TotalViolationsProjectUnderTestCurseNonFiltered}{2,233}

\DefMacro{TotalViolationsPythonSourceAstNonFiltered}{25}
\DefMacro{TotalViolationsSitePackagesAstNonFiltered}{4090}
\DefMacro{TotalViolationsProjectUnderTestAstNonFiltered}{1556}
\DefMacro{text_TotalViolationsPythonSourceAstNonFiltered}{25}
\DefMacro{text_TotalViolationsSitePackagesAstNonFiltered}{4,090}
\DefMacro{text_TotalViolationsProjectUnderTestAstNonFiltered}{1,556}

\DefMacro{TotalViolationsPythonSourceMonkeypatchingNonFiltered}{15}
\DefMacro{TotalViolationsSitePackagesMonkeypatchingNonFiltered}{351}
\DefMacro{TotalViolationsProjectUnderTestMonkeypatchingNonFiltered}{624}
\DefMacro{text_TotalViolationsPythonSourceMonkeypatchingNonFiltered}{15}
\DefMacro{text_TotalViolationsSitePackagesMonkeypatchingNonFiltered}{351}
\DefMacro{text_TotalViolationsProjectUnderTestMonkeypatchingNonFiltered}{624}

\DefMacro{TotalViolationsPythonSourceCurse}{68}
\DefMacro{TotalViolationsSitePackagesCurse}{4059}
\DefMacro{TotalViolationsProjectUnderTestCurse}{1686}
\DefMacro{text_TotalViolationsPythonSourceCurse}{68}
\DefMacro{text_TotalViolationsSitePackagesCurse}{4,059}
\DefMacro{text_TotalViolationsProjectUnderTestCurse}{1,686}

\DefMacro{TotalViolationsPythonSourceAst}{22}
\DefMacro{TotalViolationsSitePackagesAst}{3575}
\DefMacro{TotalViolationsProjectUnderTestAst}{1335}
\DefMacro{text_TotalViolationsPythonSourceAst}{22}
\DefMacro{text_TotalViolationsSitePackagesAst}{3,575}
\DefMacro{text_TotalViolationsProjectUnderTestAst}{1,335}

\DefMacro{TotalViolationsPythonSourceMonkeypatching}{14}
\DefMacro{TotalViolationsSitePackagesMonkeypatching}{263}
\DefMacro{TotalViolationsProjectUnderTestMonkeypatching}{465}
\DefMacro{text_TotalViolationsPythonSourceMonkeypatching}{14}
\DefMacro{text_TotalViolationsSitePackagesMonkeypatching}{263}
\DefMacro{text_TotalViolationsProjectUnderTestMonkeypatching}{465}

\DefMacro{ProjectsFastestMonkeypatching0to10}{3}
\DefMacro{ProjectsFastestCurse0to10}{120}
\DefMacro{ProjectsFastestAst0to10}{0}

\DefMacro{ProjectsFastestMonkeypatching10to20}{7}
\DefMacro{ProjectsFastestCurse10to20}{115}
\DefMacro{ProjectsFastestAst10to20}{0}

\DefMacro{ProjectsFastestMonkeypatching20to30}{12}
\DefMacro{ProjectsFastestCurse20to30}{110}
\DefMacro{ProjectsFastestAst20to30}{0}

\DefMacro{ProjectsFastestMonkeypatching30to40}{7}
\DefMacro{ProjectsFastestCurse30to40}{115}
\DefMacro{ProjectsFastestAst30to40}{0}

\DefMacro{ProjectsFastestMonkeypatching40to50}{7}
\DefMacro{ProjectsFastestCurse40to50}{116}
\DefMacro{ProjectsFastestAst40to50}{0}

\DefMacro{ProjectsFastestMonkeypatching50to60}{10}
\DefMacro{ProjectsFastestCurse50to60}{112}
\DefMacro{ProjectsFastestAst50to60}{0}

\DefMacro{ProjectsFastestMonkeypatching60to70}{18}
\DefMacro{ProjectsFastestCurse60to70}{104}
\DefMacro{ProjectsFastestAst60to70}{0}

\DefMacro{ProjectsFastestMonkeypatching70to80}{13}
\DefMacro{ProjectsFastestCurse70to80}{109}
\DefMacro{ProjectsFastestAst70to80}{0}

\DefMacro{ProjectsFastestMonkeypatching80to90}{14}
\DefMacro{ProjectsFastestCurse80to90}{108}
\DefMacro{ProjectsFastestAst80to90}{0}

\DefMacro{ProjectsFastestMonkeypatching90to100}{21}
\DefMacro{ProjectsFastestCurse90to100}{100}
\DefMacro{ProjectsFastestAst90to100}{2}

\DefMacro{InstrTimeOriginalLowerWhisker}{0.0}
\DefMacro{InstrTimeOriginalLowerQuartile}{0.0}
\DefMacro{InstrTimeOriginalMedian}{0.0}
\DefMacro{InstrTimeOriginalUpperQuartile}{0.0}
\DefMacro{InstrTimeOriginalUpperWhisker}{0.0}

\DefMacro{InstrTimeMonkeypatchingLowerWhisker}{0.417851448059082}
\DefMacro{InstrTimeMonkeypatchingLowerQuartile}{0.44130486249923706}
\DefMacro{InstrTimeMonkeypatchingMedian}{0.4697245359420775}
\DefMacro{InstrTimeMonkeypatchingUpperQuartile}{0.5382553339004515}
\DefMacro{InstrTimeMonkeypatchingUpperWhisker}{0.6771855354309082}

\DefMacro{InstrTimeCurseLowerWhisker}{0.37151551246643055}
\DefMacro{InstrTimeCurseLowerQuartile}{0.39481484889984125}
\DefMacro{InstrTimeCurseMedian}{0.4236346483230591}
\DefMacro{InstrTimeCurseUpperQuartile}{0.4866252541542053}
\DefMacro{InstrTimeCurseUpperWhisker}{0.6177060604095458}

\DefMacro{InstrTimeAstLowerWhisker}{3.5053038739471436}
\DefMacro{InstrTimeAstLowerQuartile}{3.753040765117645}
\DefMacro{InstrTimeAstMedian}{4.77014525478363}
\DefMacro{InstrTimeAstUpperQuartile}{8.377432830272674}
\DefMacro{InstrTimeAstUpperWhisker}{15.285018771347046}

\DefMacro{TestDurationOriginalLowerWhisker}{0.1585264205932617}
\DefMacro{TestDurationOriginalLowerQuartile}{0.28760939836502075}
\DefMacro{TestDurationOriginalMedian}{0.5760029554367065}
\DefMacro{TestDurationOriginalUpperQuartile}{1.4959235191345215}
\DefMacro{TestDurationOriginalUpperWhisker}{3.3043806552886963}

\DefMacro{TestDurationMonkeypatchingLowerWhisker}{1.1150732040405271}
\DefMacro{TestDurationMonkeypatchingLowerQuartile}{1.455992579460144}
\DefMacro{TestDurationMonkeypatchingMedian}{2.2628111839294434}
\DefMacro{TestDurationMonkeypatchingUpperQuartile}{5.656879544258118}
\DefMacro{TestDurationMonkeypatchingUpperWhisker}{11.946579933166504}

\DefMacro{TestDurationCurseLowerWhisker}{1.059140920639038}
\DefMacro{TestDurationCurseLowerQuartile}{1.3337087631225586}
\DefMacro{TestDurationCurseMedian}{1.9433839321136477}
\DefMacro{TestDurationCurseUpperQuartile}{4.451291084289551}
\DefMacro{TestDurationCurseUpperWhisker}{9.074173927307127}

\DefMacro{TestDurationAstLowerWhisker}{-0.8918529439849854}
\DefMacro{TestDurationAstLowerQuartile}{1.266510690246582}
\DefMacro{TestDurationAstMedian}{1.9967665277099607}
\DefMacro{TestDurationAstUpperQuartile}{5.450596074382782}
\DefMacro{TestDurationAstUpperWhisker}{11.524944589126587}

\DefMacro{E2ETimeOriginalLowerWhisker}{0.1585264205932617}
\DefMacro{E2ETimeOriginalLowerQuartile}{0.28760939836502075}
\DefMacro{E2ETimeOriginalMedian}{0.5760029554367065}
\DefMacro{E2ETimeOriginalUpperQuartile}{1.4959235191345215}
\DefMacro{E2ETimeOriginalUpperWhisker}{3.3043806552886963}

\DefMacro{E2ETimeMonkeypatchingLowerWhisker}{1.5539827346801758}
\DefMacro{E2ETimeMonkeypatchingLowerQuartile}{1.9458866119384766}
\DefMacro{E2ETimeMonkeypatchingMedian}{2.80076265335083}
\DefMacro{E2ETimeMonkeypatchingUpperQuartile}{6.239062190055847}
\DefMacro{E2ETimeMonkeypatchingUpperWhisker}{12.567496061325071}

\DefMacro{E2ETimeCurseLowerWhisker}{1.4575846195220947}
\DefMacro{E2ETimeCurseLowerQuartile}{1.7706674337387085}
\DefMacro{E2ETimeCurseMedian}{2.404392719268799}
\DefMacro{E2ETimeCurseUpperQuartile}{4.930524110794067}
\DefMacro{E2ETimeCurseUpperWhisker}{9.559628009796144}

\DefMacro{E2ETimeAstLowerWhisker}{4.6119396686553955}
\DefMacro{E2ETimeAstLowerQuartile}{5.3981834053993225}
\DefMacro{E2ETimeAstMedian}{8.091234683990479}
\DefMacro{E2ETimeAstUpperQuartile}{15.631664454936981}
\DefMacro{E2ETimeAstUpperWhisker}{30.724265575408936}

\DefMacro{TotalViolationsMonkeypatchingLowerWhisker}{0.0}
\DefMacro{TotalViolationsMonkeypatchingLowerQuartile}{0.0}
\DefMacro{TotalViolationsMonkeypatchingMedian}{0.0}
\DefMacro{TotalViolationsMonkeypatchingUpperQuartile}{0.0}
\DefMacro{TotalViolationsMonkeypatchingUpperWhisker}{0.0}

\DefMacro{TotalViolationsCurseLowerWhisker}{14.0}
\DefMacro{TotalViolationsCurseLowerQuartile}{70.0}
\DefMacro{TotalViolationsCurseMedian}{359.0}
\DefMacro{TotalViolationsCurseUpperQuartile}{1555.5}
\DefMacro{TotalViolationsCurseUpperWhisker}{3716.0}

\DefMacro{TotalViolationsAstLowerWhisker}{0.0}
\DefMacro{TotalViolationsAstLowerQuartile}{0.0}
\DefMacro{TotalViolationsAstMedian}{1.0}
\DefMacro{TotalViolationsAstUpperQuartile}{224.0}
\DefMacro{TotalViolationsAstUpperWhisker}{555.0}

\DefMacro{PassedTestsOriginalLowerWhisker}{0}
\DefMacro{PassedTestsOriginalLowerQuartile}{1.0}
\DefMacro{PassedTestsOriginalMedian}{6.0}
\DefMacro{PassedTestsOriginalUpperQuartile}{23.0}
\DefMacro{PassedTestsOriginalUpperWhisker}{56}

\DefMacro{PassedTestsMonkeypatchingLowerWhisker}{0}
\DefMacro{PassedTestsMonkeypatchingLowerQuartile}{1.0}
\DefMacro{PassedTestsMonkeypatchingMedian}{6.0}
\DefMacro{PassedTestsMonkeypatchingUpperQuartile}{23.0}
\DefMacro{PassedTestsMonkeypatchingUpperWhisker}{56}

\DefMacro{PassedTestsCurseLowerWhisker}{0}
\DefMacro{PassedTestsCurseLowerQuartile}{1.0}
\DefMacro{PassedTestsCurseMedian}{6.0}
\DefMacro{PassedTestsCurseUpperQuartile}{23.0}
\DefMacro{PassedTestsCurseUpperWhisker}{56}

\DefMacro{PassedTestsAstLowerWhisker}{0}
\DefMacro{PassedTestsAstLowerQuartile}{1.0}
\DefMacro{PassedTestsAstMedian}{5.0}
\DefMacro{PassedTestsAstUpperQuartile}{22.5}
\DefMacro{PassedTestsAstUpperWhisker}{54}

\DefMacro{FailedTestsOriginalLowerWhisker}{0}
\DefMacro{FailedTestsOriginalLowerQuartile}{0.0}
\DefMacro{FailedTestsOriginalMedian}{0.0}
\DefMacro{FailedTestsOriginalUpperQuartile}{2.0}
\DefMacro{FailedTestsOriginalUpperWhisker}{5}

\DefMacro{FailedTestsMonkeypatchingLowerWhisker}{0}
\DefMacro{FailedTestsMonkeypatchingLowerQuartile}{0.0}
\DefMacro{FailedTestsMonkeypatchingMedian}{0.0}
\DefMacro{FailedTestsMonkeypatchingUpperQuartile}{2.0}
\DefMacro{FailedTestsMonkeypatchingUpperWhisker}{5}

\DefMacro{FailedTestsCurseLowerWhisker}{0}
\DefMacro{FailedTestsCurseLowerQuartile}{0.0}
\DefMacro{FailedTestsCurseMedian}{0.0}
\DefMacro{FailedTestsCurseUpperQuartile}{2.0}
\DefMacro{FailedTestsCurseUpperWhisker}{5}

\DefMacro{FailedTestsAstLowerWhisker}{0}
\DefMacro{FailedTestsAstLowerQuartile}{0.0}
\DefMacro{FailedTestsAstMedian}{0.0}
\DefMacro{FailedTestsAstUpperQuartile}{2.0}
\DefMacro{FailedTestsAstUpperWhisker}{5}

\DefMacro{ViolationsPythonsourceMonkeypatchingLowerWhisker}{0}
\DefMacro{ViolationsPythonsourceMonkeypatchingLowerQuartile}{0.0}
\DefMacro{ViolationsPythonsourceMonkeypatchingMedian}{0.0}
\DefMacro{ViolationsPythonsourceMonkeypatchingUpperQuartile}{0.0}
\DefMacro{ViolationsPythonsourceMonkeypatchingUpperWhisker}{0}

\DefMacro{ViolationsSitepackagesMonkeypatchingLowerWhisker}{0}
\DefMacro{ViolationsSitepackagesMonkeypatchingLowerQuartile}{0.0}
\DefMacro{ViolationsSitepackagesMonkeypatchingMedian}{0.0}
\DefMacro{ViolationsSitepackagesMonkeypatchingUpperQuartile}{0.0}
\DefMacro{ViolationsSitepackagesMonkeypatchingUpperWhisker}{0}

\DefMacro{ViolationsProjectundertestMonkeypatchingLowerWhisker}{0}
\DefMacro{ViolationsProjectundertestMonkeypatchingLowerQuartile}{0.0}
\DefMacro{ViolationsProjectundertestMonkeypatchingMedian}{0.0}
\DefMacro{ViolationsProjectundertestMonkeypatchingUpperQuartile}{0.0}
\DefMacro{ViolationsProjectundertestMonkeypatchingUpperWhisker}{0}

\DefMacro{ViolationsPythonsourceCurseLowerWhisker}{0}
\DefMacro{ViolationsPythonsourceCurseLowerQuartile}{1.0}
\DefMacro{ViolationsPythonsourceCurseMedian}{2.0}
\DefMacro{ViolationsPythonsourceCurseUpperQuartile}{4.0}
\DefMacro{ViolationsPythonsourceCurseUpperWhisker}{8}

\DefMacro{ViolationsSitepackagesCurseLowerWhisker}{0}
\DefMacro{ViolationsSitepackagesCurseLowerQuartile}{0.0}
\DefMacro{ViolationsSitepackagesCurseMedian}{0.0}
\DefMacro{ViolationsSitepackagesCurseUpperQuartile}{3.0}
\DefMacro{ViolationsSitepackagesCurseUpperWhisker}{7}

\DefMacro{ViolationsProjectundertestCurseLowerWhisker}{0}
\DefMacro{ViolationsProjectundertestCurseLowerQuartile}{0.0}
\DefMacro{ViolationsProjectundertestCurseMedian}{0.0}
\DefMacro{ViolationsProjectundertestCurseUpperQuartile}{1.0}
\DefMacro{ViolationsProjectundertestCurseUpperWhisker}{2}

\DefMacro{ViolationsPythonsourceAstLowerWhisker}{0}
\DefMacro{ViolationsPythonsourceAstLowerQuartile}{0.0}
\DefMacro{ViolationsPythonsourceAstMedian}{0.0}
\DefMacro{ViolationsPythonsourceAstUpperQuartile}{0.0}
\DefMacro{ViolationsPythonsourceAstUpperWhisker}{0}

\DefMacro{ViolationsSitepackagesAstLowerWhisker}{0}
\DefMacro{ViolationsSitepackagesAstLowerQuartile}{0.0}
\DefMacro{ViolationsSitepackagesAstMedian}{0.0}
\DefMacro{ViolationsSitepackagesAstUpperQuartile}{2.0}
\DefMacro{ViolationsSitepackagesAstUpperWhisker}{5}

\DefMacro{ViolationsProjectundertestAstLowerWhisker}{0}
\DefMacro{ViolationsProjectundertestAstLowerQuartile}{0.0}
\DefMacro{ViolationsProjectundertestAstMedian}{0.0}
\DefMacro{ViolationsProjectundertestAstUpperQuartile}{0.0}
\DefMacro{ViolationsProjectundertestAstUpperWhisker}{0}

\DefMacro{TotalViolationsBuiltinMonkeyFastestAst}{3946556}
\DefMacro{TotalViolationsNonBuiltinMonkeyFastestAst}{29064}

\DefMacro{TotalViolationsBuiltinMonkeyFastestCurse}{14203160}
\DefMacro{TotalViolationsNonBuiltinMonkeyFastestCurse}{28965}

\DefMacro{TotalViolationsBuiltinMonkeyFastestMonkeypatching}{53412}
\DefMacro{TotalViolationsNonBuiltinMonkeyFastestMonkeypatching}{29166}

\DefMacro{TotalViolationsBuiltinCurseFastestAst}{774706}
\DefMacro{TotalViolationsNonBuiltinCurseFastestAst}{216499}

\DefMacro{TotalViolationsBuiltinCurseFastestCurse}{2012525}
\DefMacro{TotalViolationsNonBuiltinCurseFastestCurse}{213039}

\DefMacro{TotalViolationsBuiltinCurseFastestMonkeypatching}{14162}
\DefMacro{TotalViolationsNonBuiltinCurseFastestMonkeypatching}{213467}

\DefMacro{TotalViolationsBuiltinASTFastestAst}{228}
\DefMacro{TotalViolationsNonBuiltinASTFastestAst}{33}

\DefMacro{TotalViolationsBuiltinASTFastestCurse}{1837}
\DefMacro{TotalViolationsNonBuiltinASTFastestCurse}{31}

\DefMacro{TotalViolationsBuiltinASTFastestMonkeypatching}{2}
\DefMacro{TotalViolationsNonBuiltinASTFastestMonkeypatching}{31}

\DefMacro{algos_comparison_B_consistent_count}{1,437}
\DefMacro{algos_comparison_C_consistent_count}{1,461}
\DefMacro{algos_comparison_C+_consistent_count}{1,462}
\DefMacro{algos_comparison_D_consistent_count}{1,463}
\DefMacro{algos_comparison_B_mean_relative_overhead}{33.96}
\DefMacro{algos_comparison_C_mean_relative_overhead}{17.15}
\DefMacro{algos_comparison_C+_mean_relative_overhead}{16.70}
\DefMacro{algos_comparison_D_mean_relative_overhead}{12.33}
\DefMacro{algos_comparison_B_median_relative_overhead}{9.08}
\DefMacro{algos_comparison_C_median_relative_overhead}{6.61}
\DefMacro{algos_comparison_C+_median_relative_overhead}{6.60}
\DefMacro{algos_comparison_D_median_relative_overhead}{5.78}
\DefMacro{algos_comparison_B_max_relative_overhead}{5,500.27}
\DefMacro{algos_comparison_C_max_relative_overhead}{879.01}
\DefMacro{algos_comparison_C+_max_relative_overhead}{879.71}
\DefMacro{algos_comparison_D_max_relative_overhead}{848.76}
\DefMacro{algos_comparison_B_min_relative_overhead}{0.25}
\DefMacro{algos_comparison_C_min_relative_overhead}{0.20}
\DefMacro{algos_comparison_C+_min_relative_overhead}{0.19}
\DefMacro{algos_comparison_D_min_relative_overhead}{0.18}
\DefMacro{algos_comparison_B_sum_relative_overhead}{48,807.68}
\DefMacro{algos_comparison_C_sum_relative_overhead}{25,054.97}
\DefMacro{algos_comparison_C+_sum_relative_overhead}{24,421.44}
\DefMacro{algos_comparison_D_sum_relative_overhead}{18,041.01}
\DefMacro{algos_comparison_B_mean_absolute_overhead}{283.53}
\DefMacro{algos_comparison_C_mean_absolute_overhead}{207.36}
\DefMacro{algos_comparison_C+_mean_absolute_overhead}{201.69}
\DefMacro{algos_comparison_D_mean_absolute_overhead}{114.70}
\DefMacro{algos_comparison_B_median_absolute_overhead}{9.32}
\DefMacro{algos_comparison_C_median_absolute_overhead}{6.91}
\DefMacro{algos_comparison_C+_median_absolute_overhead}{6.95}
\DefMacro{algos_comparison_D_median_absolute_overhead}{6.18}
\DefMacro{algos_comparison_B_max_absolute_overhead}{17,272.45}
\DefMacro{algos_comparison_C_max_absolute_overhead}{17,356.85}
\DefMacro{algos_comparison_C+_max_absolute_overhead}{17,114.94}
\DefMacro{algos_comparison_D_max_absolute_overhead}{13,279.03}
\DefMacro{algos_comparison_B_min_absolute_overhead}{0.62}
\DefMacro{algos_comparison_C_min_absolute_overhead}{0.57}
\DefMacro{algos_comparison_C+_min_absolute_overhead}{0.54}
\DefMacro{algos_comparison_D_min_absolute_overhead}{0.55}
\DefMacro{algos_comparison_B_sum_absolute_overhead}{407,435.16}
\DefMacro{algos_comparison_C_sum_absolute_overhead}{302,947.12}
\DefMacro{algos_comparison_C+_sum_absolute_overhead}{294,874.34}
\DefMacro{algos_comparison_D_sum_absolute_overhead}{167,812.93}
\DefMacro{algos_comparison_projects_consistent_all_algos}{1,437}

\DefMacro{memory-total-p1-b-rel}{132.03}
\DefMacro{memory-mean-p1-b-rel}{0.92}
\DefMacro{memory-max-p1-b-rel}{1.05}
\DefMacro{memory-min-p1-b-rel}{0.0}
\DefMacro{memory-median-p1-b-rel}{1.0}
\DefMacro{memory-total-p1-c-rel}{132.39}
\DefMacro{memory-mean-p1-c-rel}{0.92}
\DefMacro{memory-max-p1-c-rel}{1.03}
\DefMacro{memory-min-p1-c-rel}{0.0}
\DefMacro{memory-median-p1-c-rel}{1.0}
\DefMacro{memory-total-p1-cplus-rel}{132.7}
\DefMacro{memory-mean-p1-cplus-rel}{0.92}
\DefMacro{memory-max-p1-cplus-rel}{1.03}
\DefMacro{memory-min-p1-cplus-rel}{0.0}
\DefMacro{memory-median-p1-cplus-rel}{1.0}
\DefMacro{memory-total-p1-d-rel}{132.49}
\DefMacro{memory-mean-p1-d-rel}{0.92}
\DefMacro{memory-max-p1-d-rel}{1.0}
\DefMacro{memory-min-p1-d-rel}{0.0}
\DefMacro{memory-median-p1-d-rel}{1.0}
\DefMacro{memory-total-p2-b-rel}{143.0}
\DefMacro{memory-mean-p2-b-rel}{1.0}
\DefMacro{memory-max-p2-b-rel}{1.0}
\DefMacro{memory-min-p2-b-rel}{1.0}
\DefMacro{memory-median-p2-b-rel}{1.0}
\DefMacro{memory-total-p2-c-rel}{143.2}
\DefMacro{memory-mean-p2-c-rel}{1.0}
\DefMacro{memory-max-p2-c-rel}{1.2}
\DefMacro{memory-min-p2-c-rel}{1.0}
\DefMacro{memory-median-p2-c-rel}{1.0}
\DefMacro{memory-total-p2-cplus-rel}{143.04}
\DefMacro{memory-mean-p2-cplus-rel}{1.0}
\DefMacro{memory-max-p2-cplus-rel}{1.04}
\DefMacro{memory-min-p2-cplus-rel}{1.0}
\DefMacro{memory-median-p2-cplus-rel}{1.0}
\DefMacro{memory-total-p2-d-rel}{143.0}
\DefMacro{memory-mean-p2-d-rel}{1.0}
\DefMacro{memory-max-p2-d-rel}{1.0}
\DefMacro{memory-min-p2-d-rel}{1.0}
\DefMacro{memory-median-p2-d-rel}{1.0}
\DefMacro{memory-total-p3-b-rel}{143.98}
\DefMacro{memory-mean-p3-b-rel}{1.0}
\DefMacro{memory-max-p3-b-rel}{1.0}
\DefMacro{memory-min-p3-b-rel}{0.98}
\DefMacro{memory-median-p3-b-rel}{1.0}
\DefMacro{memory-total-p3-c-rel}{143.95}
\DefMacro{memory-mean-p3-c-rel}{1.0}
\DefMacro{memory-max-p3-c-rel}{1.02}
\DefMacro{memory-min-p3-c-rel}{0.92}
\DefMacro{memory-median-p3-c-rel}{1.0}
\DefMacro{memory-total-p3-cplus-rel}{143.98}
\DefMacro{memory-mean-p3-cplus-rel}{1.0}
\DefMacro{memory-max-p3-cplus-rel}{1.0}
\DefMacro{memory-min-p3-cplus-rel}{0.98}
\DefMacro{memory-median-p3-cplus-rel}{1.0}
\DefMacro{memory-total-p3-d-rel}{144.0}
\DefMacro{memory-mean-p3-d-rel}{1.0}
\DefMacro{memory-max-p3-d-rel}{1.0}
\DefMacro{memory-min-p3-d-rel}{1.0}
\DefMacro{memory-median-p3-d-rel}{1.0}
\DefMacro{memory-total-p4-b-rel}{181.02}
\DefMacro{memory-mean-p4-b-rel}{1.26}
\DefMacro{memory-max-p4-b-rel}{1.41}
\DefMacro{memory-min-p4-b-rel}{0.99}
\DefMacro{memory-median-p4-b-rel}{1.36}
\DefMacro{memory-total-p4-c-rel}{180.89}
\DefMacro{memory-mean-p4-c-rel}{1.26}
\DefMacro{memory-max-p4-c-rel}{1.41}
\DefMacro{memory-min-p4-c-rel}{1.0}
\DefMacro{memory-median-p4-c-rel}{1.36}
\DefMacro{memory-total-p4-cplus-rel}{180.88}
\DefMacro{memory-mean-p4-cplus-rel}{1.26}
\DefMacro{memory-max-p4-cplus-rel}{1.41}
\DefMacro{memory-min-p4-cplus-rel}{1.0}
\DefMacro{memory-median-p4-cplus-rel}{1.36}
\DefMacro{memory-total-p4-d-rel}{180.74}
\DefMacro{memory-mean-p4-d-rel}{1.26}
\DefMacro{memory-max-p4-d-rel}{1.41}
\DefMacro{memory-min-p4-d-rel}{1.0}
\DefMacro{memory-median-p4-d-rel}{1.36}
\DefMacro{memory-total-p5-b-rel}{179.99}
\DefMacro{memory-mean-p5-b-rel}{1.26}
\DefMacro{memory-max-p5-b-rel}{2.57}
\DefMacro{memory-min-p5-b-rel}{0.7}
\DefMacro{memory-median-p5-b-rel}{1.23}
\DefMacro{memory-total-p5-c-rel}{180.29}
\DefMacro{memory-mean-p5-c-rel}{1.26}
\DefMacro{memory-max-p5-c-rel}{2.79}
\DefMacro{memory-min-p5-c-rel}{0.75}
\DefMacro{memory-median-p5-c-rel}{1.23}
\DefMacro{memory-total-p5-cplus-rel}{180.37}
\DefMacro{memory-mean-p5-cplus-rel}{1.26}
\DefMacro{memory-max-p5-cplus-rel}{2.79}
\DefMacro{memory-min-p5-cplus-rel}{0.67}
\DefMacro{memory-median-p5-cplus-rel}{1.23}
\DefMacro{memory-total-p5-d-rel}{180.02}
\DefMacro{memory-mean-p5-d-rel}{1.26}
\DefMacro{memory-max-p5-d-rel}{2.5}
\DefMacro{memory-min-p5-d-rel}{1.0}
\DefMacro{memory-median-p5-d-rel}{1.23}
\DefMacro{memory-total-p6-b-rel}{859.66}
\DefMacro{memory-mean-p6-b-rel}{5.97}
\DefMacro{memory-max-p6-b-rel}{50.3}
\DefMacro{memory-min-p6-b-rel}{0.6}
\DefMacro{memory-median-p6-b-rel}{2.88}
\DefMacro{memory-total-p6-c-rel}{960.8}
\DefMacro{memory-mean-p6-c-rel}{6.67}
\DefMacro{memory-max-p6-c-rel}{67.43}
\DefMacro{memory-min-p6-c-rel}{0.87}
\DefMacro{memory-median-p6-c-rel}{3.36}
\DefMacro{memory-total-p6-cplus-rel}{962.51}
\DefMacro{memory-mean-p6-cplus-rel}{6.68}
\DefMacro{memory-max-p6-cplus-rel}{77.0}
\DefMacro{memory-min-p6-cplus-rel}{0.85}
\DefMacro{memory-median-p6-cplus-rel}{3.24}
\DefMacro{memory-total-p6-d-rel}{1087.62}
\DefMacro{memory-mean-p6-d-rel}{7.55}
\DefMacro{memory-max-p6-d-rel}{49.4}
\DefMacro{memory-min-p6-d-rel}{1.0}
\DefMacro{memory-median-p6-d-rel}{3.55}
\DefMacro{memory-total-p7-b-rel}{1188.52}
\DefMacro{memory-mean-p7-b-rel}{8.25}
\DefMacro{memory-max-p7-b-rel}{127.5}
\DefMacro{memory-min-p7-b-rel}{0.87}
\DefMacro{memory-median-p7-b-rel}{3.79}
\DefMacro{memory-total-p7-c-rel}{1576.31}
\DefMacro{memory-mean-p7-c-rel}{10.95}
\DefMacro{memory-max-p7-c-rel}{182.13}
\DefMacro{memory-min-p7-c-rel}{0.99}
\DefMacro{memory-median-p7-c-rel}{4.28}
\DefMacro{memory-total-p7-cplus-rel}{1451.24}
\DefMacro{memory-mean-p7-cplus-rel}{10.08}
\DefMacro{memory-max-p7-cplus-rel}{182.0}
\DefMacro{memory-min-p7-cplus-rel}{0.98}
\DefMacro{memory-median-p7-cplus-rel}{4.32}
\DefMacro{memory-total-p7-d-rel}{1618.93}
\DefMacro{memory-mean-p7-d-rel}{11.24}
\DefMacro{memory-max-p7-d-rel}{273.83}
\DefMacro{memory-min-p7-d-rel}{1.01}
\DefMacro{memory-median-p7-d-rel}{4.62}
\DefMacro{memory-total-p8-b-rel}{877.08}
\DefMacro{memory-mean-p8-b-rel}{6.09}
\DefMacro{memory-max-p8-b-rel}{76.69}
\DefMacro{memory-min-p8-b-rel}{0.73}
\DefMacro{memory-median-p8-b-rel}{3.37}
\DefMacro{memory-total-p8-c-rel}{1246.41}
\DefMacro{memory-mean-p8-c-rel}{8.66}
\DefMacro{memory-max-p8-c-rel}{311.27}
\DefMacro{memory-min-p8-c-rel}{0.83}
\DefMacro{memory-median-p8-c-rel}{3.64}
\DefMacro{memory-total-p8-cplus-rel}{1132.19}
\DefMacro{memory-mean-p8-cplus-rel}{7.86}
\DefMacro{memory-max-p8-cplus-rel}{193.49}
\DefMacro{memory-min-p8-cplus-rel}{0.83}
\DefMacro{memory-median-p8-cplus-rel}{3.65}
\DefMacro{memory-total-p8-d-rel}{1187.78}
\DefMacro{memory-mean-p8-d-rel}{8.25}
\DefMacro{memory-max-p8-d-rel}{159.19}
\DefMacro{memory-min-p8-d-rel}{1.01}
\DefMacro{memory-median-p8-d-rel}{4.05}
\DefMacro{memory-total-p9-b-rel}{922.42}
\DefMacro{memory-mean-p9-b-rel}{6.45}
\DefMacro{memory-max-p9-b-rel}{101.2}
\DefMacro{memory-min-p9-b-rel}{0.93}
\DefMacro{memory-median-p9-b-rel}{3.35}
\DefMacro{memory-total-p9-c-rel}{1114.36}
\DefMacro{memory-mean-p9-c-rel}{7.79}
\DefMacro{memory-max-p9-c-rel}{101.2}
\DefMacro{memory-min-p9-c-rel}{1.01}
\DefMacro{memory-median-p9-c-rel}{3.99}
\DefMacro{memory-total-p9-cplus-rel}{1053.46}
\DefMacro{memory-mean-p9-cplus-rel}{7.37}
\DefMacro{memory-max-p9-cplus-rel}{101.2}
\DefMacro{memory-min-p9-cplus-rel}{1.01}
\DefMacro{memory-median-p9-cplus-rel}{3.88}
\DefMacro{memory-total-p9-d-rel}{1189.69}
\DefMacro{memory-mean-p9-d-rel}{8.32}
\DefMacro{memory-max-p9-d-rel}{101.2}
\DefMacro{memory-min-p9-d-rel}{1.01}
\DefMacro{memory-median-p9-d-rel}{4.01}
\DefMacro{memory-total-p10-b-rel}{1676.11}
\DefMacro{memory-mean-p10-b-rel}{11.64}
\DefMacro{memory-max-p10-b-rel}{309.95}
\DefMacro{memory-min-p10-b-rel}{0.58}
\DefMacro{memory-median-p10-b-rel}{4.14}
\DefMacro{memory-total-p10-c-rel}{2499.88}
\DefMacro{memory-mean-p10-c-rel}{17.36}
\DefMacro{memory-max-p10-c-rel}{373.69}
\DefMacro{memory-min-p10-c-rel}{0.65}
\DefMacro{memory-median-p10-c-rel}{5.47}
\DefMacro{memory-total-p10-cplus-rel}{2434.75}
\DefMacro{memory-mean-p10-cplus-rel}{16.91}
\DefMacro{memory-max-p10-cplus-rel}{378.09}
\DefMacro{memory-min-p10-cplus-rel}{0.65}
\DefMacro{memory-median-p10-cplus-rel}{5.63}
\DefMacro{memory-total-p10-d-rel}{3386.22}
\DefMacro{memory-mean-p10-d-rel}{23.52}
\DefMacro{memory-max-p10-d-rel}{1022.07}
\DefMacro{memory-min-p10-d-rel}{1.01}
\DefMacro{memory-median-p10-d-rel}{6.33}

\DefMacro{gc_comparison_macros_total_projects}{1363}
\DefMacro{gc_comparison_macros_no_gc_faster_endtoend_projects}{718}
\DefMacro{gc_comparison_macros_gc_faster_endtoend_projects}{645}
\DefMacro{gc_comparison_macros_no_gc_faster_test_projects}{705}
\DefMacro{gc_comparison_macros_gc_faster_test_projects}{658}
\DefMacro{gc_comparison_macros_gc_faster_0.5s_projects}{34}
\DefMacro{gc_comparison_macros_no_gc_faster_0.5s_projects}{39}
\DefMacro{gc_comparison_macros_same_speed_0.5s_projects}{1290}
\DefMacro{gc_comparison_macros_different_speed_0.5s_projects}{73}
\DefMacro{gc_comparison_macros_gc_faster_0.5s_specs_mean}{4.86}
\DefMacro{gc_comparison_macros_no_gc_faster_0.5s_specs_mean}{3.45}
\DefMacro{gc_comparison_macros_gc_faster_0.5s_events_count_mean}{42747.03}
\DefMacro{gc_comparison_macros_no_gc_faster_0.5s_events_count_mean}{299065.55}
\DefMacro{gc_comparison_macros_gc_faster_0.5s_monitors_count_mean}{2852.43}
\DefMacro{gc_comparison_macros_no_gc_faster_0.5s_monitors_count_mean}{1660.82}
\DefMacro{gc_comparison_macros_gc_faster_0.5s_multiple_objects_specs_mean}{2.16}
\DefMacro{gc_comparison_macros_no_gc_faster_0.5s_multiple_objects_specs_mean}{1.37}
\DefMacro{gc_comparison_macros_gc_faster_0.5s_multiple_objects_events_count_mean}{7688.52}
\DefMacro{gc_comparison_macros_no_gc_faster_0.5s_multiple_objects_events_count_mean}{5549.86}
\DefMacro{gc_comparison_macros_gc_faster_0.5s_multiple_objects_monitors_count_mean}{1809.93}
\DefMacro{gc_comparison_macros_no_gc_faster_0.5s_multiple_objects_monitors_count_mean}{84.44}

\DefMacro{algo_a_comparison_total_projects}{68}
\DefMacro{algo_a_comparison_A_relative_overhead_mean}{95.19}
\DefMacro{algo_a_comparison_B_relative_overhead_mean}{23.58}
\DefMacro{algo_a_comparison_C_relative_overhead_mean}{12.92}
\DefMacro{algo_a_comparison_C+_relative_overhead_mean}{13.01}
\DefMacro{algo_a_comparison_D_relative_overhead_mean}{11.11}
\DefMacro{algo_a_comparison_A_relative_overhead_median}{95.19}
\DefMacro{algo_a_comparison_B_relative_overhead_median}{4.31}
\DefMacro{algo_a_comparison_C_relative_overhead_median}{3.30}
\DefMacro{algo_a_comparison_C+_relative_overhead_median}{3.36}
\DefMacro{algo_a_comparison_D_relative_overhead_median}{3.27}
\DefMacro{algo_a_comparison_A_relative_overhead_min}{1.28}
\DefMacro{algo_a_comparison_B_relative_overhead_min}{1.26}
\DefMacro{algo_a_comparison_C_relative_overhead_min}{1.20}
\DefMacro{algo_a_comparison_C+_relative_overhead_min}{1.20}
\DefMacro{algo_a_comparison_D_relative_overhead_min}{1.15}
\DefMacro{algo_a_comparison_A_relative_overhead_max}{3197.20}
\DefMacro{algo_a_comparison_B_relative_overhead_max}{294.39}
\DefMacro{algo_a_comparison_C_relative_overhead_max}{264.61}
\DefMacro{algo_a_comparison_C+_relative_overhead_max}{265.08}
\DefMacro{algo_a_comparison_D_relative_overhead_max}{258.96}
\DefMacro{algo_a_comparison_A_relative_overhead_sum}{6473.23}
\DefMacro{algo_a_comparison_B_relative_overhead_sum}{1603.61}
\DefMacro{algo_a_comparison_C_relative_overhead_sum}{878.52}
\DefMacro{algo_a_comparison_C+_relative_overhead_sum}{884.92}
\DefMacro{algo_a_comparison_D_relative_overhead_sum}{755.78}
\DefMacro{algo_a_comparison_A_absolute_overhead_mean}{1578.18}
\DefMacro{algo_a_comparison_B_absolute_overhead_mean}{548.28}
\DefMacro{algo_a_comparison_C_absolute_overhead_mean}{262.29}
\DefMacro{algo_a_comparison_C+_absolute_overhead_mean}{263.04}
\DefMacro{algo_a_comparison_D_absolute_overhead_mean}{212.85}
\DefMacro{algo_a_comparison_A_absolute_overhead_median}{185.96}
\DefMacro{algo_a_comparison_B_absolute_overhead_median}{109.44}
\DefMacro{algo_a_comparison_C_absolute_overhead_median}{62.84}
\DefMacro{algo_a_comparison_C+_absolute_overhead_median}{60.72}
\DefMacro{algo_a_comparison_D_absolute_overhead_median}{61.20}
\DefMacro{algo_a_comparison_A_absolute_overhead_min}{10.12}
\DefMacro{algo_a_comparison_B_absolute_overhead_min}{8.36}
\DefMacro{algo_a_comparison_C_absolute_overhead_min}{6.19}
\DefMacro{algo_a_comparison_C+_absolute_overhead_min}{4.27}
\DefMacro{algo_a_comparison_D_absolute_overhead_min}{6.44}
\DefMacro{algo_a_comparison_A_absolute_overhead_max}{16627.10}
\DefMacro{algo_a_comparison_B_absolute_overhead_max}{5688.48}
\DefMacro{algo_a_comparison_C_absolute_overhead_max}{3974.41}
\DefMacro{algo_a_comparison_C+_absolute_overhead_max}{3981.48}
\DefMacro{algo_a_comparison_D_absolute_overhead_max}{3889.18}
\DefMacro{algo_a_comparison_A_absolute_overhead_sum}{107316.23}
\DefMacro{algo_a_comparison_B_absolute_overhead_sum}{37283.33}
\DefMacro{algo_a_comparison_C_absolute_overhead_sum}{17835.93}
\DefMacro{algo_a_comparison_C+_absolute_overhead_sum}{17887.04}
\DefMacro{algo_a_comparison_D_absolute_overhead_sum}{14473.72}

\pgfmathtruncatemacro{\InstrSpeedupFactor}{\UseMacro{InstrTimeSumAst} / ((\UseMacro{InstrTimeSumMonkeypatching} + \UseMacro{InstrTimeSumCurse}) / 2)}
\pgfmathtruncatemacro{\FastestTenCount}{\UseMacro{fastest-count-p10-b} + \UseMacro{fastest-count-p10-c} + \UseMacro{fastest-count-p10-cplus} + \UseMacro{fastest-count-p10-d}}
\pgfmathtruncatemacro{\FastestNineCount}{\UseMacro{fastest-count-p9-b} + \UseMacro{fastest-count-p9-c} + \UseMacro{fastest-count-p9-cplus} + \UseMacro{fastest-count-p9-d}}
\pgfmathtruncatemacro{\FastestTenNineCount}{\FastestTenCount + \FastestNineCount}
\pgfmathtruncatemacro{\FastestOneCount}{\UseMacro{fastest-count-p1-b} + \UseMacro{fastest-count-p1-c} + \UseMacro{fastest-count-p1-cplus} + \UseMacro{fastest-count-p1-d}}
\pgfmathtruncatemacro{\FastestTwoCount}{\UseMacro{fastest-count-p2-b} + \UseMacro{fastest-count-p2-c} + \UseMacro{fastest-count-p2-cplus} + \UseMacro{fastest-count-p2-d}}
\pgfmathtruncatemacro{\FastestOneTwoCount}{\FastestOneCount + \FastestTwoCount}
\pgfmathtruncatemacro{\FastestDOneTwoCount}{\UseMacro{fastest-count-p1-d} + \UseMacro{fastest-count-p2-d}}

\begin{document}
\newcommand{\Title}{A Generic and Efficient Python Runtime Verification
  System and its Large-scale Evaluation}

\title{\Title}

\author{Zhuohang Shen\inst{1}, Mohammed Yaseen, Denini Silva\inst{2}, Kevin Guan\inst{1},\\ Junho Lee\inst{3}, Marcelo d'Amorim\inst{2,4}, Owolabi Legunsen\inst{1}}

\authorrunning{Shen et al.}
\institute{Cornell University, Ithaca, New York, USA\and
  Federal University of Pernambuco, Recife, Pernambuco, Brazil\and
  University of Virginia, Charlottesville, Virginia, USA\and
  North Carolina State University, Raleigh, North Carolina, USA  
}
\maketitle              %

\thispagestyle{plain}
\pagestyle{plain}

\vspace{-4ex}

\begin{abstract}

Runtime verification (\RV) now scales for testing thousands of
open-source Java projects, helping find hundreds of bugs. The popular
Python ecosystem could use such benefits. But, today's Python \RV
systems are limited to a domain or specification logic, or slow. We
propose \Tool, a generic, extensible, and efficient \RV system for
Python. \Tool supports five logics, implements five existing
monitoring algorithms, ships with \NumSpecs{} API \specs of Python and
widely-used libraries, supports three instrumentation strategies, and
users can easily add more of these. On \NumTestsSum{} unit tests
in \NumProjects{} GitHub projects, we find mainly that (i)~the default
monitoring algorithm for Java is often not the fastest for Python;
(ii)~\Tool is up
to \num{\UseMacro{comparison-All_4-dynapyt_libs-max-relative-speedup}}\x
faster\Space{ and
finds\Space{ \UseMacro{comparison-statistics-D-Dynapyt-D-violations}}
more violations} than two recent dynamic analysis systems; and
(iii)~\NumTotalIssuesAndPrAccepted{} of \NumTruePositiveUniq{} bugs
that \Tool helped find so far were fixed by developers.\Space{ These
results suggest that} \Tool's generality and efficiency position it
well as an excellent platform for the next advances on \RV for Python.

\end{abstract}

\vspace{-8ex}

\section{Introduction}
\label{sec:intro}

\vspace{-2ex}

Runtime verification (\RV)~\cite{schneiderSecurityAutomata,
  kim1999formally,
  HavelundRosuASE2001MonitoringProgramsUsingRewriting} monitors
  program runs against formal specifications (\specs); it is now used in
  deployed systems~\cite{ArtcatWebPage,
  efficient-rv-linux-kernel}. \RV\ also amplifies the bug-finding
  ability of tests.\Space{ In recent research,} Monitoring
\emph{passing} tests in thousands of Java projects\Space{ executions}
against \specs of JDK APIs found hundreds of bugs\Space{ that testing
  alone
  missed}~\cite{JavedAndBinderLargeScaleEvaluationOfRVToolsAPSEC2018,
  MirandaETAL19RVPrio, LegunsenETAL19SpecEval,
  GuanAndLegunsenRVStudyISSTA2024}.

Despite these advances, \rv systems are limited for Python, an
increasingly important software ecosystem~\cite{TiobeIndex}.
For example,\Space{ we note that}
\pythonrv~\cite{Renberg2014PythonRVKTHThesis}\Space{ offers few
  primitives for writing \specs, only} supports a fraction of Python
syntax, and its\Space{ runtime or memory} overheads were not
evaluated. LogScope~\cite{barringer2010formal} and
PyContract~\cite{dams2022python} analyze logs offline; they do not
work online, \eg,\Space{ to monitor tests} in continuous
integration~(CI). Domain-specific \hatrv~\cite{zhou2022improving} and
\vypr~\cite{dawes2019vypr2, javed2020perfci, dawes2019specification}
only target embedded systems and performance monitoring of\Space{
  Python} web services, respectively.
\dynapyt~\cite{eghbali2022dynapyt} and \dylin~\cite{Eghbali2025DyLin}
are dynamic analysis systems; they perform \rv if their manually
written ``checkers'' are seen as monitors.

In sum, today's Python \rv\ systems have one or more of these
limitations: (i)~They are hard-wired to a \spec logic, but no logic
can express all\Space{ \RV} \specs~\cite{rosu2012safety} and users may
want other logics~\cite{teixeira2021demystifying}. (ii)~Those that use
formal logic, \eg, linear temporal logic (LTL) can only monitor
regular patterns\Space{ (in the automata-theoretic sense). So, \eg,};
they cannot monitor, \eg,
context-free~\cite{MeredithETALMOPContextFreePatterns08} or
Turing-complete~\cite{MeredithRosuMonitoringStringRewritingASE13}
patterns. (iii)~Those that require manually writing monitors cannot
use algorithms for\Space{ automatically} synthesizing efficient
monitors from \specs~\cite{HavelundRosuTACAS2002SynthesizingMonitors,
LTLMonitorSynthesis, AllenLTLMonitorSynthesis,
MeredithETALMOPContextFreePatterns08,
MeredithRosuMonitoringStringRewritingASE13,
PrasannaAndRosuMTLMonitoringRV2004, schneider2019formally,
ho2014online}. (iv)~None uses parametric monitoring
algorithms~\cite{chen2009efficient, BarringerETALJLC2010EagleToRuler,
LuoETAlRVMonitor14} that help \RV scale for testing in Java. (v)~None
was evaluated at scale with dozens of \specs and thousands of projects
like for
Java~\cite{JavedAndBinderLargeScaleEvaluationOfRVToolsAPSEC2018,
GuanAndLegunsenRVStudyISSTA2024}. (vi)~None supports online
\emph{and} offline monitoring.

We propose \textbf{\Tool} to address these limitations; it is a
\emph{generic} and \emph{efficient} \RV system for Python and the
first Python instance of\Space{ well-known and widely-cited}
{Monitoring-Oriented Programming} (MOP)\Space{ that was proposed over
  the last two decades}~\cite{ChenAndRosuTowardsMOP03}. MOP scaled
well for simultaneous monitoring of multiple \specs\Space{ in many
  \oss}~\cite{chen2009parametric, chen2009efficient,
  LuoETAlRVMonitor14,
  DeckerETALTACAS2016MonitoringWithUnionFind}. Also,
\javamop~\cite{JinEtAlJavaMOPToolPaperICSE12, LuoETAlRVMonitor14} (the
MOP instance for Java)\Space{ and its \rvmonitor
  backend~\cite{LuoETAlRVMonitor14}} spurred a lot of \RV research on \RV for
Java~\cite{ChenAndRosuMOPOOPSLA2007,
  ChenEtAlFormalMonitoringICFEM2004,
  ChenEtAlCheckingAndCorrectingWithJavaMOPRV2006, BoddenMOPBOxRV11,
  JinEtAlJavaMOPToolPaperICSE12,
  JinEtAlGarbageCollectionPLDI2011,JavaMOPWebPage, rv-monitor-webpage,
  mopbox-webpage}.

\Tool is \emph{generic} in four ways. First,\Space{ \Tool uses a logic
  plugins to support multiple \spec logics. So,} it is not hard-wired
to a logic,\Space{ supports a ``bring your own logic'' approach:} so
users can add plugins for more logics. \Tool ships with five logic
plugins (more planned in the future): past- and future-time LTL,
extended regular expressions (ERE), FSM, and context-free grammars
(CFG). Second, \Tool does not require a DSL; users can write \specs
directly in Python or in a JSON-like frontend. Third, \Tool has an API
to support different instrumentation strategies; it currently supports
three. Lastly, \Tool is not restricted to a domain---the API \specs
that we check concern Python and widely used libraries, \eg,
Tensorflow (machine learning), SciPy (scientific\Space{ and technical}
computing), and Pandas (data analysis\Space{ and manipulation}).

\Tool's \emph{efficiency} comes from two sources: it
(i)~automatically synthesizes efficient monitors from \specs; and
(ii)~uses parametric trace-slicing based monitoring
algorithms~\cite{chen2009parametric, chen2009efficient}
(\S\ref{sec:tool:prelim} has a trace slicing primer).

We implement \Tool as a plugin for \CodeIn{pytest}~\cite{Pytest}, the
most popular Python testing framework. We also conduct a large-scale
evaluation, using (i)~\NumTestsSum{} unit tests in \NumProjects{}
projects\Space{, totaling \Fix{C} SLOC}; (ii)~three tools: \Tool,
\dylin, and \dynapyt; (iii)~\NumSpecs{} \specs of correct API usage in
Python and widely-used libraries, \NumSpecsMultObj{} of which relate
multiple object types; (iv)~five\Space{ trace-slicing based}
monitoring algorithms; and (v)~three instrumentation
strategies.\Space{For context on these numbers, The prior largest
  evaluation used 11 projects and 15 \specs~\cite{Eghbali2025DyLin}.}
The\Space{ evaluated} projects come from many\Space{ application}
domains: machine learning, data processing, infrastructure-as-code,
blockchains, etc.

We find that the most complex online monitoring algorithm (and the
only one in \javamop) is the fastest in \UseMacro{fastest-all-percent-d}\% 
projects by \UseMacro{fastest-D-vs-second-fastest-diff-mean} seconds on 
average (median: \UseMacro{fastest-D-vs-second-fastest-diff-median}
seconds). But, times for all four online monitoring algorithms are
within 2 seconds in \UseMacro{fastest-count-2s-all-all} projects, and 
within 5 seconds in \UseMacro{fastest-count-5s-all-all} projects. That is, 
if users do not mind 2s or 5s slowdowns, the choice of algorithm does not 
matter for \UseMacro{fastest-count-2s-all-percent-all}\% and 
\UseMacro{fastest-count-5s-all-percent-all}\% of these projects,
respectively, using these \specs.

We also compare \Tool with \dynapyt~\cite{eghbali2022dynapyt} and
\dylin~\cite{Eghbali2025DyLin}---two recent Python dynamic
analyzers---using \NumSpecsComparingTools{} \specs that \dynapyt\ and
\dylin\ support.\Space{ To do so, we first manually convert \Fix{20}
  \Tool specs into \dylin and \dynapyt checkers and convert \Fix{17}
  \dylin and \dynapyt checkers into \Tool \specs.} All three tools
monitor the code under test (CUT). \Tool \emph{always} monitors
3rd-party libraries (necessary for \RV during
testing~\cite{GuanAndLegunsenRVStudyISSTA2024}) \emph{and} the Python
runtime (to check conformance with its own APIs). But, \dylin and
\dynapyt cannot monitor the Python runtime, so we evaluate them
without libraries (as their authors did~\cite{eghbali2022dynapyt,
  Eghbali2025DyLin}), and with libraries.
\Tool with libraries and Python-runtime monitoring is up to
\UseMacro{comparison-All_4-dynapyt-max-relative-speedup}\x (avg:
\UseMacro{comparison-All_4-dynapyt-relative-speedup}\x) faster than running
\dylin and \dynapyt \emph{without} monitoring libraries. Those
speedups translate to up to \num{\UseMacro{comparison-All_4-dynapyt-max-absolute-speedup}} seconds
(avg: \UseMacro{comparison-All_4-dynapyt-absolute-speedup} seconds). \dynapyt
\emph{with} libraries succeeds only in
\UseMacro{comparison-statistics-Dynapyt_libs-total-projects} projects.
In this mode, \Tool is up to
\num{\UseMacro{comparison-All_4-dynapyt_libs-max-relative-speedup}}\x (avg:
\num{\UseMacro{comparison-All_4-dynapyt_libs-relative-speedup}}\x) faster. Those
speedups translate to up to \num{\UseMacro{comparison-All_4-dynapyt_libs-max-absolute-speedup}} seconds
(avg: \num{\UseMacro{comparison-All_4-dynapyt_libs-absolute-speedup}} seconds).

Despite being generic and monitoring more code, \Tool scales better
than \dylin and \dynapyt\Space{ as the number of monitored events
  grows}.
We so far reported \NumTotalBugsFound{} of \NumTruePositiveUniq{} bugs that \Tool
helped us find to developers\Space{ of the monitored Python projects};
\NumTotalIssuesAndPrAccepted{} were confirmed or fixed;
\NumTotalPRIssuesRejected{} were rejected.\Space{ Like with previous
  experiments with \javamop~\cite{LegunsenETALASE2016SpecEval,
    LegunsenETAL19SpecEval}, we also find false alarms due to bugs in
  our \specs that we have since fixed.}
We also discovered four bugs in the Python interpreter itself~\cite{cpython-repo}. One has been 
fixed in a newer version~\cite{cpython125397}, one is open~\cite{cpython132372}, and the other two were confirmed\Space{. For two issues, a core Python developer acknowledged the 
problem and submitted a pull request, but it was ultimately rejected by 
another maintainer}~\cite{cpython130902,cpython130850}.

These results are promising:\Space{ \Tool's generality does not come
  at the expense of efficiency or bug finding utility.}  \Tool's
generality, and the degree to which it outperforms specialized
techniques, suggest that it is better than current alternatives as a
platform for future research and development on \RV for Python.

This paper makes the following contributions:

\begin{itemize}[topsep=0ex,itemsep=0pt,leftmargin=1em]

\item[\Contrib{}]\textbf{System.} \Tool is a generic and efficient
  \RV system for Python, and the first MOP instance for Python.

\item[\Contrib{}]\textbf{\Specs.} \Tool ships with \NumSpecs{} API
  \specs of\Space{ correct API usage in Python and} popular libraries;
  including \NumSpecsFromOtherTools{} from \dylin and \dynapyt.

\item[\Contrib{}]\textbf{Comparisons.} We compare all online trace-slicing based
  monitoring algorithms, Python \RV systems, and Python
  instrumentation strategies for \RV.

\item[\Contrib{}]\textbf{Large-scale Evaluation.} We conduct the largest
  evaluation of \RV for Python, and of \dylin and \dynapyt, till
  date\Space{ in terms of numbers of projects and \specs}.
  
\end{itemize}

\noindent
\Tool and our data are at \ArtifactLink.

\begin{figure}[t!]
  \begin{minipage}{\linewidth}
  \begin{lstlisting}[language=json, basicstyle=\tiny\ttfamily, numbers=left, numbersep=1pt, numberstyle=\tiny\ttfamily]
{ "Description": "Detects TOCTOU...",
  "Variables": {"checked_files": "set"},
  "Formalism": "fsm",
  "Formula": "s0 [use -> s1, check -> s2] s2 [use -> s3, check -> s2] s3 [use -> s3] alias Violation = s3",
  "Creation_Events": ["check"],
  "Events": {
    "After": {
      "check": [["os", "access"]], "use": [["builtins", "open"]]
  }},
  "Event_Actions": {
    "After": {
      "check": "self.checked_files.add(file)", "use": "return file in self.checked_files"
  }},
  "Handlers": {"Violation": "Security threat! ..."  } }
  \end{lstlisting}
  \end{minipage}
  \vspace{-4ex}
  \caption{\label{lst:access_spec}\label{subfig:access_spec}\label{fig:toctou_combined}\toctou \spec in \Tool's JSON-like frontend; user can also write in Python.}
  \vspace{-5.5ex}
\end{figure}

\begin{wrapfigure}{r}{.30\textwidth}
  \begin{minipage}{.29\textwidth}
    \vspace{-10ex}
    \centering
    \hspace{-2ex}
    \begin{tikzpicture}[
        scale=0.7,
        every node/.style={scale=0.75},
        ->,
        >=stealth,
        node distance=1cm,
        every state/.style={draw,circle,minimum size=15pt,inner sep=1pt, font=\scriptsize},
        auto
      ]
      \node[state,initial] (s0) {$s_0$};
      \node[state] (s1) [right=of s0] {$s_1$};
      \node[state] (s2) [below=of s0] {$s_2$};
      \node[state,accepting] (s3) [right=of s2] {Vio};
      
      \path (s0) edge node {use} (s1)
      (s0) edge node {check} (s2)
      (s1) edge[loop above] node {default} (s1)
      (s2) edge node {use} (s3)
      (s2) edge[loop below] node {check} (s2)
      (s3) edge[loop above] node {use} (s3);
    \end{tikzpicture}
    \vspace{-2ex}
    \caption{\label{fig:toctou_fsm}\toctou's FSM.}
    \vspace{-5ex}
  \end{minipage}
\end{wrapfigure}

\vspace{-3ex}
\section{Illustrative Example}
\label{sec:example}
\vspace{-1ex}

\Space{The dangerous }``Time-Of-Check to Time-Of-Use'' (\toctou)
vulnerabilities~\cite{cwe367} involve the trace <check permission for
file \CodeIn{f}>$\leadsto$<use \CodeIn{f}>. Attackers can exploit
the\Space{ short} interval between these two events, \eg, to replace
\CodeIn{f} with a symbolic link or changing its contents or
permissions~\cite{bishop2003,wei2005tocttou}. \toctou is described in
Python's documentation for
\CodeIn{os.access}~\cite{python-docs-access}: \emph{``Using access()
to check if a user is authorized to e.g. open a file before actually
doing so using open() creates a security hole''}.

\MyPara{\Tool's \toctou \spec}
Figure~\ref{lst:access_spec} shows this \toctou \spec in \Tool{}'s
JSON-like frontend (users can also write \specs directly in Python).
There, \texttt{\seqsplit{Description}} summarizes
the \spec, \CodeIn{Variables} defines the set \CodeIn{checked\_files}
to track files accessed at runtime, \CodeIn{Formalism} indicates that
the \spec is formalized as an FSM, \CodeIn{Formula} is the FSM in
textual form (Figure~\ref{fig:toctou_fsm} shows it in visual form)
\CodeIn{Creation\_Events} identifies the events that
should trigger monitor creation (\CodeIn{check}, in our case),
 \CodeIn{Events} maps \CodeIn{check} events to
\CodeIn{os.access()} calls and \CodeIn{use} events to
\CodeIn{builtins.open()} calls,
\CodeIn{Event\_Actions} defines the actions the monitor takes when events occur
(\eg, add \CodeIn{f} to
\CodeIn{checked\_files} on \CodeIn{check} events).
Finally, \CodeIn{Handler} contains code to run if the FSM reaches a
violating state; it can be any, \eg error recovery, code. But, for use in testing, we simply
print a message with sufficient information to help debug each
violation.

\begin{wrapfigure}{l}{.35\textwidth}
  \centering
  \vspace{-6.5ex}
    \begin{lstlisting}[language=Python, basicstyle=\tiny\ttfamily, numbersep=3pt, numberstyle=\tiny\ttfamily,escapeinside={(*@}{@*)}]
# Vulnerable version:(*@\label{toctou:vulnerable:start}@*)
if os.path.isfile(password_file) and
  os.access(password_file, os.R_OK):
    with open(password_file) as fp:
        return fp.readline().strip()(*@\label{toctou:vulnerable:end}@*)
# Fixed version:(*@\label{toctou:fix:start}@*)
try:
    with open(password_file) as fp:
        return fp.readline().strip()
except FileNotFoundError: ...(*@\label{toctou:fix:end}@*)
    \end{lstlisting}
    \vspace{-3ex}
    \caption{Code that violates \toctou in \CodeIn{MyCLI}~\cite{mycli}}
    \vspace{-5.5ex}
  \label{lst:toctou_fix}
\end{wrapfigure}

\MyPara{\toctou violations found by \Tool}
We reported four \toctou violations that \Tool helped us find; two
were accepted or fixed, one was rejected, and the other is
pending. One violation was in \CodeIn{mycli}\cite{mycli}, a popular
command-line MySQL interface with 664 forks and 11.6K stars. The
violation occurred in a function that reads passwords from a file.
Figure~\ref{lst:toctou_fix} shows a simplified version of the
vulnerable code
(lines~\ref{toctou:vulnerable:start}--\ref{toctou:vulnerable:end}) and
our fix (lines~\ref{toctou:fix:start}--\ref{toctou:fix:end}). The
vulnerable code yields
trace \CodeIn{os.access(f)}$\leadsto$\CodeIn{open(f)} that violates
the FSM.  Our fix, which was merged by
\CodeIn{mycli} developers~\cite{mycli-pr1203}, eliminates the explicit access checks and
instead handles access errors more securely with exception
handling,\Space{ avoiding the race condition entirely} aligning with
best practices for secure file handling~\cite{bishop2003}.  It is
worth noting that verifying behavioral properties like \toctou
statically can be challenging, one violation that we
find~\cite{livediffer-pr3} involves
\CodeIn{check} and \CodeIn{use} being signaled from different functions. An
inter-procedural static analysis would be needed to verify
\toctou in that scenario. \Specs with multiple
object types as parameters, plus the need to reason about library code
would make such static analysis even harder.

\vspace{-3ex}
\section{Framework}
\label{sec:framework}
\vspace{-2ex}

\begin{wrapfigure}{r}{.52\textwidth}
  \centering
  \vspace{-5ex}
  \includegraphics[scale=0.19,trim={0 9cm 0 0},clip]{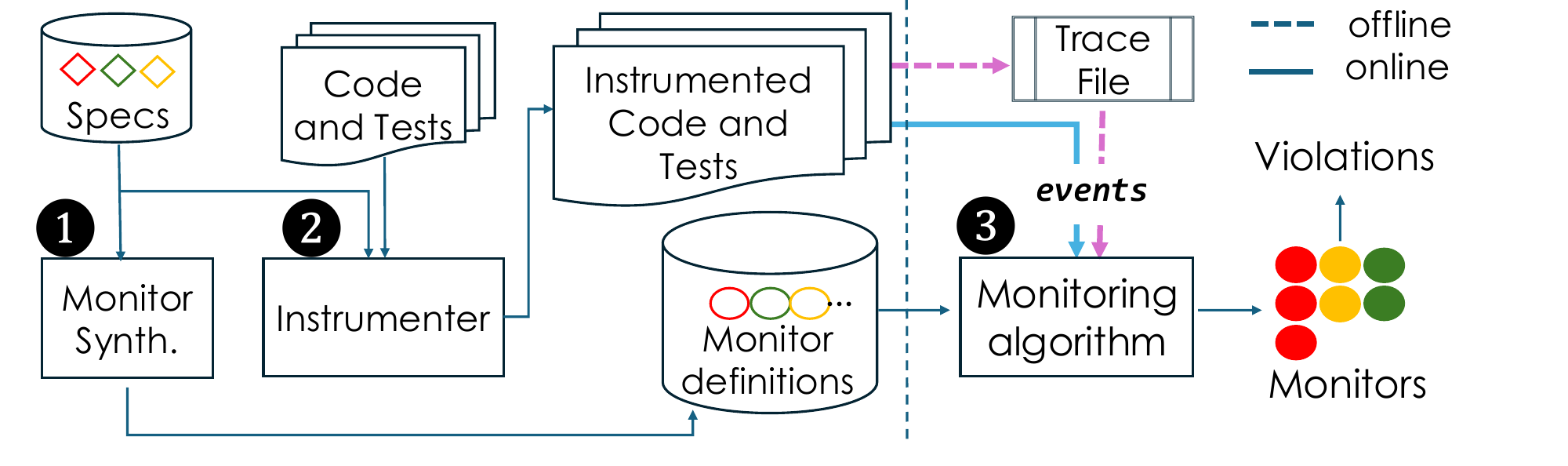}
  \vspace{-5ex}
  \caption{\Tool's Architecture.}
  \label{fig:architecture-diagram}
  \vspace{-5ex}
\end{wrapfigure}

\MyPara{Architecture}
Figure~\ref{fig:architecture-diagram} shows \Tool's architecture,
supporting offline and online monitoring. Offline
\RV~\cite{colombo2009offline, dams2022python, barringer2010formal}
logs signaled events for \emph{a posteriori} analysis. Online \RV
checks (but does not log) events as they are signaled.
\Tool's
\noindent
architecture has three main components.
\MonitorCodeGenerator (\circled{1}) uses logic plugins to compile
\specs (diamonds) into code (circles) from which the \MonitoringEngine
(\circled{3}) instantiates monitors (filled circles) at
runtime.\Space{ For example, given a \spec formalized as an ERE,
the \MonitorCodeGenerator produces code from which a
the \MonitoringEngine can instantiate finite state machine (FSM)
monitors at runtime.} \InstrumentationEngine (\circled{2}) rewrites
the code and tests to signal \spec-related events at
runtime. Finally, \MonitoringEngine (\circled{3}) instantiates,
dispatches events to, and garbage collects monitors.
\Tool's modularity allows components to be extended independently.

\Space{Lastly, by supporting multiple \spec languages (in \circled{1})
  and trace-slicing algorithms (in \circled{3}), \Tool{} can support a
  wide range of users' \RV requirements, making it flexible for \RV of
  Python code.}

\vspace{-3ex}
\subsection{Background on parametric trace slicing}
\label{sec:tool:prelim}
\vspace{-1ex}

A \emph{trace} is a sequence of events~\cite{reger2016trace}. A
\emph{parametric event} is one whose abstract parameters are bound to
concrete data like heap objects at runtime. A \emph{parametric trace}
consists of parametric events. Conceptually, a \spec encodes a
\emph{property} that maps traces to categories (e.g.,
\CodeIn{violating}, \CodeIn{not-violating}, or \CodeIn{dont-know}).
\Specs in this paper encode \emph{parametric properties} that map
parametric traces to categories.
Say, at program end, the \toctou related trace is $\tau =
\CodeIn{os.access(f_2)}\leadsto\CodeIn{open(f_1)}\leadsto\CodeIn{os.access(f_2)}\leadsto\CodeIn{os.access(f_1)}$.
Here, events \CodeIn{os.access} and \CodeIn{open} are parametric---the
parameter $f$ is bound to different objects $f_1$ and $f_2$ at
runtime. So, $\tau$ is a parametric trace. Non-parametric \RV might
wrongly map $\tau$ to
\CodeIn{violating}: \CodeIn{open} comes after
\CodeIn{os.access}. But,
$\tau$ is safe: these events bind different parameters.
Existing \RV algorithms treat parameters in different ways.
\Tool uses \emph{parametric trace slicing}~\cite{chen2009parametric, chen2009efficient}.

Say a \spec has events $e_1(a, b)$ and $e_2(b)$, where $a$ and $b$ are
two types, and these runtime events are signaled: $e_1(a_1, b_1)$,
$e_2(a_1)$, $e_3(b_2)$. The \emph{parameter instance}---a partial
function from parameter types to objects---for $e_1$ is
$\theta_1 = a
\rightarrow a_1, b \rightarrow b_1$; or, $\bm{\theta_1 = a_{1}b_{1}}$, for
short, and $\bm{\theta_2 = a_{1}}$ and $\bm{\theta_3 = b_2}$, for
$e_2$ and $e_3$, respectively.
The definitions from~\cite{slicing-tech-report,
  chen2009parametric, chen2009efficient, rosu2012semantics}, presented
 in the following, help
understand trace slicing, and the monitoring algorithms it enables.

\begin{definition}
\vspace{-1ex}
  
Parameter instances $\theta$ and $\theta'$ are \textbf{compatible} if
for any parameter $x \in \Dom(\theta) \cap \Dom(\theta')$, $\theta(x)
= \theta'(x)$, where $\Dom(\theta)$ denotes $\theta$'s domain. If $\theta$
and $\theta'$ are compatible, we can \textbf{combine} them, written
$\theta \sqcup \theta'$, as follows:

$(\theta \sqcup \theta') (x) =
\left\{
\begin{array}{ll}
\theta(x)
  & \mbox{ when } \theta(x) \mbox{ is defined} \\
\theta'(x)
  & \mbox{ when } \theta'(x) \mbox{ is defined} \\
\mbox{ undefined}
  & \mbox{ otherwise} \\
\end{array}
\right.
$

\vspace{1ex}

\noindent $\theta'$ is \textbf{less informative} than $\theta$,
denoted as $\theta'\sqsubseteq\theta$, iff. for any $x\in X$, if
$\theta'(x)$ is defined then $\theta(x)$ is also defined and
$\theta'(x)=\theta(x)$.  $\sqsubseteq$ is a partial order.

\vspace{-1ex}
\end{definition}

In our example, $\theta_1$ is compatible with $\theta_2$, since they
``agree'' on the concrete object $a_1$ that they bind to $a$. Also,
$\theta_2$ is compatible with $\theta_3$ as they do not
``disagree'' on any parameter that they bind. But, $\theta_1$ is not
compatible with $\theta_3$.

\begin{definition}
  \label{defn:trace-slicing}
  \vspace{-1ex}
  
For parametric trace $\tau$ and parameter instance $\theta$, the
\textbf{$\theta$-trace slice} $\tau\!\upharpoonright_\theta$ is a
non-parametric trace defined recursively as:
\begin{itemize}[topsep=0ex,itemsep=0pt,leftmargin=1em]
\item $\epsilon\!\upharpoonright_\theta = \epsilon$, where $\epsilon$
  is the empty trace/word, and
\item $(\tau\,e\langle\theta'\rangle)\!\upharpoonright_\theta =
\left\{
\begin{array}{ll}
(\tau\!\upharpoonright_\theta)\,e
  & \mbox{when } \theta' \mbox{ is less informative than}\ \theta\mbox{, i.e.,}\ \theta' \sqsubseteq \theta \\
\tau\!\upharpoonright_\theta
  & \mbox{when } \theta' \mbox{ is \emph{not} less informative than}\ \theta\mbox{, i.e.,}\ \theta' \not\sqsubseteq \theta
\end{array}
\right.
$
\end{itemize}

\vspace{-1ex}
\end{definition}

A \emph{trace slice} is a projection of a parametric trace onto a set
of related parameter instances.  \emph{Trace slicing} decomposes a
parametric trace $\tau$ into non-parametric slices for each related
set of parameter instances $\theta$. Trace slice
$\tau\!\upharpoonright_\theta$ first projects $\tau$ onto $\theta$,
discarding events whose parameters are unrelated to $\theta$. Then,
$(\tau\!\upharpoonright_\theta)\,e$ ``forgets'' the parameters to
obtain a non-parametric trace slice---a sequence of event names (with
no parameters). So, monitors do not have to spend time reasoning about
parameters when checking non-parametric trace slices.

Trace slicing decomposes the parametric trace $\tau$ above into slices
$\tau_1 = \tau\!\upharpoonright_{f_1}
= \CodeIn{open(f_1)}\leadsto\CodeIn{os.access(f_1)}$ and $\tau_2
= \tau\!\upharpoonright_{f_2} =
\CodeIn{os.access(f_2)}\leadsto\CodeIn{os.access(f_2)}$. After
``forgetting'' the parameters, the resulting non-parametric slices are
$\tau_1' = \CodeIn{open}\leadsto\CodeIn{os.access}$ and $\tau_2' =
\CodeIn{os.access}\leadsto\CodeIn{os.access}$.

\begin{definition}
  \label{defn:monitor}
  \vspace{-1ex}
  
\textbf{Monitor} $M$ is a tuple $(S,{\cal E}, {\cal C}, \i,
\sigma:S\times{\cal E}\rightarrow S, \gamma:S\rightarrow{\cal C})$:
$S$ is a set of states, $\cal E$ is a set of non-parametric events,
$\cal C$ is a set of categories, $\i\in S$ is the initial state,
$\sigma$ is the transition function, and $\gamma$ maps states to
categories.

\vspace{-1ex}
\end{definition}

Definition~\ref{defn:monitor} is for online and offline monitors that check
non-parametric traces.
Figure~\ref{fig:toctou_fsm} shows a non-parametric monitor; neither
$\tau_1'$ nor $\tau_2'$ causes it to reach the violating state. Had
the parameters in that trace been ``forgotten'' before slicing, that
FSM would wrongly reach the violating state.

\vspace{-3ex}
\subsection{Monitoring Algorithms}
\label{sec:tool:algos}
\vspace{-1.5ex}

We summarize five trace-slicing based monitoring algorithms for
completeness, and to clarify \Tool (\S\ref{sec:tool:implementation});
correctness/complexity proofs are in~\cite{chen2009parametric,
chen2009efficient, rosu2012semantics}.

\Space{ Finally, we describe the
monitor-synthesis algorithms in \Tool.}

\MyPara{1. Offline Algorithm $\A$}
Algorithm \ref{fig:A} shows algorithm $\A$ for \emph{offline}
monitoring. Line~\ref{algo:a:main:init} initializes $\T$ to the empty
map, the slice of the empty binding ($\bot$) to the empty trace
($\epsilon$), and $\Theta$ to $\{\bot\}$.
For each $e\langle\theta\rangle\in\tau$,
lines~\ref{algo:a:main:inner:start}--\ref{algo:a:main:inner:end}
update $\T$ as follows. For each $\theta'$ that results from combining
$\theta$ with parameter
\begin{wrapfigure}[8]{l}{.4\textwidth}
\begin{minipage}{0.4\textwidth}  
\vspace{-11ex}
\begin{algorithm}[H]
  \scriptsize
  \begin{algorithmic}[1]
  \item[\textbf{Inputs:} $\tau$ : a parametric trace]\label{algo:a:inputs}
  \item[\textbf{Outputs:} $\T$ : maps $\theta$ to a trace slice]\label{algo:tinymop:outputs}
  \item[\textbf{Globals:} $\Theta$ : all $\theta$ seen so far]\label{algo:tinymop:globals}

    \Procedure{$\mathsf{main}$}{$\tau$}\label{algo:a:main:start}
    \State $\T \gets \{\}$;\ $\T(\bot)\gets\epsilon$; \ $\Theta \gets \{\bot$\}\label{algo:a:main:init}
    \ForAll{$\mathsf{e}\langle{\theta}\rangle$ \textbf{in} $\tau$}\Space{\AlgoComment{in order from first to last event}}\label{algo:a:main:outer:start}
    \ForAll{$\theta' \in \{\theta\} \sqcup \Theta$}\label{algo:a:main:inner:start}
    \State $\T(\theta') \leftarrow\T(\max\,(\theta']_\Theta)\,e$\label{algo:a:main:slice}
    \EndFor\label{algo:a:main:inner:end}
    \State $\Theta \leftarrow \{\bot,\theta\}\sqcup\Theta$\label{algo:a:main:update}
    \EndFor\label{algo:a:main:outer:end}
    \EndProcedure\label{algo:a:main:end}
  \caption{\Space{Offline }Algorithm $\A$.}
  \label{fig:A}
  \end{algorithmic}
\end{algorithm}
\vspace{-6ex}
\end{minipage}
\end{wrapfigure}
\noindent
instances in $\Theta$ that are compatible with $\theta$,
non-parametric event $e$ is appended to $\T(\max\,(\theta']_\Theta)$,
the slice of the most informative binding in $\T$ that is less
informative than $\theta'$\Space{ (including $\theta$)}. Finally,
$\Theta$ is updated as shown on line~\ref{algo:a:main:update}\Space{,
before processing the next event}.
Algorithm $\A$ is expensive; its time complexity is $O(n \times m)$,
where $n$ is the length of $\tau$ and $m$ is the number of possible
parameter combinations~\cite{chen2009parametric}.\Space{ \Fix{Give
actual times or back-of-envelope calculations to give more context.}}
As we show in \S\ref{sec:discussion}, $\A$ is unlikely to be used
as-is during testing in CI. But, our implementation of $\A$ in \Tool
makes it support offline monitoring (so, \eg, future work can
compare \Tool with offline Python \RV tools like
LogScope~\cite{barringer2010formal} and
PyContract~\cite{dams2022python}). The online algorithms that we
present next build on $\A$. The simplicity of $\A$ also makes it a
good baseline for testing the more complex (online) algorithms, which
we do.

\MyPara{Online Monitoring Algorithms}
\Tool's \MonitoringEngine (\circled{3}\Space{ in
  Figure~\ref{fig:architecture-diagram}}) has four online monitoring
algorithms: $\B$, $\CX$, $\CXX$, and $\DX$; they differ in trade-offs
among implementation complexity, runtime overhead, and memory
\begin{wrapfigure}[10]{r}{.45\textwidth}
\begin{minipage}{0.45\textwidth}    
\vspace{-11ex}
\begin{algorithm}[H]
  \scriptsize
  \begin{algorithmic}[1]
  \item[\textbf{Inputs:} $\mathit{P}$ : instrumented code,]\label{algo:b:inputs}
  \item[\hspace{3.8em} $\mathit{M}$: monitor template]\label{algo:b:inputs}
  \item[\textbf{Outputs:} $\Gamma$ : maps $\theta$ to monitor states]\label{algo:tinymop:outputs}
  \item[\textbf{Globals:} $\Theta$ : all $\theta$ seen so far,]\label{algo:tinymop:globals}
  \item[\hspace{3.8em} $\Delta$ : maps $\theta$ to categories]
    
    \Procedure{$\mathsf{main}$}{$\mathit{P, M}$}\label{algo:b:main:start}
    \State $\Gamma \gets \{\}$; $\Delta \gets \{\}$;\ $\Delta(\bot)\gets\iLMCS$; \ $\Theta \gets \{\bot$\}\label{algo:b:main:init}
    \While{$\mathsf{e}\langle{\theta}\rangle$ is signaled from P}\Space{\AlgoComment{run until $\mathit{P}$ terminates}}\label{algo:b:main:outer:start}
    \ForAll{$\theta' \in \{\theta\} \sqcup \Theta$}\label{algo:b:main:inner:start}
    \State $\Delta(\theta') \leftarrow \sigma(\Delta(\max\,(\theta']_\Theta),e)$\label{algo:b:main:slice}
    \State $\Gamma(\theta') \leftarrow\gamma(\Delta(\theta'))$\Space{\AlgoComment{raise violation if $\Delta(\theta')$ is in error state}}\label{algo:b:main:categorize}
    \EndFor\label{algo:b:main:inner:end}
    \State $\Theta \leftarrow \{\bot,\theta\}\sqcup\Theta$\label{algo:b:main:update}
    \EndWhile\label{algo:b:main:outer:end}
    \EndProcedure\label{algo:b:main:end}

  \caption{\Space{Online }Algorithm $\B$\Space{$\B(M=(S,{\cal E},{\cal C},\iLMCS,\sigma,\gamma))$}.}
  \label{fig:B}
  \end{algorithmic}
\end{algorithm}
\vspace{-6ex}
\end{minipage}
\end{wrapfigure}
\noindent
overhead. The most complex one, $\DX$, is the only one implemented in
\javamop~\cite{JinEtAlJavaMOPToolPaperICSE12, LuoETAlRVMonitor14}, the
only MOP instance for Java.\Space{ Although all these algorithms are
  from the literature~\cite{chen2009parametric, chen2009efficient}}
Further, \Tool is the only one today that combines these algorithms
into a single framework. Due to space limits, we describe $\B$\Space{
  in detail}, and summarize how $\CX$, $\CXX$ and $\DX$ differ.

\MyParaOnly{2. Algorithm $\B$}
works event-by-event, without storing traces, and tracks only
encountered parameter instances (Line~\ref{algo:a:main:inner:start} in
Algorithm~\ref{fig:A} can cause $\A$ to track parameter
instances \emph{not} in the trace). These features make $\B$ more
time- and space-efficient than $\A$. Algorithm \ref{fig:B} shows $\B$;
it takes instrumented code $P$ and monitor template $M$.\Space{ for
the \spec being monitored. (\Tool monitors multiple \specs
simultaneously, but we describe $\B$ for one
\spec for simplicity.)} A main difference with $\A$ is that, in $\B$,
traces are sliced and monitored ``on the fly'' as each parametric
event is signaled from $P$ (line~\ref{algo:b:main:outer:start}). $\B$
tracks the state of the {\em monitor instance} (obtained from $M$) for
$\theta'$ in $\Delta$, and transitions that instance based on $\sigma$
by $e$ (line~\ref{algo:b:main:slice}). Next, $\B$ categorizes the
current state of the monitor instance for $\theta'$ according to
$\gamma$ and stores it in
$\Gamma$~(line~\ref{algo:b:main:categorize}).  A \CodeIn{Handler} is
triggered if $\theta'$ slice violates a \spec.

\MyPara{3. Algorithm $\CX$}
$\B$'s search for compatible parameters and combination with $\theta$
(line~\ref{algo:b:main:inner:start} in Algorithm~\ref{fig:B}) can be
expensive for long traces. To optimize, $\CX$ uses an auxiliary data
structure, $\mathcal{U}$, with $\Delta$ and $\Gamma$\Space{ that $\B$
  uses}.  $\mathcal{U}$ maps each parameter instance $\theta$ to the
set of parameter instances in $\Delta$ that are strictly more
informative than $\theta$.
So, for each $e\langle{\theta}\rangle$, $\CX$ can more quickly find
compatible parameters by looking up $\theta$ in $\mathcal{U}$, instead
of repeatedly searching through all elements of (often very large)
$\Theta$ like $\B$ does.  Also, unlike $\B$, $\CX$ does {\em not}
explicitly store $\Theta$, but implicitly captures it as $\Delta$'s
domain. Doing so reduces the time to update $\Delta$ after each
$e\langle{\theta}\rangle$, and saves memory. But, in projects where
traces are mostly short, $\CX$'s extra machinery can add extra costs.

\MyPara{4. Algorithm $\CXX$}
For \specs with multiple parameter types, violations often cannot
occur until a ``creation event''\Space{ that binds more than one of
those types} has occurred. $\CXX$ leverages this observation: monitors
are only instantiated upon creation events. If many non-creation
events occur before the creation event or if a creation event never
occurs, $\CXX$ can be faster and use less memory than $\CX$\Space{,
which does not use creation events}.\Space{ The cost of $\CXX$ is that
the \spec syntax and the \spec-compilation process must be extended to
allow one specify which events in a \spec are creation events.} In a
sense, $\CX$ is a special case of $\CXX$ where all events are creation
events~\cite{BoddenMOPBOxRV11}. So, $\CX$ can be as efficient at
$\CXX$ for \specs that do not designate a creation event.

\MyPara{5. Algorithm $\DX$}
This algorithm uses information about the \spec to avoid creating
unneeded monitors.\Space{ \Fix{Use example in \S\ref{sec:example} to
explain unneeded monitors.}} The information that $\DX$ uses is
called \emph{enable sets} and it is computed via static analysis of
each \spec at compile time. Intuitively, the enable sets of a \spec
maps each non-parametric event $e$ to the set of parameters that must
have been bound by the current slice $\tau$ for $\tau.e$ or its future
extensions to reach a category of interest. \Tool implements the
algorithms in~\cite{chen2009efficient} for computing enable-sets for
regular and context-free \specs.  $\B$, $\CX$, and $\CXX$ do not take
such information into account, so they can create monitors for slices
that do make semantic sense, or those that cannot reach a violating
state. $\DX$ is the most complex among MOP-style parametric monitoring
algorithms and it can be more efficient than others for monitoring
multi-object \specs. But, enable sets can be an unnecessary overhead
for single-object \specs.

\vspace{-3ex}
\subsection{Monitor Garbage-Collection Algorithm}
\label{sec:tool:gc:algorithm}
\vspace{-1ex}

Monitor garbage collection
(MGC)~\cite{JinEtAlGarbageCollectionPLDI2011} makes \RV more efficient
by proactively eliminating monitors that can no longer reach a verdict
category. The MGC algorithm uses \emph{coenable set} to determine when
to remove monitors; it is set of events that should still be possible
(depending on alive parameters) to reach a category. \Tool\ implements
coenable sets: each event received by a monitor triggers an MGC check
of that monitor's coenable set. If parameter instances associated with
events in the coenable set have been garbage collected, then \Tool
concludes that the monitor can be safely garbage collected.

\vspace{-3ex}
\subsection{Implementation}
\label{sec:tool:implementation}
\vspace{-1ex}

We implement \Tool as a plugin for \CodeIn{pytest}\Space{, the most
popular unit testing framework for Python.}; users can invoke it like
so:
\texttt{pytest -p pythonmop}. We briefly discuss salient parts of
\Tool's implementation\Space{; circled numbers refer to
Figure~\ref{fig:architecture-diagram}}.\Space{ Then we briefly
  describe how to provide inputs and process outputs.}

\MyPara{Monitor Synthesizer \circled{1}}
\Tool invokes \javamop's mature and well-tested monitor-synthesis
plugins for ERE, FSM, past- and future-time LTL \specs. Doing so sped
up our \Tool development. We also provide a new CFG monitor-synthesis
plugin as a proof of concept on how to add other logic plugins.
All monitor-synthesis plugins produce monitor templates (circles in
Figure~\ref{fig:architecture-diagram}) from which runtime monitors are
instantiated. For ERE, FSM, and LTL, that template defines an FSM. For
CFG, that template defines a parser
derivative~\cite{MeredithETALMOPContextFreePatterns08}.

\MyPara{Instrumenter \circled{2}}
\Tool{} supports three instrumentation strategies: (i)~Python-level
Monkey Patching only~\cite{hunt2023monkey}, (ii)~Python-level Monkey
Patching plus C-level Monkey Patching via the curses function
in \CodeIn{forbiddenfruit}~\cite{clarete_forbiddenfruit}, and
(iii)~Python-level Monkey Patching plus AST
transformation~\cite{pythonAst}. Using these strategies, \Tool{}
rewrites the CUT,\Space{ third-party} libraries, and the Python
runtime to signal events ``before'' or ``after'' \spec-related events
at runtime. \Tool also has an API for adding more strategies in the
future. Our appendix discusses in great detail the tradeoffs in
capability and cost among the currently supported strategies.

\MyPara{Monitoring Algorithms \circled{3}}
MOP monitoring algorithms can be implemented as (i)~libraries (easier
for humans to work with but slower) or (ii)~\spec-compiler generated
code (harder for humans to work with but faster).
Different from prior work~\cite{BoddenMOPBOxRV11, mopbox-webpage,
  JinEtAlJavaMOPToolPaperICSE12, LuoETAlRVMonitor14}, we implement
  $\A$, $\B$, $\CX$, $\CXX$, and $\DX$ as libraries in \Tool\ to help
  others understand and extend them. \Tool offers command-line options
  for choosing among these algorithms (\eg, \texttt{pytest -p
  pythonmop --algo=C+} for $\CXX$). Future work can further
  optimize \Tool by implementing these algorithms via code generation.

\MyPara{\Specs}
\Tool ships with \NumSpecs{} \specs (\S\ref{sec:eval:setup}) that use all
its logic plugins. Users can add more \specs by writing them in
\Tool's logic syntax, or via a JSON-like syntax (\eg,
Figure~\ref{lst:access_spec}) that \Tool translates to its logic
syntax.

\vspace{-3ex}

\section{Evaluation}
\label{sec:eval}

\vspace{-1ex}

Our evaluation addresses four research questions:

\begin{itemize}[topsep=0ex,itemsep=0pt,leftmargin=1em]
  
\item{\textbf{\rqAlgos{}.}} How do monitoring algorithms impact
  \Tool's time overhead?

\item{\textbf{\rqInstrumentation{}.}} How do instrumentation
  strategies impact \Tool's time overhead and violations
  detected?

\item{\textbf{\rqEffectiveness{}.}} How effective is \Tool\ for finding bugs in \oss?

\item{\textbf{\rqEfficiency{}.}} How efficient is \Tool\ compared to
  \dylin and \dynapyt?

\end{itemize}

\noindent
\rqAlgos and \rqInstrumentation evaluate the efficiency of \Tool's
monitoring algorithms and instrumentation strategies. \rqEffectiveness
evaluates \Tool's effectiveness for finding bugs, and \rqEfficiency
compares \Tool's efficiency and effectiveness
with \dynapyt~\cite{eghbali2022dynapyt}
and \dylin~\cite{Eghbali2025DyLin}, two recent frameworks for writing
dynamic analyzers in Python.

\vspace{-3ex}

\subsection{Experimental Setup}
\label{sec:eval:setup}

\vspace{-1ex}

\MyPara{Specifications}~We manually write \NumSpecs{} API-level \specs
of the Python language and of Python libraries, which we refer to as
\emph{Python specs} and \emph{library specs}, respectively. To do so,
we follow the procedure that Lee et al.~\cite{FormalizingJavaAPI,
  LuoETAlRVMonitor14} used to obtain JDK API \specs. Specifically,
we first search API documentation
for hints on API usage constraints, \ie, we look for sentences with keywords like
\emph{``should''}, \emph{``only''}, \emph{``must''}, \emph{``note''},
etc. (Our artifacts contain the complete list of keywords and
regexes.) We then manually inspect these sentences to decide whether
to write a \spec. Of \NumSpecs{} \specs{} that we write, \NumPySpecs{}
are Python specs and \NumLibSpecs{} are library specs. We manually
translate \NumSpecsFromOtherTools{} of these \NumSpecs{} \specs from
\dylin.  We plan to translate more in the future.\Space{ We plan to
  translate 5 more in the future, and the other 2 work in \dylin but
  not \dynapyt. Table~\ref{table:spec-breakdown} shows the source of
  all \NumSpecs{} \specs.}  Some of those Python and Library \specs\
  are: (i)~\emph{UnsafeDictIterator} (ERE; Python) detects calls to
\CodeIn{next} on an \CodeIn{Iterator} after the \CodeIn{Dict} was modified (\eg, via \CodeIn{update},
\CodeIn{pop}, or \CodeIn{setitem}). Doing so can cause unpredictable
behavior~\cite{pythonDict}\Space{ It tracks multiple dict-iterator
  pairs with the ERE: \CodeIn{createDict updateDict* createIter next*
    updateDict+ next}}; (ii)~\emph{UselessFileOpen} (FSM; Python)
detects files that are opened but never used, causing resource
leaks~\cite{pythonFileIO}\Space{ Using an FSM (\CodeIn{s0: open → s1;
    s1: read|write|readline|writelines → s2, close|end → s3}), It
  flagging unused files at close or program end for efficient resource
  management.}; (iii)~\emph{\seqsplit{ArraysSortBeforeBinarySearch}}
(LTL; Python) checks if an array is sorted before operations
like \CodeIn{bisect\_left} or \CodeIn{bisect\_right}\Space{, as
  required by the Python \CodeIn{bisect}
  module}~\cite{pythonBisect}\Space{ Using the LTL formula
  \CodeIn{[](binsearch => (*)(not modify S sort))}, it enforces that
  no \CodeIn{append} (i.e., modification) occurs after a \CodeIn{sort}
  and before a binary search, ensuring correctness in binary search
  application.}; (iv)~\emph{\seqsplit{RequestsPreparedInit}} (FSM;
Lib)--checks if \CodeIn {PreparedRequest}s are initialized correctly
via a \CodeIn{Request} instances, not manually\Space{. Otherwise
  requests may not be properly prepared}~\cite{pythonRequests};
(v)~\emph{NLTKProbSum} (ERE; Lib) checks that updates to
\CodeIn{nltk.MutableProbDist} probabilities maintain a sum of one to
avoid invalid probability distributions~\cite{nltkProbability}\Space{
  Using ERE \CodeIn{(prob | logprob | max | discount | generate)}+, it
  monitors probability-related methods, flagging violations if the sum
  exceeds one, ensuring valid probability distributions.};
(vi)~\emph{\seqsplit{TornadoNoAdditionalOutput}} (CFG; Lib) checks
that Tornado's \CodeIn{RequestHandler}s do not call output methods
(\CodeIn {set\_header, add\_header, clear\_header, set\_status}) after
\CodeIn {finish}, which can invalidate responses~\cite{tornadoRender}.

\MyPara{Evaluation Subjects}
We use \NumProjects{} projects, obtained in two ways. First, we use
GitHub's REST API to find projects whose manifest files\Space{ (e.g.,
  \CodeIn{requirements.txt}, \CodeIn{pyproject.toml})} mention
libraries that we have \specs for: \CodeIn{nltk}~\cite{nltk},
\CodeIn{requests}~\cite{requests},
\CodeIn{tensorflow}~\cite{tensorflow}, \CodeIn{flask}~\cite{flask},
\CodeIn{tornado}~\cite{tornado}, and \CodeIn{scipy}~\cite{scipy}. We
find \NumProjectsSearch{} projects.  Then, we use GitHub's Search to
find \spec-related projects that use
\CodeIn{pytest}.\Space{ To do so, we pass the list of APIs in our
  query as parameter.}  We find \NumProjectsFromAPISearch{} more
projects.  Table~\ref{tab:program-charac} shows summary statistics
about all \NumProjects{} projects.

\begin{table}[t!]
  \vspace{-1ex}
  \scriptsize
  \caption{Stats on
    \NumProjects{}
    evaluated projects: no. of tests (\#Tests), time w/o \RV in
    seconds (time), size (SLOC), line cov. ($\mathit{cov_s}$),
    branch cov.  ($\mathit{cov_b}$), no. of commits (\#Sha), age in
    years (Age), no. of GitHub stars ($\star$), no. of events
    ($\mathit{evt}$) and monitors ($\mathit{mon}$).}
  \label{tab:program-charac}
  \setlength{\tabcolsep}{3pt}
  \centering
  \begin{tabular}{lrrrrrrrrrr}
    \toprule
     & \#Tests & time & SLOC & $\mathit{cov_s}$ & $\mathit{cov_b}$ & \#Sha & Age & $\star$ & $\mathit{evt}$ & $\mathit{mon}$ \\
    \midrule
    Mean & 226.0 & 19.5 & 29,342 & 35.3 & 29.8 & 1,937.1 & 5.3 & 376.1 & 12,342.1 & 3,845.1 \\
    Med & 17.5 & 1.1 & 2,317.5 & 25.0 & 16.7 & 107 & 4.9 & 6.0 & 2,221 & 1,016 \\
    Min & 1.0 & 0.1 & 2.0 & 1.0 & 0.1 & 1.0 & 0.3 & 1 & 209 & 157 \\
    Max & 45,876 & 3,354.9 & 4,118,502 & 100 & 100 & 176,925 & 15.5 & 41,433 & 1,132,305 & 115,865 \\
    Sum & 290,133 & 25,072.6 & 37,675,099 & n/a & n/a & n/a & n/a & n/a & 18,254,008 & 5,686,846 \\
    \bottomrule
  \end{tabular}
  \vspace{-7ex}
\end{table}

\MyPara{Running Experiments}
We run all experiments on a server with AMD\textregistered{}
EPYC\textregistered{} 7763 64-Core Processor with 4 virtual CPUs
and 15GB of RAM, and Ubuntu 24.04.2 LTS. Each experiment runs in a
Docker container to ease dependency management, and improve
reproducibility.%

\begin{figure*}
	\vspace{-2ex}
    \captionsetup[subfigure]{justification=centering}
    \begin{subfigure}[b]{\textwidth}
        \centering
        \begin{tikzpicture} [scale=0.52]
            \begin{axis}[
                width=11cm,
                height=5cm,
                ybar,
                bar width=4pt,
                enlargelimits=0.05,
                legend style={
                    at={(0.02,0.95)},
                    anchor=north west,
                    legend columns=2,
                    font=\footnotesize,
                    draw=black,
                    fill=white
                },
                ylabel={Relative Overhead},
                ylabel style={font=\normalsize},
                ymax  = 25000,
                symbolic x coords={10, 20, 30, 40, 50, 60, 70, 80, 90, 100},
                xtick=data,
                x tick label style={font=\normalsize, anchor=north}
            ]
        
            \addplot+[ybar, black, fill=pink!70] plot coordinates {
                (10,\UseMacro{overhead-total-p1-b-rel})
                (20,\UseMacro{overhead-total-p2-b-rel})
                (30,\UseMacro{overhead-total-p3-b-rel})
                (40,\UseMacro{overhead-total-p4-b-rel})
                (50,\UseMacro{overhead-total-p5-b-rel})
                (60,\UseMacro{overhead-total-p6-b-rel})
                (70,\UseMacro{overhead-total-p7-b-rel})
                (80,\UseMacro{overhead-total-p8-b-rel})
                (90,\UseMacro{overhead-total-p9-b-rel})
                (100,\UseMacro{overhead-total-p10-b-rel})
            };
            \addplot+[ybar, black, fill=blue!70] plot coordinates {
                (10,\UseMacro{overhead-total-p1-c-rel})
                (20,\UseMacro{overhead-total-p2-c-rel})
                (30,\UseMacro{overhead-total-p3-c-rel})
                (40,\UseMacro{overhead-total-p4-c-rel})
                (50,\UseMacro{overhead-total-p5-c-rel})
                (60,\UseMacro{overhead-total-p6-c-rel})
                (70,\UseMacro{overhead-total-p7-c-rel})
                (80,\UseMacro{overhead-total-p8-c-rel})
                (90,\UseMacro{overhead-total-p9-c-rel})
                (100,\UseMacro{overhead-total-p10-c-rel})
            };
            \addplot+[ybar, black, fill=orange!70] plot coordinates {
                (10,\UseMacro{overhead-total-p1-cplus-rel})
                (20,\UseMacro{overhead-total-p2-cplus-rel})
                (30,\UseMacro{overhead-total-p3-cplus-rel})
                (40,\UseMacro{overhead-total-p4-cplus-rel})
                (50,\UseMacro{overhead-total-p5-cplus-rel})
                (60,\UseMacro{overhead-total-p6-cplus-rel})
                (70,\UseMacro{overhead-total-p7-cplus-rel})
                (80,\UseMacro{overhead-total-p8-cplus-rel})
                (90,\UseMacro{overhead-total-p9-cplus-rel})
                (100,\UseMacro{overhead-total-p10-cplus-rel})
            };
            \addplot+[ybar, black, fill=green!70] plot coordinates {
                (10,\UseMacro{overhead-total-p1-d-rel})
                (20,\UseMacro{overhead-total-p2-d-rel})
                (30,\UseMacro{overhead-total-p3-d-rel})
                (40,\UseMacro{overhead-total-p4-d-rel})
                (50,\UseMacro{overhead-total-p5-d-rel})
                (60,\UseMacro{overhead-total-p6-d-rel})
                (70,\UseMacro{overhead-total-p7-d-rel})
                (80,\UseMacro{overhead-total-p8-d-rel})
                (90,\UseMacro{overhead-total-p9-d-rel})
                (100,\UseMacro{overhead-total-p10-d-rel})
            };
        
            \legend{
                \strut B,
                \strut C,
                \strut C+,
                \strut D
            }
            \end{axis}
        \end{tikzpicture}
            \label{fig:algo-bar-chart}
        \begin{tikzpicture} [scale=0.52]
            \begin{axis}[
                width=11cm,
                height=5cm,
                ybar,
                bar width=4pt,
                enlargelimits=0.05,
                legend style={
                    at={(0.02,0.95)},
                    anchor=north west,
                    legend columns=2,
                    font=\footnotesize,
                    draw=black,
                    fill=white
                },
                ylabel={Number of projects},
                ylabel style={font=\normalsize},
                ymax  = 160,
                symbolic x coords={10, 20, 30, 40, 50, 60, 70, 80, 90, 100},
                xtick=data,
                x tick label style={font=\normalsize, anchor=north},
                nodes near coords,
                every node near coord/.append style={font=\scriptsize, yshift=1pt, rotate=90, anchor=west}
            ]

            \addplot+[ybar, black, fill=pink!70] plot coordinates {
              (10,\UseMacro{fastest-count-p1-b})
              (20,\UseMacro{fastest-count-p2-b})
              (30,\UseMacro{fastest-count-p3-b})
              (40,\UseMacro{fastest-count-p4-b})
              (50,\UseMacro{fastest-count-p5-b})
              (60,\UseMacro{fastest-count-p6-b})
              (70,\UseMacro{fastest-count-p7-b})
              (80,\UseMacro{fastest-count-p8-b})
              (90,\UseMacro{fastest-count-p9-b})
              (100,\UseMacro{fastest-count-p10-b})
          };
            \addplot+[ybar, black, fill=blue!70] plot coordinates {
                (10,\UseMacro{fastest-count-p1-c})
                (20,\UseMacro{fastest-count-p2-c})
                (30,\UseMacro{fastest-count-p3-c})
                (40,\UseMacro{fastest-count-p4-c})
                (50,\UseMacro{fastest-count-p5-c})
                (60,\UseMacro{fastest-count-p6-c})
                (70,\UseMacro{fastest-count-p7-c})
                (80,\UseMacro{fastest-count-p8-c})
                (90,\UseMacro{fastest-count-p9-c})
                (100,\UseMacro{fastest-count-p10-c})
            };
            \addplot+[ybar, black, fill=orange!70] plot coordinates {
                (10,\UseMacro{fastest-count-p1-cplus})
                (20,\UseMacro{fastest-count-p2-cplus})
                (30,\UseMacro{fastest-count-p3-cplus})
                (40,\UseMacro{fastest-count-p4-cplus})
                (50,\UseMacro{fastest-count-p5-cplus})
                (60,\UseMacro{fastest-count-p6-cplus})
                (70,\UseMacro{fastest-count-p7-cplus})
                (80,\UseMacro{fastest-count-p8-cplus})
                (90,\UseMacro{fastest-count-p9-cplus})
                (100,\UseMacro{fastest-count-p10-cplus})
            };
            \addplot+[ybar, black, fill=green!70] plot coordinates {
                (10,\UseMacro{fastest-count-p1-d})
                (20,\UseMacro{fastest-count-p2-d})
                (30,\UseMacro{fastest-count-p3-d})
                (40,\UseMacro{fastest-count-p4-d})
                (50,\UseMacro{fastest-count-p5-d})
                (60,\UseMacro{fastest-count-p6-d})
                (70,\UseMacro{fastest-count-p7-d})
                (80,\UseMacro{fastest-count-p8-d})
                (90,\UseMacro{fastest-count-p9-d})
                (100,\UseMacro{fastest-count-p10-d})
            };
        
            \legend{
                \strut B,
                \strut C,
                \strut C+,
                \strut D
            }
            \end{axis}
        \end{tikzpicture}
            \label{fig:algo-fastest-bar-chart}
        \begin{tikzpicture} [scale=0.52]
              \begin{axis}[
                width=11cm,
                height=5cm,
                  ybar,
                  bar width=4pt,
                  enlargelimits=0.05,
                  legend style={
                      at={(0.61,0.97)},
                      anchor=north west,
                      legend columns=4,
                      font=\footnotesize,
                      draw=black,
                      fill=white
                  },
                  ylabel={Number of projects},
                  ylabel style={font=\normalsize},
                  ymax  = 215,
                  symbolic x coords={10, 20, 30, 40, 50, 60, 70, 80, 90, 100},
                  xtick=data,
                  x tick label style={font=\normalsize, anchor=north},
                  nodes near coords,
                  every node near coord/.append style={font=\scriptsize, yshift=1pt, rotate=90, anchor=west}
              ]
  
              \addplot+[ybar, black, fill=pink!70] plot coordinates {
                (10,\UseMacro{fastest-count-2s-p1-b})
                (20,\UseMacro{fastest-count-2s-p2-b})
                (30,\UseMacro{fastest-count-2s-p3-b})
                (40,\UseMacro{fastest-count-2s-p4-b})
                (50,\UseMacro{fastest-count-2s-p5-b})
                (60,\UseMacro{fastest-count-2s-p6-b})
                (70,\UseMacro{fastest-count-2s-p7-b})
                (80,\UseMacro{fastest-count-2s-p8-b})
                (90,\UseMacro{fastest-count-2s-p9-b})
                (100,\UseMacro{fastest-count-2s-p10-b})
            };
              \addplot+[ybar, black, fill=blue!70] plot coordinates {
                  (10,\UseMacro{fastest-count-2s-p1-c})
                  (20,\UseMacro{fastest-count-2s-p2-c})
                  (30,\UseMacro{fastest-count-2s-p3-c})
                  (40,\UseMacro{fastest-count-2s-p4-c})
                  (50,\UseMacro{fastest-count-2s-p5-c})
                  (60,\UseMacro{fastest-count-2s-p6-c})
                  (70,\UseMacro{fastest-count-2s-p7-c})
                  (80,\UseMacro{fastest-count-2s-p8-c})
                  (90,\UseMacro{fastest-count-2s-p9-c})
                  (100,\UseMacro{fastest-count-2s-p10-c})
              };
              \addplot+[ybar, black, fill=green!70] plot coordinates {
                  (10,\UseMacro{fastest-count-2s-p1-d})
                  (20,\UseMacro{fastest-count-2s-p2-d})
                  (30,\UseMacro{fastest-count-2s-p3-d})
                  (40,\UseMacro{fastest-count-2s-p4-d})
                  (50,\UseMacro{fastest-count-2s-p5-d})
                  (60,\UseMacro{fastest-count-2s-p6-d})
                  (70,\UseMacro{fastest-count-2s-p7-d})
                  (80,\UseMacro{fastest-count-2s-p8-d})
                  (90,\UseMacro{fastest-count-2s-p9-d})
                  (100,\UseMacro{fastest-count-2s-p10-d})
              };
              \addplot+[ybar, black, fill=red!70] plot coordinates {
                  (10,\UseMacro{fastest-count-2s-p1-all})
                  (20,\UseMacro{fastest-count-2s-p2-all})
                  (30,\UseMacro{fastest-count-2s-p3-all})
                  (40,\UseMacro{fastest-count-2s-p4-all})
                  (50,\UseMacro{fastest-count-2s-p5-all})
                  (60,\UseMacro{fastest-count-2s-p6-all})
                  (70,\UseMacro{fastest-count-2s-p7-all})
                  (80,\UseMacro{fastest-count-2s-p8-all})
                  (90,\UseMacro{fastest-count-2s-p9-all})
                  (100,\UseMacro{fastest-count-2s-p10-all})
              };
              \legend{
                  \strut B,
                  \strut C,
                  \strut D,
                  \strut Same
              }
              \end{axis}
          \end{tikzpicture}
          \label{fig:algo-fastest-2s-bar-chart}
          \begin{tikzpicture} [scale=0.52]
            \begin{axis}[
                width=11cm,
                height=5cm,
                ybar,
                bar width=4pt,
                enlargelimits=0.05,
                legend style={
                    at={(0.61,0.97)},
                    anchor=north west,
                    legend columns=4,
                    font=\footnotesize,
                    draw=black,
                    fill=white
                },
                ylabel={Number of projects},
                ylabel style={font=\normalsize},
                ymax  = 215,
                symbolic x coords={10, 20, 30, 40, 50, 60, 70, 80, 90, 100},
                xtick=data,
                x tick label style={font=\normalsize, anchor=north},
                nodes near coords,
                every node near coord/.append style={font=\scriptsize, yshift=1pt, rotate=90, anchor=west}
            ]

            \addplot+[ybar, black, fill=pink!70] plot coordinates {
              (10,\UseMacro{fastest-count-5s-p1-b})
              (20,\UseMacro{fastest-count-5s-p2-b})
              (30,\UseMacro{fastest-count-5s-p3-b})
              (40,\UseMacro{fastest-count-5s-p4-b})
              (50,\UseMacro{fastest-count-5s-p5-b})
              (60,\UseMacro{fastest-count-5s-p6-b})
              (70,\UseMacro{fastest-count-5s-p7-b})
              (80,\UseMacro{fastest-count-5s-p8-b})
              (90,\UseMacro{fastest-count-5s-p9-b})
              (100,\UseMacro{fastest-count-5s-p10-b})
          };
            \addplot+[ybar, black, fill=blue!70] plot coordinates {
                (10,\UseMacro{fastest-count-5s-p1-c})
                (20,\UseMacro{fastest-count-5s-p2-c})
                (30,\UseMacro{fastest-count-5s-p3-c})
                (40,\UseMacro{fastest-count-5s-p4-c})
                (50,\UseMacro{fastest-count-5s-p5-c})
                (60,\UseMacro{fastest-count-5s-p6-c})
                (70,\UseMacro{fastest-count-5s-p7-c})
                (80,\UseMacro{fastest-count-5s-p8-c})
                (90,\UseMacro{fastest-count-5s-p9-c})
                (100,\UseMacro{fastest-count-5s-p10-c})
            };
            \addplot+[ybar, black, fill=green!70] plot coordinates {
                (10,\UseMacro{fastest-count-5s-p1-d})
                (20,\UseMacro{fastest-count-5s-p2-d})
                (30,\UseMacro{fastest-count-5s-p3-d})
                (40,\UseMacro{fastest-count-5s-p4-d})
                (50,\UseMacro{fastest-count-5s-p5-d})
                (60,\UseMacro{fastest-count-5s-p6-d})
                (70,\UseMacro{fastest-count-5s-p7-d})
                (80,\UseMacro{fastest-count-5s-p8-d})
                (90,\UseMacro{fastest-count-5s-p9-d})
                (100,\UseMacro{fastest-count-5s-p10-d})
            };
            \addplot+[ybar, black, fill=red!70] plot coordinates {
                (10,\UseMacro{fastest-count-5s-p1-all})
                (20,\UseMacro{fastest-count-5s-p2-all})
                (30,\UseMacro{fastest-count-5s-p3-all})
                (40,\UseMacro{fastest-count-5s-p4-all})
                (50,\UseMacro{fastest-count-5s-p5-all})
                (60,\UseMacro{fastest-count-5s-p6-all})
                (70,\UseMacro{fastest-count-5s-p7-all})
                (80,\UseMacro{fastest-count-5s-p8-all})
                (90,\UseMacro{fastest-count-5s-p9-all})
                (100,\UseMacro{fastest-count-5s-p10-all})
            };
            \legend{
                \strut B,
                \strut C,
                \strut D,
                \strut Same
            }
            \end{axis}
        \end{tikzpicture}
              \label{fig:algo-fastest-5s-bar-chart}
    \end{subfigure}
    \vspace{-5ex}
    \caption{\Tool’s overhead in \UseMacro{algos_comparison_projects_consistent_all_algos} 
    projects, grouped by decile. First row, from left to right: (i) overhead 
    across all four algorithms; (ii) count of how often $\B$, $\CX$, $\CXX$, 
    and $\DX$ are the fastest. Second row, from left to right: (iii) and (iv) 
    show $\B$, $\CX$, and $\DX$ with 2-second and 5-second tolerance, respectively.}
    \vspace{-6ex}
    \label{fig:algorithm_comparison}
  \end{figure*}

\vspace{-3ex}

\subsection{\rqAlgos: The overheads of \Tool's monitoring algorithms}
\label{seq:eval:algos}

\vspace{-1ex}

\MyParaOnly{1. How robust are the algorithms?}
An algorithm is considered \emph{successful} for a given project if it
completes with the same outcome (success or failure) as the baseline
(i.e., without \RV). An execution fails either because it exceeds the
time limit (4 hours) or exhausts system memory (15GB), often caused by
the creation of a large number of monitors. When monitoring
all \NumSpecs{} \specs simultaneously, each algorithm exhibits
different levels of robustness. $\DX$ was the only algorithm
successful in all \NumProjects{} projects. $\CXX$ was successful in
\UseMacro{algos_comparison_C+_consistent_count} projects, $\CX$ was
successful in \UseMacro{algos_comparison_C_consistent_count} projects,
and $\B$ was successful on \UseMacro{algos_comparison_B_consistent_count}
projects.

\MyParaOnly{2. Is there a statistical difference in the runtime
  overhead of \Tool's online monitoring algorithms?} \emph{Overhead}
is the ratio between the execution time of an \RV algorithm and the
time to run tests without \RV. $\DX$ has mean and max overheads of
\textbf{\UseMacro{algos_comparison_D_mean_relative_overhead}}\x and
\UseMacro{algos_comparison_D_max_relative_overhead}\x,
respectively. $\CXX$'s mean and max overheads were
\textbf{\UseMacro{algos_comparison_C+_mean_relative_overhead}}\x and
\UseMacro{algos_comparison_C+_max_relative_overhead}\x, respectively.
$\CX$'s mean and max overheads were
\textbf{\UseMacro{algos_comparison_C_mean_relative_overhead}}\x and
\UseMacro{algos_comparison_C_max_relative_overhead}\x,
respectively. Lastly, $\B$ has mean and max overhead of
\textbf{\UseMacro {algos_comparison_B_mean_relative_overhead}}\x and
\UseMacro{algos_comparison_B_max_relative_overhead}\x, respectively.

We conduct Friedman and pairwise Wilcoxon signed-rank test with
Bonferroni correction for multiple
comparisons~\cite{conover1971practical}. We use these non-parametric
tests due to the lack of normal distributions in the algorithms'
times. For fairness, we only use 628 projects for these statistical
tests: starting from
\UseMacro{algos_comparison_projects_consistent_all_algos} projects
that all four algorithms succeed on, we discard 798 where tests
without \RV take less than 2s (to reduce the influence of noise from
short executions), and discard 11 whose times vary
non-deterministically (to improve measurement reliability). Our
conclusions remain the same when we use all 
\UseMacro{algos_comparison_projects_consistent_all_algos} projects.

The Friedman test shows statistically significant differences in four
algorithms' times
but it does not reveal which pairs differ. So, we perform pairwise
Wilcoxon signed-rank tests with Bonferroni correction~\cite{conover1971practical,hollander2013nonparametric}:
(i)~differences between $\CX$ and $\CXX$ are not statistically
significant; (ii)~all other pairs have high significant differences
($p < 0.001$). To see the magnitude of differences, we calculate
effect sizes using the rank-biserial correlation
coefficient~\cite{kerby2014simple}: (i)~the effect size for $\CX$ vs. $\CXX$ is
negligible; and (ii)~$\DX$ outperforms the other algorithms in our
experiments, having the largest effect against $\B$.

\MyParaOnly{3. Which\Space{ monitoring} algorithm(s) should future work improve?}
Figure~\ref{fig:algorithm_comparison} shows four plots with a
per-project view, grouped per decile by increasing overhead.
There, the top left plot shows that all algorithms have about the same
overhead in the first three deciles; it is not until the 4th decile
that one starts to notice a difference. From 4th to 10th deciles, $\B$
gets much slower and $\DX$ slightly outperforms $\CX$ and $\CXX$. Only
in the 9th and 10th deciles (\FastestTenNineCount{} projects with highest 
overheads) does $\DX$ visibly outperform $\CX$ and $\CXX$.

The top right plot in Figure~\ref{fig:algorithm_comparison} counts how
frequently per decile $\B$, $\CX$, $\CXX$, and $\DX$ is the fastest.
We make three main observations. First, across all deciles, each
algorithm\Space{ (except $\B$)} is the fastest in at least five
projects, So, even among high-overhead projects, there is no
universally ``best'' algorithm. Second, overall across all deciles,
$\B$ is the fastest in
\UseMacro{fastest-all-count-b} projects, $\CX$ is the fastest in
\UseMacro{fastest-all-count-c} projects, $\CXX$ is the fastest in
\UseMacro{fastest-all-count-cplus} projects, and $\DX$ is the fastest
in \UseMacro{fastest-all-count-d} projects. So, with the \specs that
we use, $\DX$ is only the fastest in \UseMacro{fastest-all-percent-d}\% 
of projects. Bucking expected trends, despite its complexity, $\DX$ is 
the fastest in \FastestDOneTwoCount{} projects (of \FastestOneTwoCount{} 
with the lowest overheads) in the first two deciles. So, we next analyze 
these speedups some more.

The plots at the bottom of Figure~\ref{fig:algorithm_comparison}
respectively show which of $\B$, $\CX$ and $\DX$ are the fastest
within 2s (left) and 5s (right) thresholds. That is, these two plots
consider all three algorithms to be ``same''---\ie, equally fast---if
the difference between the fastest algorithm and the mean of the three
other algorithms is less than or equal to 2s and 5s. ($\CX$ and $\CXX$
overheads are very similar, so we elide $\CXX$ from these two plots.)
Within a 2s threshold, 
\begin{minipage}[t]{.62\textwidth}
  \vspace{-2ex}
  \centering
  \scriptsize
  \setlength{\tabcolsep}{2.1pt}  
  \captionof{table}{Relative and absolute time overheads per algo}
  \vspace{-3ex}
\label{tab:overhead_analysis}
\resizebox{0.98\linewidth}{!}{
\begin{tabular}{lrrrr}
\hline
\textbf{Metric} & $\B$ & $\CX$ & $\CXX$ & $\DX$ \\
\hline
Relative Maximum (\x) & \UseMacro{algos_comparison_B_max_relative_overhead} & \UseMacro{algos_comparison_C_max_relative_overhead} & \UseMacro{algos_comparison_C+_max_relative_overhead} & \UseMacro{algos_comparison_D_max_relative_overhead} \\
Relative Minimum (\x) & \UseMacro{algos_comparison_B_min_relative_overhead} & \UseMacro{algos_comparison_C_min_relative_overhead} & \UseMacro{algos_comparison_C+_min_relative_overhead} & \UseMacro{algos_comparison_D_min_relative_overhead} \\
Relative Mean (\x) & \UseMacro{algos_comparison_B_mean_relative_overhead} & \UseMacro{algos_comparison_C_mean_relative_overhead} & \UseMacro{algos_comparison_C+_mean_relative_overhead} & \cellcolor{lightgray}\UseMacro{algos_comparison_D_mean_relative_overhead} \\
Relative Median (\x) & \UseMacro{algos_comparison_B_median_relative_overhead} & \UseMacro{algos_comparison_C_median_relative_overhead} & \UseMacro{algos_comparison_C+_median_relative_overhead} & \UseMacro{algos_comparison_D_median_relative_overhead} \\
\hline
Absolute Maximum (s) & \UseMacro{algos_comparison_B_max_absolute_overhead} & \UseMacro{algos_comparison_C_max_absolute_overhead} & \UseMacro{algos_comparison_C+_max_absolute_overhead} & \cellcolor{lightgray}\UseMacro{algos_comparison_D_max_absolute_overhead} \\
Absolute Minimum (s) & \UseMacro{algos_comparison_B_min_absolute_overhead} & \UseMacro{algos_comparison_C_min_absolute_overhead} & \UseMacro{algos_comparison_C+_min_absolute_overhead} & \UseMacro{algos_comparison_D_min_absolute_overhead} \\
Absolute Mean (s) & \UseMacro{algos_comparison_B_mean_absolute_overhead} & \UseMacro{algos_comparison_C_mean_absolute_overhead} & \UseMacro{algos_comparison_C+_mean_absolute_overhead} & \cellcolor{lightgray}\UseMacro{algos_comparison_D_mean_absolute_overhead} \\
Absolute Median (s) & \UseMacro{algos_comparison_B_median_absolute_overhead} & \UseMacro{algos_comparison_C_median_absolute_overhead} & \UseMacro{algos_comparison_C+_median_absolute_overhead} & \UseMacro{algos_comparison_D_median_absolute_overhead} \\
\hline
\end{tabular}
}
\vspace{1ex}
\end{minipage}
\begin{minipage}[t]{0.37\textwidth}
  $\B$, $\CX$ and $\DX$ have the same overhead in
  \UseMacro{fastest-count-2s-all-all} projects, $\B$ is fastest in 
  \UseMacro{fastest-count-2s-all-b} projects, $\CX$ is fastest in 
  \UseMacro{fastest-count-2s-all-c} projects, and $\DX$ is fastest in
  \UseMacro{fastest-count-2s-all-d} projects. Within a 5s threshold, $\B$, $\CX$ and
 \end{minipage}

\noindent %
$\DX$ have the same overhead in
\UseMacro{fastest-count-5s-all-all} projects, $\B$ is fastest in
\UseMacro{fastest-count-5s-all-b} projects, $\CX$ is fastest in
\UseMacro{fastest-count-5s-all-c} projects, and $\DX$ is fastest in
\UseMacro{fastest-count-5s-all-d} projects. So, if users do not mind 
2s or 5s slowdowns, the choice of algorithm does not matter for 59.1\%
and 68.1\% of these projects, respectively, using these \specs.  That is,
simpler algorithms work well for most of these Python projects and
these \specs.

\MyPara{Take-away messages}
We take three main messages away from these overhead
comparisons. First, future research is needed to reduce the high \RV
overhead that \Tool incurs on some
projects. Table~\ref{tab:overhead_analysis} shows summary statistics
on relative and absolute time overheads, which are as high as
\UseMacro{algos_comparison_D_max_relative_overhead}\x, or 3.69 hours
for $\DX$. Second, future research is needed to improve all four
algorithms and to find which algorithm is best for each project and
set of \specs. Third, and most importantly, \Tool can be the platform
of choice for such future research.

\vspace{-3ex}

\subsection{\rqInstrumentation: Comparing \Tool's instrumentation strategies}
\label{seq:eval:instrumentation}

\vspace{-1ex}

We compare \Tool's instrumentation strategies---\ie, monkey patching,
monkey patching plus curses, and monkey patching plus
AST~(\S~\ref{sec:tool:implementation})\Space{---w.r.t. time and number
of violations}.

\MyPara{Time comparison}
Table~\ref{table:instr:time} shows summary statistics about
instrumentation \begin{wraptable}{l}{0.56\textwidth}  
  \addtolength{\tabcolsep}{-0.15em}  
  \tiny
  \scriptsize
  \vspace{-8ex}
  \caption{Instrumentation times (s) per strategy.}
  \vspace{-3ex}
\label{table:instr:time}
\begin{tabular}{lccccc}
\toprule
\textbf{Strategy} & \textbf{Mean} & \textbf{Med} & \textbf{Max} & \textbf{Min} & \textbf{Sum} \\
\midrule
Monkey patching & \UseMacro{InstrTimeMeanMonkeypatching} & \UseMacro{InstrTimeMedianMonkeypatching} & \UseMacro{InstrTimeMaxMonkeypatching} & \UseMacro{InstrTimeMinMonkeypatching} & \UseMacro{InstrTimeSumMonkeypatching} \\ 
Monkey patching + curses & \UseMacro{InstrTimeMeanCurse} & \UseMacro{InstrTimeMedianCurse} & \UseMacro{InstrTimeMaxCurse} & \UseMacro{InstrTimeMinCurse} & \UseMacro{InstrTimeSumCurse} \\ 
Monkey patching + AST & \UseMacro{InstrTimeMeanAst} & \UseMacro{InstrTimeMedianAst} & \UseMacro{InstrTimeMaxAst} & \UseMacro{InstrTimeMinAst} & 9,204.04 \\ 
\bottomrule
\end{tabular}
\vspace{-8ex}
\end{wraptable}
\noindent time (in
seconds) across all \UseMacro{ProjectCountFilteredRQ2} projects for
which all instrumentation strategies show consistent results (AST
fails in 239 projects).\Space{ with the original ones \mh{we had to
remove ~200 projects for fair comparison: because AST is
problematic}.} There, monkey patching and its curses extension exhibit
similar execution times, while being significantly faster than the
AST-based approach. Comparing the sums, monkey-patching and
monkey-patching plus curses strategies were approximately
\InstrSpeedupFactor{} times faster than AST. This result is expected: 
AST involves building, parsing, and modifying the abstract syntax
tree---costs not incurred by the other strategies.
Results show that the strategies are comparable in monitoring time.

\MyPara{Comparison of violations} We count violations that \Tool
detects with each instrumentation strategy and group violations into
three categories, based
\begin{wraptable}{r}{0.4\textwidth}
  \vspace{-6ex}
  \scriptsize
\caption{\label{tab:violation-per-strategy}Violations by source
  location and strategy (MP = Monkey Patching, MP+A = Monkey 
  Patching with AST, MP+C = Monkey Patching with Curses).}
\vspace{-3ex}
\centering
\begin{tabular}{lrrr}
  \toprule
  & \multicolumn{3}{c}{Instrumentation} \\
\textbf{Source} & \textbf{MP} & \textbf{MP+A} & \textbf{MP+C} \\
\midrule
Python source & \UseMacro{text_TotalViolationsPythonSourceMonkeypatching} & \UseMacro{text_TotalViolationsPythonSourceAst} & \UseMacro{text_TotalViolationsPythonSourceCurse} \\
Dependencies & \UseMacro{text_TotalViolationsSitePackagesMonkeypatching} & \UseMacro{text_TotalViolationsSitePackagesAst} & \UseMacro{text_TotalViolationsSitePackagesCurse} \\ 
Project           & \UseMacro{text_TotalViolationsProjectUnderTestMonkeypatching} & \UseMacro{text_TotalViolationsProjectUnderTestAst} & \UseMacro{text_TotalViolationsProjectUnderTestCurse} \\
\bottomrule
\end{tabular}
\vspace{-8ex}
\end{wraptable}
\noindent
on the location of the violating event: (i)~Python source,
(ii)~site-packages (dependencies), and (3)~project under
test. Table~\ref{tab:violation-per-strategy} summarizes where
violations occur per instrumentation strategy. Overall, instrumenting
with monkey patching plus curses detects the most violations. We turn
to the question of the effectiveness, in terms of false positives and
false negatives, in
\rqEffectiveness

\MyPara{Take-away messages}
We think that all three strategies are stop gaps; the \RV community
should investigate better instrumentation strategies for Python that
could become what, \eg, AspectJ~\cite{kiczales2001aspectjoverview},
DiSL~\cite{marek2012disl}, or BISM~\cite{soueidi2023bridging,
soueidi2023instrumentation, soueidi2023efficient, soueidi2023dynamic}
are in the \RV for Java community. In the meantime, monkey patching
$+$ curses strikes a good compromise among the strategies; it is fast
and finds the most violations.  Also, monkey patching $+$ AST is most
useful in projects with high
\RV overheads, where the efficiency of the instrumented code offsets 
its instrumentation time, and where the goal is to find violations 
in the code under test.

\vspace{-3ex}

\subsection{\rqEffectiveness: \Tool's effectiveness for finding bugs}
\label{sec:eval:effectiveness}

\vspace{-1ex}

We report on our ongoing manual inspection of violations that
\Tool finds.

\MyPara{Inspection Process}
Manual inspection can take up to one person hour per \spec
violation~\cite{LegunsenETAL19SpecEval,
LegunsenETALASE2016SpecEval}. So, we select to
inspect \NumUniqViolations{} unique violations in
Table~\ref{tab:violation-per-strategy} (they are
from \NumProjectInspection{} projects); our goal is to confirm true
bugs with developers. Then, using procedure and classification scheme
from prior work on \RV for Java~\cite{LegunsenETALASE2016SpecEval,
LegunsenETAL19SpecEval, MirandaETAL19RVPrio}, we inspect each selected
violation to see if it is a ``True bug'', or ``False alarm''---\ie,
the \spec is violated, but that violation is not a bug in the CUT. We
inspect violations in CUT, 3rd-party library, or Python. Also,
\begin{wraptable}{r}{.45\textwidth}
\scriptsize
\centering
\vspace{-7ex}
\caption{\label{tab:inspection}Manual Inspection Summary.}
\vspace{-3ex}
\begin{tabular}{lccr}
\toprule
\textbf{Classification} & \textbf{Count} & \textbf{Unique} & \% \\
\midrule
True Positive        & \NumTruePositive{} & \NumTruePositiveUniq{} & 
\PercTruePositiveUniq{} \\
False Positive       & \NumFalsePositive{} & \NumFalsePositiveUniq{} & 
\PercFalsePositiveUniq{} \\
Difficult to Inspect & \NumDifficultInspect{} & \NumDifficultInspectUniq{} & 
\PercDifficultInspectUniq{} \\
$\Sigma$             & \NumAllViolations{} & \NumUniqViolations{} & 
100\% \\
\bottomrule
\end{tabular}
\vspace{-9ex}
\end{wraptable}
\noindent
at least two co-authors inspected each violation: one reviews the
other's work. For true bugs, we open a pull request (PR) if we know
how to fix the violations. Otherwise, we open a GitHub issue. We
estimate that we spent 200 person hours on average to inspect, repair,
and report these violations.

\MyPara{Inspection Results}
Table~\ref{tab:inspection} shows results of our inspections so far. We
inspect \NumUniqViolations{} unique violations that occur
\NumAllViolations{} times, \eg, if it is in a library that multiple
projects use. ``Difficult to Inspect'' means non-trivial effort beyond
our time budget is needed, or we cannot find the root cause. We open
PRs or issues for \NumTotalBugsFound{} unique ``True bugs''; we do not
report
\NumTrueBugsNotReported{}: the project is no longer maintained (\NumNotMaintained); the code changed
since the version we inspect (\NumCodeChanged); the violation was
already fixed (\NumFixedInLatest); the fix has no effect on program
behavior, performance, or correctness, \eg,
the \CodeIn{KeyInList} \spec monitors against checking membership of a
key in a list, which is less efficient than doing so with a
set~\cite{pythonMembershipTests}, but there is negligible effect with
a three-element list (\NumNoEffect); the violation is
intentional---\eg, as part of a ``negative''
test---(\NumCommittedOnPurpose); or a fix would require significant
code re-design (\NumDifficultToChange).
\begin{wraptable}{l}{0.48\textwidth}  
  \tiny
  \setlength{\tabcolsep}{1.5pt}
  \vspace{-9ex}
  \caption{\label{tab:pr-issues}Pull requests and issues reported per
    \spec. Highlighted rows show library \specs; others rows show Python
      \specs. O=Opened; A=Accepted; R=Rejected.}
  \vspace{-5ex}
  \centering
    \begin{tabular}{lccc}
      \toprule
      Spec & O & A & R \\
      \midrule
      KeyInList & 35+2 & 17 & 3+1 \\
      RandomMustUseSeed & 19+1 & 7 & 2 \\
      FileMustClose & 7+1 & 4+1 & 0 \\
      \UseMacro{PydocsMustShutdownBeforeCloseSocket} & 5+4 & 3+1 & 1+3 \\
      XMLParserParseMustFinalize & 3+1 & 3+1 & 0 \\
      \UseMacro{PydocsShouldUseStreamWriterCorrectly} & 2+1 & 1 & 0+1 \\
      \UseMacro{PydocsShouldNotInstantiateStreamWriter} & 2 & 1 & 0 \\
      \toctou\ & 2 & 1 & 0 \\
      \cellcolor{lightgray}HostnamesTerminatesWithSlash & 1+1 & 1+1 & 0 \\
      \UseMacro{PydocsMustShutdownThreadPoolExecutor} & 0+1 & 0+1 & 0 \\
      \cellcolor{lightgray}TfFunctionNoSideEffect & 1 & 1 & 0 \\
      PyDocsMustSortBeforeGroupBy & 1 & 0 & 1 \\
      PydocsMustCloseSocket & 1 & 0 & 0 \\
      BuiltinAllAnalysis & 2 & 0 & 0 \\
      \midrule
      \textbf{Total} & \textbf{81+12} & \textbf{\NumTotalPRAccepted{}+5} & \textbf{7+5} \\
      \bottomrule
    \end{tabular}
    \vspace{-9ex}
\end{wraptable}
\noindent
Table~\ref{tab:pr-issues} shows the pull requests we created and
issues we opened, grouped by the related \spec.\Space{ It is not
uncommon for the number of violated \specs to be a small fraction of
all \specs, \eg, only \Fix{A} of \Fix{180} \specs were violated
in~\cite{LegunsenETAL19SpecEval, LegunsenETALASE2016SpecEval}.} Our
appendix has a table with each violation, the project, the \spec, and
a link to a PR or issue.

\MyPara{Qualitative Analysis}
The \PercFalsePositiveUniq{} false positive rate in Table~\ref{tab:inspection} is
$\sim$2\x lower than the $\sim$90\% for
Java~\cite{LegunsenETAL19SpecEval, LegunsenETALASE2016SpecEval}. All
\NumFalsePositiveUniq{} false positives are due to one of three
reasons: (1)~Misunderstanding valid use cases\Space{ in the
project}~(e.g., setting a seed in a random generator is unnecessary as
reproducibility is not needed); (2)~Limited runtime visibility (e.g.,
missed monitoring file accesses from third-party library);
(3)~Overly-strict \spec\ because of lack of domain knowledge (e.g.,
custom sockets may not need to call shutdown before closing). These
problems are similar to those found in \RV \specs for
Java~\cite{LegunsenETALASE2016SpecEval, LegunsenETAL19SpecEval}; we
will continue to improve \Tool's \specs.

\vspace{-3ex}

\subsection{\rqEfficiency: \Tool vs. \dynapyt and \dylin}
\label{seq:eval:efficiency}

\vspace{-1ex}

We compare \Tool, \dynapyt (with, w/o libraries), and \dylin (w/o
libraries)\Space{ w.r.t. overhead and violations}.

\MyPara{Setup}
We use $\DX$ with monkey-patching plus curses for \Tool, and 22
\specs. 19 of these are single-parameter specs that we can easily
translate to \dynapyt and \dylin; others cannot be translated due to
usage of idioms that are incompatible with \dylin, or usage of
multiple parameters and formal logic. We have contributed these 19
\specs to \dylin~\cite{dylin-pr2}. So far, we have translated 3 of 
\dylin's 8 available checkers (out of their total 15 checkers) that 
are suitable for this experiment (\dylin was only recently published). 
The reasons for not considering the other 7 are: they rely on new \dylin 
functionality that is not supported in \dynapyt (2 checkers), they induce 
circular references because the checkers use 3rd-party libraries (4 checkers), 
and we cannot implement them consistently between \Tool and \dylin 
(1 checker).

\dynapyt and \dylin were only previously evaluated on
CUT~\cite{eghbali2022dynapyt, Eghbali2025DyLin}, since they target
dynamic linting of developer code. With input from their
authors~\cite{AryazDynaPytLib2025}, we implement support for
using \dynapyt to also check 3rd-party libraries. Neither \dynapyt
nor \dylin can monitor Python sources like \Tool does. The reason is
that both tools rely on source code instrumentation, which is risky
for Python source code: the modifications are not reversible and often
lead to unexpected errors and infinite circular references. Also, we
have so far been unable to add 3rd-party library support to \dylin,
because the
\dynapyt API that we use for doing so is now used internally by \dylin
in such a way that the suggested \dylin workflow no longer exposes
this API to the user.

\begin{figure}[!t]
    \centering
    \begin{tikzpicture}
      \begin{axis}[
          width=12cm,
          height=4cm,
          ybar,
          bar width=2.7pt,
          enlargelimits=0.05,
          legend style={
              at={(0.01,1.10)},
              anchor=north west,
              legend columns=4,
              font=\scriptsize,
              draw=black,
              fill=white,
              nodes={scale=0.7, transform shape}
          },
          ylabel={Absolute Overhead},
          ylabel style={font=\scriptsize},
          ymax  = 300000,
          ymode=log,
          symbolic x coords={10, 20, 30, 40, 50, 60, 70, 80, 90, 100},
          xtick=data,
          x tick label style={font=\scriptsize, anchor=north},
          ymajorgrids=true,
          yminorgrids=true
        ]
  
        \addplot+[ybar, black, fill=pink!70] plot coordinates {
          (10,\UseMacro{Comparison-7-algo-D-p1-overhead-sum})
          (20,\UseMacro{Comparison-7-algo-D-p2-overhead-sum})
          (30,\UseMacro{Comparison-7-algo-D-p3-overhead-sum})
          (40,\UseMacro{Comparison-7-algo-D-p4-overhead-sum})
          (50,\UseMacro{Comparison-7-algo-D-p5-overhead-sum})
          (60,\UseMacro{Comparison-7-algo-D-p6-overhead-sum})
          (70,\UseMacro{Comparison-7-algo-D-p7-overhead-sum})
          (80,\UseMacro{Comparison-7-algo-D-p8-overhead-sum})
          (90,\UseMacro{Comparison-7-algo-D-p9-overhead-sum})
          (100,\UseMacro{Comparison-7-algo-D-p10-overhead-sum})
        };
        \addplot+[ybar, black, fill=blue!70] plot coordinates {
          (10,\UseMacro{Comparison-7-algo-dynapyt-p1-overhead-sum})
          (20,\UseMacro{Comparison-7-algo-dynapyt-p2-overhead-sum})
          (30,\UseMacro{Comparison-7-algo-dynapyt-p3-overhead-sum})
          (40,\UseMacro{Comparison-7-algo-dynapyt-p4-overhead-sum})
          (50,\UseMacro{Comparison-7-algo-dynapyt-p5-overhead-sum})
          (60,\UseMacro{Comparison-7-algo-dynapyt-p6-overhead-sum})
          (70,\UseMacro{Comparison-7-algo-dynapyt-p7-overhead-sum})
          (80,\UseMacro{Comparison-7-algo-dynapyt-p8-overhead-sum})
          (90,\UseMacro{Comparison-7-algo-dynapyt-p9-overhead-sum})
          (100,\UseMacro{Comparison-7-algo-dynapyt-p10-overhead-sum})
        };
        \addplot+[ybar, black, fill=green!70] plot coordinates {
          (10,\UseMacro{Comparison-7-algo-dylin-p1-overhead-sum})
          (20,\UseMacro{Comparison-7-algo-dylin-p2-overhead-sum})
          (30,\UseMacro{Comparison-7-algo-dylin-p3-overhead-sum})
          (40,\UseMacro{Comparison-7-algo-dylin-p4-overhead-sum})
          (50,\UseMacro{Comparison-7-algo-dylin-p5-overhead-sum})
          (60,\UseMacro{Comparison-7-algo-dylin-p6-overhead-sum})
          (70,\UseMacro{Comparison-7-algo-dylin-p7-overhead-sum})
          (80,\UseMacro{Comparison-7-algo-dylin-p8-overhead-sum})
          (90,\UseMacro{Comparison-7-algo-dylin-p9-overhead-sum})
          (100,\UseMacro{Comparison-7-algo-dylin-p10-overhead-sum})
        };
        \addplot+[ybar, black, fill=red!70] plot coordinates {
          (10,\UseMacro{Comparison-7-algo-dynapyt_libs-p1-overhead-sum})
          (20,\UseMacro{Comparison-7-algo-dynapyt_libs-p2-overhead-sum})
          (30,\UseMacro{Comparison-7-algo-dynapyt_libs-p3-overhead-sum})
          (40,\UseMacro{Comparison-7-algo-dynapyt_libs-p4-overhead-sum})
          (50,\UseMacro{Comparison-7-algo-dynapyt_libs-p5-overhead-sum})
          (60,\UseMacro{Comparison-7-algo-dynapyt_libs-p6-overhead-sum})
          (70,\UseMacro{Comparison-7-algo-dynapyt_libs-p7-overhead-sum})
          (80,\UseMacro{Comparison-7-algo-dynapyt_libs-p8-overhead-sum})
          (90,\UseMacro{Comparison-7-algo-dynapyt_libs-p9-overhead-sum})
          (100,\UseMacro{Comparison-7-algo-dynapyt_libs-p10-overhead-sum})
        };
  
        \legend{
          \strut \Tool,
          \strut \dynapyt,
          \strut \dylin,
          \strut \dynapyt (with lib),
        }
      \end{axis}
    \end{tikzpicture}
    \vspace{-2ex}
    \caption{Log scale (base 10) plot of distributions of time
      overhead per decile for \UseMacro{Comparison-7-total-projects}
      projects using \Tool, \dylin w/o libraries, and \dynapyt (with
      and w/o libraries).}
    \vspace{-6ex}
      \label{fig:dynapyt_dylin_pymop_distribution_plot}
  \end{figure}

\MyPara{Results}
Table~\ref{tab:dynapyt_dylin_pymop_time_table} shows absolute times
summed across all \UseMacro{Comparison-7-total-projects} projects that
we use in this RQ. These are the projects where all four
configurations in this RQ
\begin{wraptable}{r}{0.63\textwidth}
  \centering
  \scriptsize
  \vspace{-8ex}
  \caption{\Tool, \dynapyt, and \dylin time (s).}
  \vspace{-3ex}
  \begin{tabular}{lcccccc}
  \toprule
  & \Tool & \dynapyt & \dylin & \dynapyt (libs) \\
  \midrule
  \textbf{Instrumentation} & \pgfmathprintnumber[precision=0]{\UseMacro{Comparison-7-algo-D-instrumentation_time-sum}} & \pgfmathprintnumber[precision=0]{\UseMacro{Comparison-7-algo-dynapyt-instrumentation_time-sum}} & \pgfmathprintnumber[precision=0]{\UseMacro{Comparison-7-algo-dylin-instrumentation_time-sum}} & \pgfmathprintnumber[precision=0]{\UseMacro{Comparison-7-algo-dynapyt_libs-instrumentation_time-sum}} \\
  \textbf{Monitoring} & \pgfmathprintnumber[precision=0]{\UseMacro{Comparison-7-algo-D-test_duration-sum}} & \pgfmathprintnumber[precision=0]{\UseMacro{Comparison-7-algo-dynapyt-test_duration-sum}} & \pgfmathprintnumber[precision=0]{\UseMacro{Comparison-7-algo-dylin-test_duration-sum}} & \pgfmathprintnumber[precision=0]{\UseMacro{Comparison-7-algo-dynapyt_libs-test_duration-sum}} \\
  \textbf{End-to-End} & \pgfmathprintnumber[precision=0]{\UseMacro{Comparison-7-algo-D-end_to_end_time-sum}} & \pgfmathprintnumber[precision=0]{\UseMacro{Comparison-7-algo-dynapyt-end_to_end_time-sum}} & \pgfmathprintnumber[precision=0]{\UseMacro{Comparison-7-algo-dylin-end_to_end_time-sum}} & \pgfmathprintnumber[precision=0]{\UseMacro{Comparison-7-algo-dynapyt_libs-end_to_end_time-sum}} \\
  \bottomrule
  \end{tabular}
  \label{tab:dynapyt_dylin_pymop_time_table}
  \vspace{-9ex}
\end{wraptable}  
\noindent
succeed; \dynapyt with libraries fails on the others.
In Table~\ref{tab:dynapyt_dylin_pymop_time_table}, we show
instrumentation-only, \seqsplit{monitoring}-only, and end-to-end
times; the later is the sum of the former two. \Tool has the lowest
instrumentation-only and end-to-end times. Also, direct comparison of
monitoring-only times is unfair to \Tool, since it checks more code
than \dynapyt and \dylin in all their configurations. Yet, \Tool's
monitoring-only time is faster than both configurations of \dynapyt,
and is only \UseMacro{Comparison-7-D-DyLin-test-duration-mean-diff} 
seconds slower on average than the monitoring-only time of \dylin 
(which monitors only CUT).

Figure~\ref{fig:dynapyt_dylin_pymop_distribution_plot} shows the
distribution of end-to-end time overheads among these four techniques,
grouped by decile in ascending order of \dynapyt with libraries
overhead.  There, the y-axis is shown in log scale (base 10). \Tool
has the lowest absolute overhead across all deciles. Even without
monitoring libraries and Python source code, \dynapyt and \dylin have
similar overheads that are much higher than \Tool's. That slowdown
gets much worse when \dynapyt is used to check 3rd-party libraries (but not
Python source). Overall, \Tool is much more efficient than \dynapyt
and \dylin, and our use of generic monitoring algorithms and
lightweight instrumentation seems to be paying off.

\MyPara{\Tool vs. \dylin comparison of violations}
The Venn diagrams in
Figure~\ref{fig:venn-pymop-dynapyt-dylin-violation-level} show sets of
violations from \Tool, \dylin, and \dynapyt in this RQ. When \dylin
and \dynapyt do not check libraries, the first plot shows all
violations, while the second shows violations after removing
violations that \Tool finds in libraries and the Python
source. Here, \Tool
finds \UseMacro{comparison-statistics-D-Dynapyt-D-violations}
violations that these tools miss, and seven of those missed
violations are in the CUT. The third plot shows all violations from
\Tool and \dynapyt with library (but not the Python
source) checking, and the fourth plot shows the result of filtering
out those violations that \Tool finds in the Python source. Here, each
tool finds violations that the other misses. \Tool finds 
\UseMacro{comparison-statistics-D-Dynapyt_libs-D-violations}
violations that \dynapyt with libraries miss, four of which are
in Python sources. But, \dynapyt with library checking also
finds \UseMacro{comparison-statistics-D-Dynapyt_libs-DynaPyt_libs-violations} 
violations that \Tool misses.

\dylin and \dynapyt miss violations because of a bug: they do not record some
violations. We have reported this bug~\cite{dylin-issue3}. All
violations that \Tool misses occur because \Tool only starts
monitoring after it is loaded by \CodeIn{pytest}, so it misses events
occurring before that; a \CodeIn{pytest} limitation. In the future, we
plan to switch from \Tool as a \CodeIn{pytest} plugin to
using \CodeIn{sitecustomize.py}~\cite{pythonSitecustomize}. Doing so 
will trigger monitoring earlier, and enable \Tool to work with any test 
framework or even in deployment.
7 of all \UseMacro{comparison-statistics-D-Dynapyt-D-violations}
violations that \Tool finds but \dynapyt or \dylin miss are true bugs;
6 of \UseMacro{comparison-statistics-D-Dynapyt_libs-D-violations}
violations that these tools find but \Tool misses are true
bugs. Overall, better instrumentation strategies and multiple
Python \RV systems are needed: without the ``differential testing''
that we do, we would not have found these missed violations.

\begin{figure}[t!]
    \captionsetup[subfigure]{justification=centering}
    \scriptsize
    \vspace{-6ex}
    \begin{subfigure}[b]{\textwidth}
        \centering
        \begin{tikzpicture} [scale=0.25]
            \draw[fill=blue!20, opacity=0.6] (0,0) circle (2.0cm);
            \draw[fill=orange!20, opacity=0.6] (2,0) circle (2.0cm);
            \node at (-1.8,3) {\textbf{\Tool}};
            \node at (4.0,3.5) {\textbf{\dynapyt}};
            \node at (4.0,2.5) {\textbf{\dylin}};
            \node at (-0.7,0) {\UseMacro{comparison-statistics-D-Dynapyt-D-violations}};
            \node at (2.7,0) {\UseMacro{comparison-statistics-D-Dynapyt-DynaPyt-violations}};
            \node at (1,0) {\UseMacro{comparison-statistics-D-Dynapyt-same-violations}};
        \end{tikzpicture}
        \label{fig:venn-pymop-dynapyt-dylin-violation-level}
        \begin{tikzpicture} [scale=0.25]
            \draw[fill=blue!20, opacity=0.6] (0,0) circle (2.0cm);
            \draw[fill=orange!20, opacity=0.6] (2,0) circle (2.0cm);
            \node at (-1.8,3) {\textbf{\Tool}};
            \node at (4.0,3.5) {\textbf{\dynapyt}};
            \node at (4.0,2.5) {\textbf{\dylin}};
            \node at (-0.7,0) {\UseMacro{comparison-statistics-D-Dynapyt-D-violations-filtered}};
            \node at (2.7,0) {\UseMacro{comparison-statistics-D-Dynapyt-DynaPyt-violations-filtered}};
            \node at (1,0) {\UseMacro{comparison-statistics-D-Dynapyt-same-violations-filtered}};
        \end{tikzpicture}
        \label{fig:venn-pymop-dynapyt-dylin-violation-level-filtered}
        \begin{tikzpicture} [scale=0.25]
            \draw[fill=blue!20, opacity=0.6] (0,0) circle (2.0cm);
            \draw[fill=orange!20, opacity=0.6] (2,0) circle (2.0cm);
            \node at (-1.8,3) {\textbf{\Tool}};
            \node at (4.0,3.5) {\textbf{\dynapyt}};
            \node at (4.0,2.5) {\textbf{(libs)}};
            \node at (-0.7,0) {\UseMacro{comparison-statistics-D-Dynapyt_libs-D-violations}};
            \node at (2.7,0) {\UseMacro{comparison-statistics-D-Dynapyt_libs-DynaPyt_libs-violations}};
            \node at (1,0) {\UseMacro{comparison-statistics-D-Dynapyt_libs-same-violations}};
        \end{tikzpicture}
        \label{fig:venn-pymop-dynapyt-libs-violation-level}
        \begin{tikzpicture} [scale=0.25]
            \draw[fill=blue!20, opacity=0.6] (0,0) circle (2.0cm);
            \draw[fill=orange!20, opacity=0.6] (2,0) circle (2.0cm);
            \node at (-1.8,3) {\textbf{\Tool}};
            \node at (4.0,3.5) {\textbf{\dynapyt}};
            \node at (4.0,2.5) {\textbf{(libs)}};
            \node at (-0.7,0) {\UseMacro{comparison-statistics-D-Dynapyt_libs-D-violations-filtered}};
            \node at (2.7,0) {\UseMacro{comparison-statistics-D-Dynapyt_libs-DynaPyt_libs-violations-filtered}};
            \node at (1,0) {\UseMacro{comparison-statistics-D-Dynapyt_libs-same-violations-filtered}};
        \end{tikzpicture}
        \label{fig:venn-pymop-dynapyt-libs-violation-level-filtered}
    \end{subfigure}
    \vspace{-5ex}
        \caption{Sets of \Tool, \dynapyt, and \dylin violations in
          \rqEfficiency, explained in text.}
    \vspace{-7ex}
    \label{fig:venn-pymop-dynapyt-dylin-violation-level}
\end{figure}

\vspace{-3ex}

\section{Discussion}
\label{sec:discussion}

\vspace{-1ex}

\MyParaOnly{Is \Tool the only needed Python \RV system?}
No! We develop \Tool because (i)~despite recent success of work that
used \javamop for simultaneous monitoring of scores of \specs during
testing in many projects, we cannot find a similar system for Python;
(ii)~the lack of such Python \RV systems is a big gap in today's
AI-driven and data-intensive computing landscape; and (iii)~we
developed other MOP-based \RV systems~\cite{YorihiroETAleMOP2023, TraceMOP,
GuanAndLegunsenTraceMOPFSEDemo2025, eMOPWebPage}. But, \emph{there are
many other styles of \RV beyond MOP}, \eg,~\cite{EventBasedRV, PQL,
bonakdarpour2013time, WuETALReducingMonitoringOverheadRV2013,
RegerETALTACAS15MarQ, HavelundETALFMCAD2017MonitoringWithBDDs,
BarringerETALVMCAI2004Eagle, BarringerETALJLC2010EagleToRuler,
DeckerETALTACAS2016MonitoringWithUnionFind, ho2014online, d2005lola,
basin2017runtime, bataineh2019efficient, ForejtETALIncrementalRV2012,
Renberg2014PythonRVKTHThesis, decker2013junit, kim1999formally,
ErlingssonAndSchneiderIRM2000}. \RV's practical impact will be greatly enhanced
by having more robust and open-source Python systems for these
other \RV styles as well.

\MyPara{\Tool's offline \RV support}
\Tool enables offline \RV, but our main evaluation focuses on online
algorithms for use in testing. We conduct a small experiment to get a
sense of \Tool's offline \RV capability. To do so, we compare the time
of \Tool's offline $\A$ algorithm (\S\ref{sec:tool:algos}) with those
of online algorithms implemented in \Tool ($\B$, $\CX$, $\CXX$, and
$\DX$), using \NumSpecs{} \specs
and \UseMacro{algo_a_comparison_total_projects} projects from the last
decile in \S\ref{seq:eval:algos}, where the online algorithms show the
largest overhead and noticeable differences among them. The results
show that the average relative overhead of $\A$
is \UseMacro{algo_a_comparison_A_relative_overhead_mean}\x, while the
average relative overhead of $\B$
is \UseMacro{algo_a_comparison_B_relative_overhead_mean}\x, $\CX$
is \UseMacro{algo_a_comparison_C_relative_overhead_mean}\x, $\CXX$
is \UseMacro{algo_a_comparison_C+_relative_overhead_mean}\x, and $\DX$
is \UseMacro{algo_a_comparison_D_relative_overhead_mean}\x. So, online
algorithms are more efficient than \Tool's offline algorithm.

\MyPara{Impact of monitor garbage collection (MGC)}
We compare $\DX$'s times with and without MGC
(\S\ref{sec:tool:gc:algorithm}),
using \num{\UseMacro{gc_comparison_macros_total_projects}} projects\Space{
and \NumSpecsFromPyMOP\ \specs}. \UseMacro{gc_comparison_macros_gc_faster_endtoend_projects}
of these projects are faster with MGC, the
other \UseMacro{gc_comparison_macros_no_gc_faster_endtoend_projects}
are slower. With a 0.5 seconds tolerance for time equality, MGC has no
impact on \num{\UseMacro{gc_comparison_macros_same_speed_0.5s_projects}}
projects. Among the
other \UseMacro{gc_comparison_macros_different_speed_0.5s_projects}
projects, only \UseMacro{gc_comparison_macros_gc_faster_0.5s_projects}
are faster with MGC. As expected, projects in which many monitors for
multi-object \specs are instantiated at runtime tend to see more
speedups from MGC. So, as more of these kinds of \specs are added
to \Tool in the future, MGC will be important for reducing its
overhead.

\Space{
\MyParaOnly{Impact of test coverage}
Guan and Legunsen~\cite{GuanAndLegunsenRVStudyISSTA2024} found no
strong correlation between \RV overhead for Java and coverage. We use
PynGuin~\Fix{cite} to generate additional unit tests for \Fix{A}
projects from Table~\ref{tab:program-charac} with low statement (avg:
\Fix{B}\%) and branch (avg: \Fix{C}\%) coverage. Then, we re-monitor
all \Fix{82} \specs on developer written and automatically generated
tests. Average statement and branch coverage became \Fix{D}\% and
\Fix{E}\%, respectively. In turn, \Tool's overhead grew by an average
of \Fix{F}\%, finding \Fix{G} more violations. So, \Fix{interpret
  results}. There is little research beyond Artho et al.'s early
work~\cite{artho2005combining, artho2003experiments} on test
generation for \RV, and none on Python or in CI. \Tool can be the
basis of such future work.
}

\MyParaOnly{Why not static analysis?}
\S\ref{sec:example} discusses how static analysis may not detect
violations that require inter-procedural analysis or multi-type
reasoning. Also, some violations had traces with some events in a
program and some events in a 3rd-party library; static analysis may
not scale well for these. But, static analysis and \RV are orthogonal
and complementary~\cite{ChenAndRosuMOPOOPSLA2007,
BoddenEtAlAheadOfTime08, DwyerETALOptimizing2010,
PurandareEtAlStutterEquivalentLoops2010,
BoddenETAlStagedProgramAnalysisECOOP2007, RVCryptoAPIs,
hinrichs2014model}. In fact, \dylin authors show that Python static
and dynamic analysis are complementary~\cite{Eghbali2025DyLin}.

\MyPara{Limitations and future work}
We only evaluate \Tool during testing. Future work should investigate
the suitability of \Tool for monitoring deployed Python
programs. \Tool currently does not yet have monitor-synthesis plugins
for many \spec logics, \eg,~\cite{AllenLTLMonitorSynthesis,
MeredithRosuMonitoringStringRewritingASE13,
PrasannaAndRosuMTLMonitoringRV2004, schneider2019formally,
ho2014online, d2005lola, basin2017runtime}. We plan to continue adding
more plugins. \Tool is efficient, but there is room to optimize it,
given the high overheads in some projects. Two potential directions:
(i)~optimized code generation of monitoring algorithms (as done
in \javamop) could provide speedups; and (ii)~evolution-aware
techniques for Python could speed up \RV by focusing on code affected
by changes during CI~\cite{LegunsenETALeMOPICST2019,
LegunsenEtAleMOP15, YorihiroETAleMOP2023,
GuanAndLegunsenRVStudyISSTA2024, GuanAndLegunsenIMOPICSE2025}. We do
not yet study the ease of writing \Tool specs in Python or its
JSON-like frontend. Future work can conduct user studies like we did
for Java \RV \specs~\cite{teixeira2021demystifying}.

\MyPara{Threats to validity}
To mitigate the threat due to the kinds of \specs, we follow prior \RV
work's approach to writing \specs in different logics, APIs, domains,
and frameworks (Python or libraries). There may be bugs in our \specs,
so we peer review all our \specs and write at least two tests
per \spec---one violating and another not. These tests
are part of \Tool. Another set of projects could show different
results. To mitigate this threat, we use a large corpus to
evaluate \Tool, with \NumProjects{} GitHub \oss (instead of benchmarks
similar to DaCapo~\cite{BlackburnETALDaCaPoOOPSLA2006}). We lack
domain expertise on the evaluated projects, so we open pull requests
and issues to check our findings about bugs with developers.

\vspace{-3ex}

\section{Related Work}
\label{sec:related}

\vspace{-1ex}

Several surveys and competitions showcase the impressive progress on
\RV research over the last few decades~\cite{BartocciEtAlIntroToRV2018,
  FalconeEtAlTutorialOnRV2013, FalconeEtAlTaxonomyOfRVToolsRV2018,
  LeuckerAndSchallhartBriefAccountofRVFLACOS2007, RVCompetition2015,
  RVCompetition2016, RVCompetitionTACAS,
  RVCompetitionJournalPaper2019, taleb2023uncertainty,
  sanchez2019survey}. Some main directions in \RV research are
  (i)~developing monitoring
  algorithms~\cite{MeredithRosuMonitoringStringRewritingASE13,
  MeredithETALMOPContextFreePatterns08,
  HavelundRosuASE2001MonitoringProgramsUsingRewriting,
  HavelundRosuTACAS2002SynthesizingMonitors, ho2014online,
  PrasannaAndRosuMTLMonitoringRV2004, kim1999formally,
  LTLMonitorSynthesis, AllenLTLMonitorSynthesis,
  bauer2012decentralised, bataineh2019efficient}; (ii)~theoretical
  analysis of monitorability~\cite{rosu2012safety,
  schneiderSecurityAutomata, aceto2019operational,
  aceto2019adventures, henzinger2020monitorability,
  bauer2010monitorability, havelund2023monitorability,
  francalanza2017foundation, diekert2014topology, falcone2012can};
  (iii)~speeding up
\RV~\cite{BoddenETAlStagedProgramAnalysisECOOP2007,
  JinETALScalableParametricMonitoringTR2012,
  PurandareEtAlCompactionISSTA2013, JinEtAlGarbageCollectionPLDI2011,
  BoddenEtAlAheadOfTime08, DwyerETALOptimizing2010,
  LuoETAlRVMonitor14, WuETALReducingMonitoringOverheadRV2013,
  PurandareEtAlStutterEquivalentLoops2010,
  DeckerETALTACAS2016MonitoringWithUnionFind, chen2009efficient};
  (iv)~\RV for new domains~\cite{HusseinETAlSecurityMonitoringPLAS12,
  DeckerETALTACAS2016MonitoringWithUnionFind, basin2015failure,
  ganguly2021distributed, gangulyEtAlMTLCrossChainProtocolsICDCS2022,
  ganguly2024distributedJPDC, yaseen2020aragog, sen2004efficient,
  RVCryptoAPIs, francalanza2018runtime, danielsson2019decentralized,
  gorostiaga2018striver, shukla2019runtime}; (v)~different \RV
  styles~\cite{RegerETALTACAS15MarQ,
  HavelundETALFMCAD2017MonitoringWithBDDs,
  BarringerETALVMCAI2004Eagle, EventBasedRV, PQL,
  ErlingssonAndSchneiderIRM2000, AllanEtAlTraceMatchingOOPSLA2005,
  BarringerETALJLC2010EagleToRuler, feasibleTraceMonitors, d2005lola,
  basin2017runtime, huang2022temporal,
  RunsAndGodefoidTemporalLogicQueriesLICS2001, ogale2007detecting};
  and (vi)~systems~\cite{KaraormanAndFreemanJMonitor2004,
  YorihiroETAleMOP2023, JinEtAlJavaMOPToolPaperICSE12,
  BoddenMOPBOxRV11, ArnoldEtAlQVM2008, ChenAndRosuTowardsMOP03,
  HavelundRosuRV2001JPAX}. \Tool fills a gap as the first instance of
  MOP for Python. This huge body of work on \RV helped us
  develop \Tool, and will play a big role in its future improvement.
Many recent works are on \RV for AI-based or data-intensive systems,
\eg,~\cite{an2020uncertainty, torpmann2025runtime, desai2017combining,
  liu2024runtime, yang2024case, zapridou2020runtime, zhang2025rvllm,
  luo2018runtime, dong2015runtime, brown2025perception,
  cheng2023runtime, rahman2021run, grieser2020assuring}.\Space{ Since
  most of the systems in this contemporary space are written in
  Python, \Tool can offer a new basis and perspective for
  investigating new \RV techniques for this domain that increasingly
  important in today's world.} In a sense, \Tool brings the MOP
approach up to date for the AI age\Space{, and we are excited by the
  research that it will enable in doing so}.

\vspace{-3ex}
\section{Conclusion}
\label{sec:conclusion}
\vspace{-1ex}

\Tool is the first MOP-based \RV system for Python. It
is \emph{generic}, supporting five monitoring algorithms, three
instrumentation strategies, and five \spec logics
(with \NumSpecs{} \specs\ written), with features to add more of
these. \Tool is also \emph{efficient} as our large-scale evaluation
with
\NumProjects{} projects shows. \Tool has helped us find \NumTotalIssuesAndPrAccepted{}
confirmed bugs during testing of \oss. We outline an agenda of future
research and development that \Tool can enable.

\bibliographystyle{splncs04}
\bibliography{ref}
\end{document}